\pdfoutput=1
\documentclass[a4paper,fleqn]{cas-sc}
\usepackage[authoryear]{natbib}
\usepackage{amsmath,amssymb}
\usepackage{float}
\usepackage{placeins}
\usepackage{algorithm}
\usepackage{algpseudocode}
\usepackage{adjustbox}
\usepackage{booktabs}
\usepackage[table]{xcolor}
\usepackage{graphicx}

\begin{document}
\let\WriteBookmarks\relax
\shorttitle{Chance-Constrained Selection from Counterfactual Estimates}
\shortauthors{M. Kim and B. Jang}
\title[mode=title]{Chance-constrained selection of sequential intervention strategies from counterfactual estimates}

\author[1]{Minkyoung Kim}[orcid=0009-0001-3530-0885]
\ead{minky@yonsei.ac.kr}
\author[1]{Beakcheol Jang}[orcid=0000-0002-3911-5935]
\cormark[1]
\ead{bjang@yonsei.ac.kr}
\affiliation[1]{organization={Graduate School of Information, Yonsei University},
                addressline={50 Yonsei-ro, Seodaemun-gu},
                city={Seoul},
                postcode={03722},
                country={Republic of Korea}}
\cortext[1]{Corresponding author}
\begin{abstract}
Many operational decisions are sequences of interventions under a cumulative resource limit, such as a maintenance schedule within a crew-hour budget. Choosing among them calls for the outcome and the cumulative cost each would produce, counterfactual quantities identified from observational data. Two strategies with the same expected cost can exceed the budget at very different rates, so constraining the mean does not bound how often an overrun occurs. Prior two-step architectures, recently extended to continuous doses, constrain the mean cost rather than its tail and allocate at a single decision point. Methods that do bound a cost tail take its distribution from a specified model rather than identifying it from data. We present a predict-then-optimize framework. In the prediction step, any estimator returning an outcome value and a cost distribution supplies what the decision rule consumes, so the predictor is interchangeable. In the optimization step, a chance-constrained selection over a finite candidate set bounds the probability that the cumulative cost exceeds the budget. That tail does not decompose across stages, so each strategy is scored whole. Sweeping the tolerated violation probability traces a safety-utility frontier, and distribution-free finite-sample bounds cover violation and outcome shortfall. Four of five environments, spanning clinical treatment and equipment maintenance, supply exact counterfactual ground truth; the fifth carries real outcomes from a digital-health micro-randomized trial. Across them, the rule holds the budget where a point-estimate rule overruns it, at an outcome cost the frontier makes explicit. All code is available at \url{https://github.com/mfriendly/counterfactual-chance-selection}.
\end{abstract}
\begin{keywords}
Analytics \sep Prescriptive analytics \sep Causal inference \sep Dynamic treatment regimes \sep Chance-constrained programming
\end{keywords}

\newcommand{\num}[2]{#2}

\maketitle
\suppressfloats[t]
\section{Introduction}
\label{sec:introduction}

Many operational decisions are not single choices but sequences of interventions carried out under a cumulative resource budget. A maintenance planner schedules overhauls across an equipment fleet, a firm times retention offers to its customers, and a clinician administers treatments to a patient over a stay. In each setting, the cost of a strategy is a random variable, so a constraint on its mean leaves the probability of an overrun uncontrolled. Past interventions were chosen in response to the state, and the cost is observed only for the strategy actually followed. Choosing among the rest turns on counterfactual quantities identified from observational data. Table~\ref{tab:applications} sets out these settings and the form the budget takes in each.

\begin{table}[pos=t]
\centering
\small
\caption{Sequential intervention settings in which a cumulative treatment cost is constrained by a budget.}
\label{tab:applications}
\begin{adjustbox}{max width=\textwidth}
\begin{tabular}{@{}lllll@{}}
\toprule
Setting & Outcome & Intervention & Cumulative cost & Source of spread \\
\midrule
Maintenance & Availability & Overhaul & Crew-hours & Time-to-failure heterogeneity \\
Retention & Customer retained & Offer & Offer spend & Response heterogeneity \\
Clinical & Survival & Treatment & Cumulative exposure & Stochastic patient trajectory \\
Digital health & Engagement & Notification & Notifications sent & Randomization over the horizon \\
\bottomrule
\end{tabular}
\end{adjustbox}
\begin{flushleft}\footnotesize
Three of these settings are instantiated in Section~\ref{sec:experiments}.
\end{flushleft}
\end{table}

In each case, the resource a strategy consumes accumulates over the horizon. That resource need not be monetary. It may be a physical quantity such as crew-hours, or a cumulative exposure such as a lifetime dose limit. The interventions a strategy administers depend on the trajectory the unit follows under it. Two strategies with the same expected cost can therefore exceed the budget at very different rates. The experiments show this directly on a set of strategies whose mean costs all lie inside the budget. Several of these strategies place more than a fifth of their cost mass above the budget. The operational question is which strategy maximizes the outcome while holding the probability of an overrun within a tolerated level. Each intervention changes the state that determines the next one. A criterion on the tail, accumulation over a horizon, and identification from observational data all follow from that single fact rather than from three independent modeling choices.

Budget-constrained prescriptive analytics has posed this decision at a single point in time. \citet{devos2026uplift} allocate continuous treatment doses under a budget on total treatment cost, separating a prediction step from an optimization step. They name the extension to sequential decisions and the relaxation of their deterministic-cost assumption as the directions their framework leaves open. Both describe the setting above, in which the horizon is multi-stage and the cost accumulated over it is random.

Two further lines of work address parts of this setting, and neither covers all of it. One takes its cost distribution from a known model; the other constrains the mean rather than the tail. Sequential decision making under a resource constraint has a risk-neutral form and a risk-sensitive form. The risk-neutral form is the constrained Markov decision process, which bounds the expected cumulative cost of a policy \citep{altman1999constrained}. The risk-sensitive form replaces the expected cost with a chance constraint or a conditional-value-at-risk constraint \citep{chow2017risk}. That work assumes an online or simulator-based regime in which the transition dynamics are known or can be sampled. Here the cost distribution must instead be identified from observational trajectories in which treatment is confounded with prognosis. Policy learning under a resource constraint couples estimation with a downstream budgeted decision \citep{sakaguchi2025dewm, atheywager2021policy, liu2024cbr}. Its budgeted instances, however, are either single-stage or, when sequential, constrain the expected cost rather than its tail. In the taxonomy of the contextual-optimization survey of \citet{sadana2025contextual}, that literature is organized around an uncertain objective evaluated under risk neutrality. The decision here places the uncertainty in the constraint and the criterion on the tail, both of which the survey identifies as open.

That leaves a cumulative-budget chance constraint on the tail of the realized cumulative cost, in a sequential setting with costs identified counterfactually. The expectation of a cumulative cost decomposes across stages while its tail probability does not, so the stage-wise machinery these methods build on does not carry over. The constraint has to be read off the full-horizon cost distribution of a strategy taken whole. Such an evaluation is possible strategy by strategy but not over a parameterized policy class, so the decision is restricted to a finite set of candidates. That restriction is also what admits a guarantee holding across the set at once, at a price growing only with its logarithm.

Such a decision requires two estimates per candidate strategy, the outcome and the cumulative-cost distribution, and counterfactual prediction under time-varying treatments supplies both \citep{lim2018rmsn, bica2020crn, li2021gnet, melnychuk2022ct, xiong2024gtransformer, wu2024counterfactual}. These models predict what each strategy would do and leave the choice among strategies to the analyst.

The chance-constraint rule selects the outcome-maximizing strategy subject to a constraint on the probability that the cumulative cost exceeds the budget. It consumes one outcome value and one cost distribution per candidate strategy and nothing else. Any estimator returning those two quantities can therefore supply them, and either step can be replaced without touching the other. In the taxonomy of \citet{sadana2025contextual}, this predict-then-optimize architecture is sequential learning and optimization, estimating first and optimizing after, rather than decision rule optimization or integrated learning and optimization. The constraint sits on the resource axis rather than the outcome. Imposing risk aversion on the outcome is degenerate when the outcome is a binary terminal event such as survival. By contrast, the probability of a budget overrun is a quantity an operator can state in advance and be held to. Sweeping the tolerated violation probability traces a safety-utility frontier, and the decision maker selects an operating point on it by setting how much overrun is acceptable.

This paper makes the following advances.
\begin{itemize}
\item We formulate the selection of a sequential treatment strategy under a cumulative budget as a chance-constrained decision problem over counterfactual outcome and cost distributions. To the best of our knowledge no established method constrains the upper tail of a cost that accumulates over a horizon and is identified counterfactually from observational data.
\item We show that the tail probability of a cumulative cost does not decompose across stages (Proposition~1). The rule therefore scores each candidate strategy on its full-horizon cost distribution and consumes one outcome value and one cost distribution per strategy. It traces a safety-utility frontier as the tolerated violation is swept.
\item We establish two distribution-free finite-sample guarantees under independent cost and value samples. The first bounds the overrun probability of every admitted strategy. The second bounds each admitted strategy's outcome shortfall against the best strategy feasible at a tightened tolerance. The same estimation gap equates the rule with its distributionally robust counterpart, and both bounds grow with the candidate set only through its logarithm.
\item We evaluate by decision quality in five environments drawn from the settings of Table~\ref{tab:applications}. Four provide exact counterfactual ground truth, and the fifth provides real outcomes from a micro-randomized trial. We characterize how the decision degrades as predictor error grows and how large a cost underestimate the violation guarantee withstands.
\end{itemize}

The remainder of the paper is organized as follows. Section~\ref{sec:related} places the problem against the lines of work that bear on it. Section~\ref{sec:methodology} formalizes the decision problem and Section~\ref{sec:method-decision} develops the chance-constrained selection rule and its guarantees. Section~\ref{sec:experiments} describes the experimental design and Section~\ref{sec:results} reports the results. Section~\ref{sec:conclusion} concludes and states the limitations and the directions they indicate.

\section{Sequential Intervention under a Cumulative Budget}
\label{sec:related}

\subsection{The problem and the two axes}
In a budget-constrained sequential intervention problem, interventions are taken over a horizon and interact, so an action taken now changes the state the next decision must respond to. The resource each strategy consumes accumulates over that horizon into a random total, and an outcome is to be maximized subject to a ceiling on that total. Two axes organize the work that bears on such a problem. The first axis is the functional of the cumulative cost the risk criterion constrains, its mean or its upper tail. The second is whether the distribution that functional is taken over is supplied by a known model or identified counterfactually from observational trajectories. Table~\ref{tab:related} places the closest work along both axes.

Relevant work comes from both sides of the prediction-to-decision pipeline and from operations research. Counterfactual prediction under time-varying treatments supplies the estimates, and methods that turn such predictions into a chosen strategy consume them. A budgeted decision has been posed before in operations research, as a risk constraint over a sequential decision process and as an allocation of a limited intervention budget. Each is reviewed in turn, and the axes return at the close of the section to account for the cell no method occupies.

\subsection{Counterfactual prediction and the step to decision}
A mature line of work estimates the counterfactual trajectories that would follow a given sequence of treatments. Recurrent marginal structural networks \citep{lim2018rmsn}, the counterfactual recurrent network \citep{bica2020crn}, G-Net \citep{li2021gnet}, and the Causal Transformer \citep{melnychuk2022ct} established neural g-computation and balancing as the standard tools for this setting. The G-Transformer \citep{xiong2024gtransformer} extended g-computation to dynamic treatment regimes with a transformer architecture. These methods supply the prediction step. A g-computation model of this family can supply, for each candidate strategy, not a point prediction but the full distribution of survival and of cumulative treatment. Where this literature reports forecast accuracy, the method treats accurate forecasting as a prerequisite and shifts the evaluation to decision quality.

Several recent methods move from prediction toward decision but stop short of selecting a strategy under a resource constraint. The generative approach of \citet{wu2024counterfactual} produces high-dimensional counterfactual outcome distributions for time-varying treatments and positions itself as a tool for decision-making. It leaves the choice among treatment sequences to the analyst. DistDeD \citep{killian2023distded} learns return distributions with implicit quantile networks and uses a conditional-value-at-risk signal to flag dead-end states. The framework is an early-warning tool and does not learn a policy. Our optimization step requires the predictor to expose a distribution instead of a mean, and places that distribution inside an explicit budget-constrained selection rule. The AI Clinician \citep{komorowski2018aiclinician} learns and evaluates sequential treatment policies but relies on off-policy estimation rather than counterfactual ground truth. That reliance is a source of evaluation difficulty, and the field has developed off-policy estimators to mitigate it \citep{dudik2011dr, thomas2015hcope, thomas2016ope, kallus2022stateful}. Maximizing an estimated value to obtain a treatment rule is itself well established \citep{zhao2012owl}.

The predict-then-optimize line of work spans prescriptive analytics that maps covariates to decisions \citep{bertsimas2020prescriptive} and decision-focused losses that train a predictor for its downstream optimization \citep{elmachtoub2022spo}. \citet{sadana2025contextual} survey these methods. Within that survey's taxonomy, the uncertainty here sits in the constraint instead of the objective. The criterion is risk-averse on the cost tail instead of risk-neutral on the mean. Both placements are among the directions the survey identifies as open beyond its core of uncertain objectives under risk neutrality. Finite-sample and distributionally robust guarantees for predict-then-optimize have been established for uncertain objectives, through covariate sample-average approximation \citep{kannan2025saa} and bootstrap-robust prescription \citep{bertsimas2022bootstrap}. The guarantee of Section~\ref{sec:method-decision} is of the same finite-sample kind but controls the tail of the constraint. Separating estimation from decision follows the distinction between estimating a causal effect well and making a good decision from it \citep{fernandezloria2022causal}. The decision-quality claims are anchored to environments in which the counterfactual optimum is computable, so that regret and budget violation can be measured exactly.

\subsection{Risk- and chance-constrained sequential decision-making}
A separate line in operations research constrains risk over a sequential decision process instead of a single choice. The constrained Markov decision process bounds the expected cumulative cost of a policy and is solved through occupation measures and linear programming \citep{altman1999constrained}. A risk-sensitive branch replaces the expected cost with a tail functional. \citet{chow2017risk} constrain the conditional value at risk of the cumulative cost, or its probability of exceeding a threshold. \citet{bauerleott2011avar} minimize the average value at risk of the discounted cost through state augmentation. \citet{haskelljain2015convex} minimize the expected cumulative cost of a Markov decision process subject to a chance constraint on that same cost. They reformulate the problem as a linear program over occupation measures. \citet{angelotti2026offline} share the offline selection structure of this problem and choose among a fixed set of candidate policies. However, they maximize a risk-aware objective over a Bayesian posterior on the transition model instead of constraining a resource. The risk they control is therefore epistemic and contracts as data accumulate. By contrast, the tail we constrain belongs to the cost distribution itself and persists under exact knowledge of the dynamics. These methods constrain the same upper-tail object we target, and the chance-constrained formulation of \citet{haskelljain2015convex} is the nearest neighbor on the constraint. The difference is the regime in which the cost distribution is obtained. In this literature, the transition dynamics are known. The cost distribution of a policy is therefore generated by rolling it out in the environment or by solving the model through its occupation measure. Here the cost distribution is instead identified counterfactually from observational trajectories in which treatment is confounded with prognosis, under the assumptions of Section~\ref{sec:method-setup}. The constraint is then enforced in an optimization step that is agnostic to the predictor producing it, not inside a policy-optimization procedure. Approximate and feasibility-oriented treatments of constrained risk-sensitive Markov decision processes also appear in this operational-research setting \citep{kumar2023crsmdp}.

Closer to the guarantee we establish, \citet{rahimian2023contextual} study chance-constrained programs with the constraint imposed on a distribution conditional on observed covariates and estimated from data. They establish finite-sample feasibility of the sample problem solved at a tightened risk level. Proposition~3 is of that kind, with the conditioning on an intervention instead of on a covariate. It reaches the shifted level by the same route, a uniform deviation over the candidate set carried by a union bound. The distribution the constraint is evaluated on is therefore counterfactual and rests on the identification assumptions rather than on observed conditioning alone.

A separate simulation-optimization line shares the discrete structure of this problem more closely than the policy-optimization one does. Chance-constrained selection of the best chooses, among a finite set of alternatives, the one maximizing a primary performance measure subject to a probabilistic constraint on secondary measures \citep{hong2015ccsb}. Recent procedures achieve a guaranteed probability of correct selection without indifference-zone assumptions \citep{zhong2026ccsb}. That work poses the same discrete decision, maximizing a value over a finite candidate set subject to a probabilistic constraint, and they differ in what is scarce. There the alternatives can be simulated at will and the question is how much simulation effort each requires, so the guarantee concerns the sampling procedure. Here they cannot be run at all: the primary and secondary quantities of every strategy but the one actually followed are counterfactual. The guarantee concerns the constraint evaluated on distributions identified from observational trajectories.

The term chance-constrained Markov decision process is also used for a different problem. There, the constraint is placed on uncertain cost or reward parameters rather than on the realized cumulative cost. Tail objectives of this kind recur across operational-research decision problems, from feature-based conditional-value-at-risk minimization in the newsvendor setting \citep{liu2024newsvendor} to robust utility maximization over chance constraints \citep{wang2025robust}. The setting studied here takes the constraint as given rather than developing the optimization theory those methods contribute.

\subsection{Prescriptive analytics with budgets}
Prescriptive analytics turns estimated treatment effects into intervention decisions, and a line of operational-research work couples effect estimation with a downstream allocation under a resource limit. \citet{devos2026uplift} give the architecture we share. A prediction step estimates dose responses off the shelf, and an optimization step allocates doses subject to a budget on total treatment cost. They evaluate in regret against a full-information optimum computed on known counterfactuals. Their separation of the two steps is deliberate and is defended on the same grounds we adopt. The separation admits pre-trained or substituted predictors and allows constraints to be revised without retraining. \citet{caljon2026interference} likewise optimize treatment allocation under a resource limit when treatment of one entity affects the outcomes of others. They select which entities in a network to target under a cardinality budget. Both allocate at a single decision point under a constraint on the total taken across units. Carried to a multi-stage horizon, that architecture meets a cumulative cost that is a random variable. The constraint falls on the upper tail of that cost rather than on its total.

A separate line of prescriptive analytics has used heterogeneous treatment effects to allocate a fixed intervention budget across individuals, constraining an aggregate or single-decision budget. The framework of \citet{mcfowland2021prescriptive} converts estimated heterogeneous treatment effects into individual-level targeting decisions through an integer program and is the closest prior work in spirit. It allocates treatment across individuals under a budget at a single decision point rather than over a sequential horizon. \citet{vanderschueren2024allocating} allocate limited resources across uncertain tasks under a capacity that is itself random with a known distribution. They absorb that randomness by maximizing the expected payoff under it rather than constraining its tail. They report that a model trained directly on the allocation objective outperforms a two-stage pipeline when capacity is scarce. This result holds at a single decision point where that objective is computable from observed labels. \citet{sun2026ewm} extends empirical welfare maximization to a budget with a cost that must itself be estimated. The constraint is placed on the population expected cost of an eligibility policy at a single decision point. Under such a constraint, no statistical rule is uniformly welfare-efficient and feasible at once. Where the constraint binds exactly, the welfare loss does not vanish with sample size. The same obstacle governs the benchmark against which Section~\ref{sec:method-decision} states its objective bound. Closest on the constrained-decision axis, \citet{atheywager2021policy} learn treatment-assignment policies from observational data under application-specific constraints, a budget among them, encoded as restrictions on the admissible policy class. They give regret guarantees against the best policy in that class, for a single-decision assignment.

A parallel line constrains risk within the dynamic treatment regime itself. \citet{liu2024brdtr} learn optimal regimes subject to stagewise controls on the risk of adverse events, enforced inside a weighted classification estimator. Their decomposition into single-stage problems rests on an acute-risk condition under which the risk at a stage depends only on the most recent treatment. They identify a cumulative count of costly treatments as a case the condition excludes. \citet{liu2024cbr} take the constraint to the horizon total, maximizing the expected reward subject to a ceiling on the expected cumulative risk. A constraint on the total blocks the backward algorithms the stagewise problem admits, since the control depends on future-stage rules still to be estimated. They recover those algorithms by converting the constrained problem through a Lagrange function into a single reweighted outcome, the decision maker stating the ceiling on the expected risk. \citet{laber2018safety} bound the expected value of a harm cost defined over the whole trajectory in a dosing regime estimated from observational data. They sweep that bound to present an efficacy-against-risk curve on which a clinician sets a tolerability limit. All three constraints are placed on an expectation, of a per-stage risk, of the horizon total, and of a trajectory-level maximum. By contrast, the constraint here is placed on the upper tail of the horizon total.

In econometrics, \citet{sakaguchi2025dewm} brings budgets into the sequential setting directly, extending empirical welfare maximization to dynamic treatment regimes under intertemporal budget and capacity constraints. It gives finite-sample distribution-free bounds on both the welfare regret and the excess of the estimated regime's mean cost over the budget. The constraint there bounds a weighted sum, across stages, of the expected fraction of the population treated at each stage, with stage weights representing relative treatment costs. It therefore acts on population averages. The distribution of the cost an individual accumulates over the horizon does not enter the formulation. The constraint here is placed on the upper tail of that individual cumulative cost. Even with the cost distribution of every strategy known exactly, the cumulative treatment a strategy consumes is a random variable across units. This is because the treatments administered depend on the trajectory each unit follows. The probability that the realized cost exceeds the budget can therefore be positive, and $\varepsilon$ is the tolerated value of that tail probability. A strategy can hold its expected cost within budget while leaving a heavy tail above it, so the functionals do not coincide even with unlimited data.

\begin{table}[pos=tbp]
\centering
\small
\caption{Sequential methods with a cumulative cost, by the functional the risk criterion constrains and by where the cost distribution comes from. No entry occupies the lower-right cell.}
\label{tab:related}
\begin{adjustbox}{max width=\textwidth}
\begin{tabular}{@{}l p{4.6cm} p{4.6cm}@{}}
\toprule
& Known model, simulator, or online interaction & Observational trajectories, counterfactual identification \\
\midrule
Mean or aggregate resource use
& \citet{altman1999constrained}
& \citet{sakaguchi2025dewm}; \citet{liu2024cbr} \\[4pt]
Upper tail of the cumulative cost
& \citet{chow2017risk}; \citet{bauerleott2011avar}; \citet{haskelljain2015convex}
& \cellcolor{black!8} \\
\bottomrule
\end{tabular}
\end{adjustbox}

\vspace{2pt}
\noindent\begin{minipage}{\linewidth}
\footnotesize\raggedright
Budgeted allocation at a single decision point \citep{devos2026uplift, mcfowland2021prescriptive, vanderschueren2024allocating, caljon2026interference} constrains a total across units, not across time. \citet{liu2024brdtr} constrain a per-stage adverse-event probability where its companion in the table constrains the horizon total. Chance-constrained ranking and selection \citep{hong2015ccsb, zhong2026ccsb} shares the discrete selection form. Its constrained measure, however, is a secondary simulation output, and the resource spent is simulation effort rather than a decision cost.
\end{minipage}
\end{table}

\subsection*{The two axes together}
The two axes of Table~\ref{tab:related} are not independent, and the lower-right cell is not reached by moving along either one alone. The expected cumulative cost is additive across stages and, in the Markov-decision-process formulation, linear in the occupation measure of the policy. That allows a mean constraint to be estimated by stage-wise sample means and imposed inside backward-induction or mixed-integer-programming machinery over a parameterized policy class. A tail probability is a functional of the distribution of the full-horizon sum and decomposes in neither way, as Section~\ref{sec:method-quantities} makes precise. It is a static quantile constraint, and static quantile constraints are not time consistent. A choice that is optimal at the start of the horizon need not remain optimal in continuation. The constraint therefore admits no Bellman recursion unless the state is augmented with the cost accumulated so far \citep{bauerleott2011avar, haskelljain2015convex}. The methods in the left column obtain that augmented recursion from a specified transition model, at the price of an extra continuous dimension. When the cost distribution is instead identified counterfactually, the model that would supply the recursion is missing. The horizon distribution of each strategy therefore has to be evaluated as a whole. The optimization step then scores a finite set of candidate strategies and evaluates the constraint on the distribution of each, rather than optimizing over a parameterized policy class.

\section{Problem Formulation}
\label{sec:methodology}

\begin{figure}[pos=tbp]
\centering
\includegraphics[width=\linewidth,height=0.26\textheight,keepaspectratio]{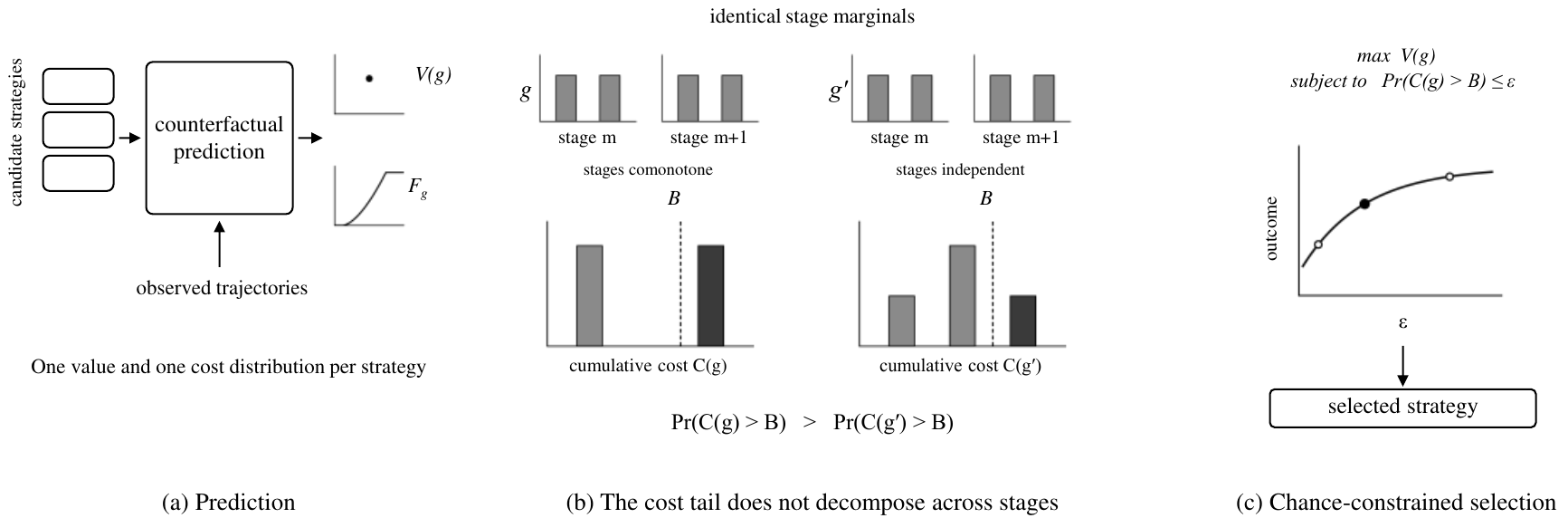}
\caption{The two steps and the fixed contract between them, one outcome value and one cost distribution per strategy. The center panel shows two strategies with identical per-stage cost distributions and identical mean cost, and cumulative costs that exceed the budget with different probability.}
\label{fig:method}
\end{figure}

Figure~\ref{fig:method} summarizes the two steps. The decision problem is formalized first, and Section~\ref{sec:method-decision} develops the optimization step that solves it. The setting is any sequential intervention with a scalar outcome to be maximized and a cumulative resource cost to be constrained. Section~\ref{sec:experiments} instantiates it on clinical, operations, and behavioral objectives. Table~\ref{tab:notation} of the supplementary material collects the notation.

\subsection{Problem setup}
\label{sec:method-setup}

We adopt the sequential potential-outcome framework for time-varying treatments, in the formulation used by the g-computation and balancing literature \citep{robins1986, robins2000msm, bica2020crn, melnychuk2022ct, li2021gnet}. A unit is observed over discrete time steps $t \in \{0, 1, \ldots, K\}$. At each step, we observe time-varying covariates $L_t$, a treatment decision $A_t$, and an outcome that is a component of the next-step covariate vector. The observed history up to step $m$ is $H_m = (\bar{L}_m, \bar{A}_{m-1})$, where $\bar{L}_m = (L_0, \ldots, L_m)$ and $\bar{A}_{m-1} = (A_0, \ldots, A_{m-1})$ denote the covariate and treatment histories. A dynamic treatment strategy $g = (g_m, \ldots, g_K)$ is a sequence of decision rules, as in the literature on dynamic treatment regimes \citep{murphy2003optimal, chakraborty2013dtr}. Each rule maps the history available at that step to a treatment, $g_t : H_t \mapsto A_t$. For a strategy $g$ we write $L_t(g)$ for the covariate vector at step $t$ that would be observed if treatment were assigned according to $g$ from step $m$ onward. We write $\bar{L}_{>t}(g)$ for the path the strategy generates after step $t$. The outcome under $g$ is the corresponding component of $L_t(g)$.

Identification of the counterfactual quantities below rests on three standard assumptions of the framework.

\textbf{Assumption 1} (Sequential ignorability). At each step, treatment is independent of the future potential covariate path given the observed history, $A_t \perp \bar{L}_{>t}(g) \mid H_t$.

\textbf{Assumption 2} (Positivity). Every treatment a strategy may assign has positive probability given any history that can occur, $\Pr\big(A_t = g_t(h_t) \mid H_t = h_t\big) > 0$ whenever $\Pr(H_t = h_t) > 0$, so that the counterfactual is supported by data.

\textbf{Assumption 3} (Consistency). On the event that the realized treatment history agrees with what the strategy assigns, the potential covariates equal the observed ones, $L_t(g) = L_t$ on $\{\bar{A}_{t-1} = \bar{a}_{t-1}\}$ with $a_s = g_s(h_s)$.

These assumptions are the sequential analogues of the ignorability (no unmeasured confounding), overlap, and consistency conditions familiar from cross-sectional treatment-effect estimation. Under them, the g-computation formula identifies the distribution of the covariate path under a strategy from observational quantities \citep{robins1986},
\begin{equation}
p\big(\bar{l}_{m:K} \mid g\big) = \prod_{t=m}^{K} p\big(l_t \mid \bar{l}_{t-1}, \bar{a}_{t-1}\big)\Big|_{a_s = g_s(h_s)},
\label{eq:gcomp}
\end{equation}
where each factor is a conditional law of the observed data and the treatments are set to those the strategy would assign. \ref{app:identification} of the supplementary material carries the assumptions through to the cost distribution. Both quantities the decision rule consumes are functionals of Equation~\eqref{eq:gcomp}. The outcome value is an expectation under it. The cost distribution is the law of the accumulated treatment along the same path,
\begin{equation}
F_g(b) = \Pr\big(C(g) \le b\big) = \sum_{\bar{l}_{m:K}} p\big(\bar{l}_{m:K} \mid g\big) \, \mathbb{1}\!\left[\sum_{t=m}^{K} c\big(g_t(h_t)\big) \le b\right].
\label{eq:costdist}
\end{equation}
The budget constraint is therefore imposed on a counterfactual object: $F_g$ is never observed for any strategy other than the one the data-generating policy followed. Equation~\eqref{eq:costdist} makes it estimable.

\subsection{Counterfactual quantities under a strategy}
\label{sec:method-quantities}

The method scores a candidate strategy by two counterfactual quantities rather than by a single point prediction. The first is the strategy's outcome value, the probability that the unit reaches the horizon in the favorable terminal state under the strategy,
\begin{equation}
V(g) = \Pr\big( \text{favorable terminal state} \mid g \big),
\label{eq:value}
\end{equation}
which is the utility the decision maker wishes to maximize. The second is the strategy's cumulative treatment cost over the horizon,
\begin{equation}
C(g) = \sum_{t=m}^{K} c\big(A_t\big), \qquad A_t \sim g,
\label{eq:cost}
\end{equation}
where $c(\cdot)$ counts the treatments administered at a step, so $C(g)$ is the cumulative quantity of treatment a strategy consumes. It is a random variable, not a fixed number, because the treatments administered depend on the trajectory the unit follows under the strategy. The method therefore works with the full distribution of $C(g)$, not only its mean.

Which functional of that distribution the budget constrains determines what machinery is available. By linearity of expectation, the mean separates across stages,
\begin{equation}
\mathbb{E}\big[C(g)\big] = \sum_{t=m}^{K} \mathbb{E}\big[c(A_t)\big],
\label{eq:meandecomp}
\end{equation}
so a constraint on $\mathbb{E}[C(g)]$ is a sum of per-stage terms, each estimable from the visits to stage $t$ alone. That permits a mean-budget constraint to be imposed inside backward induction or written as a linear constraint over occupation measures. The tail probability admits no such separation. Writing $S_t = \sum_{s=m}^{t} c(A_s)$ for the partial sum,
\begin{equation}
\Pr\big(C(g) > B\big) = \Pr\big(S_K > B\big) = 1 - F_{S_K}(B),
\label{eq:taildecomp}
\end{equation}
a functional of the distribution of the full-horizon sum. It depends on the joint law of $(c(A_m), \ldots, c(A_K))$, not on the per-stage marginals separately.

\textbf{Proposition 1} (Non-decomposability of the cost tail). There exist strategies for which the per-stage cost distributions coincide at every stage while the budget-violation probabilities differ. Consequently $\Pr\bigl(C(g) > B\bigr)$ is not a functional of the stage-wise marginal laws of $c(A_m), \ldots, c(A_K)$, and a constraint on it cannot be assembled from per-stage estimates. Equation~\eqref{eq:meandecomp}, by contrast, writes the mean as a sum of per-stage terms under every dependence structure.

The construction is a two-stage horizon, $K = m+1$, with per-stage cost valued in $\{0,1\}$. Under $g$, a single $\mathrm{Bernoulli}(1/2)$ draw $Z$ sets both stage costs, $c(A_m) = c(A_{m+1}) = Z$, so the stages are comonotone; under $g'$ they are independent $\mathrm{Bernoulli}(1/2)$ draws. Every stage marginal is $\mathrm{Bernoulli}(1/2)$ under both, so by Equation~\eqref{eq:meandecomp} the mean costs agree at $\mathbb{E}[C(g)] = \mathbb{E}[C(g')] = 1$. The totals do not: $C(g)$ takes the values $0$ and $2$ with probability $1/2$ each. The total $C(g')$ takes $0$, $1$, $2$ with probabilities $1/4$, $1/2$, $1/4$. At any budget $B \in [1, 2)$,
\begin{equation}
\Pr\bigl(C(g) > B\bigr) = \tfrac{1}{2}, \qquad \Pr\bigl(C(g') > B\bigr) = \tfrac{1}{4},
\label{eq:nondecomp}
\end{equation}
so two strategies indistinguishable stage by stage, in marginal law and in mean, differ in violation probability by a factor of two. Both are realizable in the framework of Section~\ref{sec:method-setup}. A binary state variable is drawn at the first stage and carried forward in the covariates. The strategy $g$ treats at both stages or at neither, according to that variable. The strategy $g'$ draws its treatment indicator independently at each stage. The gap widens with the horizon: extending the pair to $h$ stages leaves every stage marginal $\mathrm{Bernoulli}(1/2)$. The comonotone strategy violates a budget $B \in [h-1, h)$ with probability $1/2$, and the independent one violates it with probability $2^{-h}$.

The phenomenon is the time inconsistency of a static quantile constraint, known in the risk-sensitive control literature. There it forces the state to be augmented with the cost accumulated so far before a Bellman recursion becomes available \citep{bauerleott2011avar}. The tail probability depends on the dependence between stages, which leaves the mean unchanged. It is therefore reachable either through that augmented recursion, which needs a transition model to recurse through, or by estimating the law of the full-horizon sum directly. Where the cost distribution is identified counterfactually rather than specified, the second route is the one that remains, and the chance-constraint rule takes it.

Both $V(g)$ and the distribution of $C(g)$ are identified under Assumptions 1 to 3, and we compute them by simulating trajectories under $g$. They lie on different axes, the first a property of the outcome and the second of the treatment the strategy consumes. A constraint that moved with the outcome would never bind whereas a treatment budget trades off against it and shapes the decision.

\section{The Optimization Step: Chance-Constrained Selection}
\label{sec:method-decision}

The optimization step turns the two counterfactual quantities of Section~\ref{sec:method-quantities} into a recommendation. We state the decision rules and separate them by the functional of the cost distribution each one constrains. An operating point is then selected on the frontier the advocated rule traces. The violation and the outcome shortfall of the selected strategy are bounded in finite samples.

\subsection{The decision rules}
\label{sec:method-rules}

The decision maker seeks the strategy that maximizes the outcome while respecting a treatment budget $B$. The choice is over the finite set $\mathcal{G}$ of candidate strategies that Section~\ref{sec:related} arrives at. Section~\ref{sec:exp-strategies} describes how we obtain $\mathcal{G}$ in each evaluation environment, where it must be constructed rather than taken as given so that the counterfactual optimum is computable. The rules below differ in their robustness to the fact that, in deployment, the outcome and cost of a strategy are estimated rather than known. Let $\hat{V}(g)$ denote the estimated outcome value, $\hat{C}(g)$ the estimated mean cost, and $\hat{\sigma}_C(g)$ the estimated standard deviation (s.d.) of the cost.

The first rule acts on the estimated mean cost directly,
\begin{equation}
g^{\ast}_{\mathrm{naive}} = \arg\max_{g} \, \hat{V}(g) \quad \text{subject to} \quad \hat{C}(g) \le B.
\label{eq:naive}
\end{equation}
This point-estimate rule is the implicit one whenever a strategy is selected by comparing expected costs to a budget. Its weakness is an instance of the optimizer's curse \citep{smithwinkler2006}. Even when $\hat{C}(g)$ is unbiased for every strategy separately, restricting attention to those that pass the test $\hat{C}(g) \le B$ selects preferentially among the realizations in which the cost was underestimated. Hence $\mathbb{E}[\hat{C}(g^{\ast}_{\mathrm{naive}}) - C(g^{\ast}_{\mathrm{naive}})] < 0$, and a strategy with true expected cost above the budget can appear feasible and be recommended.

The second rule guards against this by holding out a margin proportional to the estimated cost uncertainty,
\begin{equation}
g^{\ast}_{\mathrm{ub}} = \arg\max_{g} \, \hat{V}(g) \quad \text{subject to} \quad \hat{C}(g) + \kappa \, \hat{\sigma}_C(g) \le B,
\label{eq:ub}
\end{equation}
where $\kappa \ge 0$ controls the conservatism of the margin. This rule reduces violations, but $\kappa$ carries no units in which a decision maker can reason. A margin of one standard deviation corresponds to no stated tolerance for exceeding the budget, and the value that performs well depends on the cost distributions at hand.

The third rule, which is the one we advocate, constrains the probability that the cost exceeds the budget directly, the functional Equations~\eqref{eq:meandecomp} and~\eqref{eq:taildecomp} distinguish,
\begin{equation}
g^{\ast}_{\mathrm{cc}} = \arg\max_{g} \, \hat{V}(g) \quad \text{subject to} \quad \Pr\big(C(g) > B\big) \le \varepsilon,
\label{eq:cc}
\end{equation}
where $\varepsilon \in [0, 1]$ is the violation probability the decision maker is willing to tolerate. The constraint in Equation~\eqref{eq:cc} is a chance constraint in the sense of \citet{charnes1959ccp}. Writing $F_g$ for the distribution function of $C(g)$, it restricts the budget to lie above the corresponding quantile,
\begin{equation}
\Pr\big(C(g) > B\big) \le \varepsilon
\;\Longleftrightarrow\;
F_g(B) \ge 1 - \varepsilon
\;\Longleftrightarrow\;
\mathrm{VaR}_{1-\varepsilon}\big(C(g)\big) \le B,
\label{eq:var}
\end{equation}
so the rule admits a strategy only if the upper tail of its cost distribution stays within the budget with the required probability. Replacing the value at risk by the conditional value at risk gives a constraint that is not equivalent but safe. Because $\mathrm{CVaR}_{1-\varepsilon}(C) \ge \mathrm{VaR}_{1-\varepsilon}(C)$, every strategy it admits satisfies Equation~\eqref{eq:cc}. A strategy with a cost that exceeds the budget rarely but by a wide margin can satisfy Equation~\eqref{eq:cc} and fail the conditional-value-at-risk constraint. That constraint is convex in the cost distribution and is the least conservative of the generator-based safe convex approximations of a chance constraint \citep{rockafellar2000cvar, nemirovski2006convex}. It is evaluated on the sampled distribution as the average of the worst $\lceil \varepsilon n \rceil$ realizations,
\begin{equation}
\widehat{\mathrm{CVaR}}_{1-\varepsilon}\big(C(g)\big) = \frac{1}{\lceil \varepsilon n \rceil} \sum_{j=1}^{\lceil \varepsilon n \rceil} C_{(n-j+1)}(g),
\label{eq:cvarhat}
\end{equation}
where $C_{(1)} \le \cdots \le C_{(n)}$ are the ordered cost samples of $g$. As $\varepsilon$ approaches one that average covers the whole sample and the constraint reduces to the mean-budget test of Equation~\eqref{eq:naive}. The two rules coincide in that limit. Chance constraints are in general intractable to optimize, and the standard remedy is a tractable convex approximation \citep{nemirovski2006convex}. Here the cost distribution is available in sampled form from the predictor. The risk measure sits on the constraint, not on the objective, since the lower tail of a binary terminal outcome does not separate strategies.

Each rule constrains a different functional of the cost distribution: the mean, the mean plus a margin, the upper tail quantile, and the mean of that tail. The rules therefore differ even where the cost distribution is known exactly. The last two bound the frequency and the severity of the same overrun and Section~\ref{sec:res-cvar} reports how they compare. They would coincide only if the cost were deterministic. The regret the chance-constraint rule incurs under exact counterfactuals is therefore the intrinsic price of bounding the tail instead of the mean, separate from any estimation error. Under estimation, the naive rule is additionally unreliable, admitting strategies with a mean that only appears to satisfy the budget. Sweeping $\varepsilon$ traces the safety-utility frontier, on which it positions the recommendation between admitting overruns and sacrificing outcome. Worst-case and multi-objective alternatives fit the budget less well as it is usually given;~\ref{app:dro} sets out both.

The rule is evaluated on the sampled cost distribution of each strategy, the sample-average-approximation form of the constraint \citep{luedtke2008saa, nemirovski2006convex}. For each candidate the predictor supplies an outcome value and $n$ cost samples, the empirical tail probability is the fraction of those samples above the budget. The rule returns the highest-outcome strategy with tail probability at most $\varepsilon$, or reports infeasibility when none is; Algorithm~\ref{alg:selection} in the supplementary material states the procedure.

Selection is $O(\lvert\mathcal{G}\rvert n)$ given the outcome value and the $n$ cost samples of each candidate, and a sweep over $\varepsilon$ reuses the same per-strategy tail counts. The entire frontier therefore costs one pass over the sampled costs, and the budget or the tolerated violation can be revised without re-estimating anything.

\subsection{Selecting an operating point}
\label{sec:method-operating}

Equation~\eqref{eq:cc} and the frontier it traces still leave open which operating point the decision maker should adopt. A recommender that returns the whole curve defers that choice instead of supporting it. Under exact counterfactuals, the frontier is an ordered trade-off by construction.

\textbf{Proposition 2} (Frontier monotonicity). Fix the budget $B$, and let $V^{\ast}(\varepsilon) = \max\{\hat{V}(g) : \Pr(C(g) > B) \le \varepsilon\}$ be the achievable outcome at tolerated violation $\varepsilon$. Then $V^{\ast}$ is non-decreasing in $\varepsilon$ on $[0, 1]$.

Write $\mathcal{A}(\varepsilon) = \{g \in \mathcal{G} : \Pr(C(g) > B) \le \varepsilon\}$ for the admissible set. For $\varepsilon \le \varepsilon'$ the defining inequality is weaker on the right, so $\mathcal{A}(\varepsilon) \subseteq \mathcal{A}(\varepsilon')$ and hence $V^{\ast}(\varepsilon) \le V^{\ast}(\varepsilon')$, a maximum over a larger set being no smaller. The frontier therefore exchanges tolerated violation for outcome monotonically. This monotonicity makes a target on either axis resolve to a single recommendation. Its curvature is another matter: Section~\ref{sec:res-frontier-char} shows it is too irregular for a curvature-based knee to be stable or consistent across environments. The operating point is set by operational target instead. Given a maximum tolerated tail probability $\rho$, the method returns the admissible strategy of greatest outcome. Given a minimum acceptable outcome, it returns the admissible strategy with the smallest tail probability. The decision maker therefore states a target on whichever axis is operationally meaningful instead of choosing the constraint level $\varepsilon$ directly. The selection assumes nothing about the shape of the frontier, since it optimizes one objective over the swept strategies subject to a bound on the other. The realized tail probability of a strategy can exceed its nominal tolerance once the outcome and cost are estimated rather than known. The target is then applied to the realized tail probability.

\subsection{Finite-sample guarantees}
\label{sec:method-guarantees}

The chance-constraint rule of Equation~\eqref{eq:cc} is stated in terms of the true cost distribution but in deployment is applied to an estimate of it. That substitution must not silently inflate the budget violation the rule is meant to control. Relating the violation of a sampled chance constraint to the sample size in this way is standard \citep{luedtke2008saa}. Let $F_g(b) = \Pr(C(g) \le b)$ be the true cumulative distribution of the cost of strategy $g$. Let $\hat{F}_g$ be the empirical distribution formed from $n$ independent samples of $C(g)$. The rule admits $g$ when $1 - \hat{F}_g(B) \le \varepsilon$.

\textbf{Proposition 3} (Feasibility under cost estimation). If $\sup_b \lvert \hat{F}_g(b) - F_g(b) \rvert \le \delta$, then every strategy admitted by the empirical chance-constraint rule satisfies
\begin{equation}
\Pr\big(C(g) > B\big) \le \varepsilon + \delta .
\label{eq:l0feas}
\end{equation}
When the cost samples are independent, the Dvoretzky-Kiefer-Wolfowitz inequality with the tight constant of \citet{massart1990dkw} gives $\Pr(\sup_b \lvert \hat{F}_g(b) - F_g(b) \rvert > \delta) \le 2e^{-2n\delta^2}$ for each strategy. A union bound over $\mathcal{G}$ carries it to all of them at once,
\begin{equation}
\Pr\Big(\exists\, g \in \mathcal{G} : \sup_b \big\lvert \hat{F}_g(b) - F_g(b) \big\rvert > \delta \Big) \le 2\lvert\mathcal{G}\rvert e^{-2n\delta^2},
\label{eq:dkwunion}
\end{equation}
so setting the right-hand side to $\eta$ gives $\delta = \sqrt{\log(2\lvert\mathcal{G}\rvert/\eta)/(2n)} = O(n^{-1/2})$, at which Equation~\eqref{eq:l0feas} holds with probability at least $1-\eta$ simultaneously for every admitted strategy.

Evaluating the uniform bound at the budget and chaining it with the admission rule gives
\begin{equation}
F_g(B) \;\ge\; \hat{F}_g(B) - \delta \;\ge\; (1-\varepsilon) - \delta,
\label{eq:feaschain}
\end{equation}
where the second inequality is the admission condition $\hat{F}_g(B) \ge 1-\varepsilon$; rearranging is Equation~\eqref{eq:l0feas}. Because the control is uniform over the threshold, it certifies the budget evaluation at every $B$ at once. The union bound is therefore needed only over the finite strategy set, not over the continuum of budgets. Sections~\ref{sec:res-rules} and~\ref{sec:res-frontier-char} sweep both and use this uniformity.

The two quantities in Equation~\eqref{eq:l0feas} are named for use throughout. The certified slack is $\delta$, the uniform deviation the premise allows. The certified ceiling is $\varepsilon + \delta$, the bound the proposition places on the realized tail probability of an admitted strategy. The certified slack is an estimation term that vanishes as the cost distribution is estimated from more data. The finite-sample term in the mean-cost budget bound of \citet{sakaguchi2025dewm} plays the same role. It is distinct from the tolerated violation $\varepsilon$, which is the nominal tail-risk level fixed in the constraint. The two settings differ in the constrained functional, the tail instead of the mean.

Evaluating the constant of Equation~\eqref{eq:dkwunion} in the real-outcome environment of Section~\ref{sec:exp-realdata}, with $n = 349$ independent participants, $\lvert\mathcal{G}\rvert = 17$ candidate strategies and $\eta = 0.05$, gives $\delta \approx 0.10$. A nominal tolerance $\varepsilon = 0.2$ therefore carries a certified ceiling of about $0.30$ on the realized tail probability.\footnote{The independent unit is the participant, so Monte Carlo redraws of the treatment realization do not enlarge $n$, and a distribution-free bound is conservative at a few hundred units. In the exact-counterfactual environments, the cost distribution comes from simulation, so $n$ can be taken large and $\delta$ made negligible. The regret and violation reported there are therefore not subject to this inflation.}

The same translation applies to the conditional-value-at-risk operating point, with a constant that depends on the cost support. The chance-constraint form is stated because it admits the explicit Dvoretzky-Kiefer-Wolfowitz constant. The uniform deviation in Proposition~3 also connects the rule to its distributionally robust counterpart. Constraining the worst case over a Kolmogorov ball of radius $r$ around $\hat{F}_g$ gives the same rule at tolerance $\varepsilon - r$ (\ref{app:dro}). Taking $r = \delta$ recovers Proposition~3 and its certified ceiling of $\varepsilon + \delta$.

Proposition~3 controls feasibility but leaves open how close the outcome of the selected strategy is to the budget-feasible optimum. A bound against $g^{\ast}(\varepsilon)$, the best strategy truly feasible at the nominal tolerance, cannot hold uniformly. A strategy with cost tail exactly at $\varepsilon$ is excluded by the empirical rule whenever sampling error pushes its estimated tail above the threshold. The empirical feasible set therefore need not contain the true optimum. Accordingly, the comparison is made against a benchmark defined at a tightened tolerance. This parallels the analysis of sampled chance constraints, where solving the sample problem at one risk level certifies a bound relative to a slightly different level \citep[Theorem 3]{luedtke2008saa}. The same obstacle appears in the econometric treatment of budgeted policy choice, where no statistical rule attains the highest welfare and satisfies the budget uniformly at once. The welfare loss fails to vanish with sample size exactly where the constraint binds \citep{sun2026ewm}. The tightened benchmark is thus forced by the structure of the problem, not adopted for convenience. Write $g^{\ast}(\varepsilon') = \arg\max\{V(g) : \Pr(C(g) > B) \le \varepsilon'\}$ for the best strategy truly feasible at tolerance $\varepsilon'$, with $\hat{V}(g)$ formed as an average over $n$ independent units.

\textbf{Proposition 4} (Finite-sample regret at a tightened benchmark). Suppose the per-unit contributions to $\hat{V}(g)$ lie in an interval of length $R$ for every $g \in \mathcal{G}$. Suppose also that the cost samples and the value samples are each independent across units. Fix $\eta \in (0,1)$ and set
\begin{equation}
\delta = \sqrt{\frac{\log(4\lvert\mathcal{G}\rvert/\eta)}{2n}}, \qquad
\epsilon_V = R\sqrt{\frac{\log(4\lvert\mathcal{G}\rvert/\eta)}{2n}} .
\label{eq:l1const}
\end{equation}
Let $\hat{g}$ be the strategy returned by the empirical chance-constraint rule at tolerance $\varepsilon$. Then, with probability at least $1-\eta$, simultaneously
\begin{equation}
\Pr\big(C(\hat{g}) > B\big) \le \varepsilon + \delta
\qquad\text{and}\qquad
V\big(g^{\ast}(\varepsilon - \delta)\big) - V(\hat{g}) \le 2\epsilon_V .
\label{eq:l1regret}
\end{equation}

The proof intersects two uniform deviation bounds: a Dvoretzky-Kiefer-Wolfowitz bound on the cost distributions and a Hoeffding bound \citep{hoeffding1963} on the value estimates. Each is taken over $\mathcal{G}$ at confidence $\eta/2$. On that intersection, the feasibility half follows as in Proposition~3, and $g^{\ast}(\varepsilon-\delta)$ is admitted by the empirical rule. The rule then maximizes $\hat{V}$ over a set containing it, and the regret half follows. \ref{app:prop4} of the supplementary material gives the argument.

The constants carry $\log(4\lvert\mathcal{G}\rvert/\eta)$ where Proposition~3 carries $\log(2\lvert\mathcal{G}\rvert/\eta)$ because the confidence budget is split between the two bounds; the numerical check above is unaffected to two decimals. Both terms depend on the candidate set only through $\log\lvert\mathcal{G}\rvert$, so enlarging it is inexpensive at a computational cost Section~\ref{sec:res-frontier-char} measures. A sampled chance constraint over a finite candidate set exhibits the same logarithmic dependence \citep[Theorem 5]{luedtke2008saa}. Comparing against $g^{\ast}(\varepsilon-\delta)$ rather than $g^{\ast}(\varepsilon)$ concedes whatever outcome lies between the two tolerances. The concession is small where the frontier is flat near $\varepsilon$, larger where a high-outcome strategy sits just inside it, and shrinks at the rate of the feasibility slack. The constant $R$ belongs to the value estimator rather than the decision rule and scales with the largest importance weight for an inverse-probability-weighted one. The certified margin is therefore loose at these sample sizes in the way a distribution-free bound is expected to be. Two benchmarks are in play and measure different things: Equation~\eqref{eq:l1regret} bounds the shortfall against $g^{\ast}(\varepsilon-\delta)$, feasible under the same tail constraint at a tightened tolerance. The regret of Section~\ref{sec:results} is measured instead against the risk-neutral oracle of Section~\ref{sec:exp-oracle}. That oracle is constrained on the mean cost and therefore admits strategies the chance constraint excludes. The reported regret then carries the intrinsic price of bounding the tail.

Propositions~2 to~4 take the cost distribution to be estimated directly from independent cost samples, as it is in the real-outcome environment, not produced by an estimated outcome predictor. Its error would propagate through the dynamics into the cost tail and be coupled across stages by the sequential structure. They also hold for a finite candidate set. Extending them to a parameterized policy class would therefore require a complexity measure in place of the $\log\lvert\mathcal{G}\rvert$ of Equation~\eqref{eq:l1const}. \citet{rahimian2023contextual} carry a finite-set feasibility bound to a compact decision set along those lines. Both extensions are left to future work.

\subsection{Predictor and evaluation}
\label{sec:method-eval}

The decision rule consumes the outcome value and the cost distribution of each strategy but does not prescribe how they are produced. An estimator must supply the distribution of the cumulative cost, not only its mean. A g-computation model estimated from observational data is the natural fit: simulating trajectories under a strategy yields as many draws of Equations~\eqref{eq:value} and~\eqref{eq:cost} as are wanted. Recurrent and Transformer forms are available \citep{li2021gnet, xiong2024gtransformer}. Inverse probability weighting also meets the requirement, and the real-outcome environment of Section~\ref{sec:exp-realdata} uses it. Weighting estimators, however, recover the upper tail less reliably than the mean. Any estimator honoring the same contract, one outcome value and one cost distribution per strategy, can be substituted without changing the rules.

Whether a useful safety-utility trade-off exists, and how a chance constraint traces it, is a property of the decision problem and the environment rather than of any particular estimator. It is therefore established under exact counterfactuals, where the outcome and cost distribution of each strategy can be computed without estimation error (Section~\ref{sec:experiments}). The dependence of the decision on predictor accuracy is a separate question. We address it by injecting controlled estimation error into those exact quantities, since committing to one learned predictor would tie the errors to its architecture and training. Because a predictor fit to observational data does not err symmetrically, the injected error includes a systematic cost underestimate as well as mean-zero noise. Section~\ref{sec:res-predictor} reports how large an underestimate the violation guarantee withstands.

\section{Experimental Design}
\label{sec:experiments}

We use five environments, ordered from fully synthetic to fully real. In four of them, the counterfactual outcome and treatment cost of every strategy can be computed exactly. Table~\ref{tab:environments} summarizes them by data provenance and by the cost and outcome each one measures, and Table~\ref{tab:envscale} gives their scale.

The evaluation measures decision quality against a known optimum rather than the out-of-sample cost that contextual-optimization studies typically report \citep{sadana2025contextual}. The environments exercise the features that distinguish the problem: a sequential horizon, a cumulative cost with a genuine upper tail, and identification of that cost from observational trajectories. A single-stage benchmark with uncertainty in the objective would test none of them. The environments also vary one axis at a time away from fully controlled dynamics. The simulators fix the dynamics entirely. The semi-synthetic environment replaces the covariates with real ones while keeping the outcome model that makes the oracle available. The maintenance environment replaces the clinical objective with an operations one on real degradation data. The comparison to a mean-budgeted approach is internal to every result. The naive rule of Equation~\eqref{eq:naive} imposes the expected-cost constraint, and its budget violations measure the cost of managing the mean instead of the tail. An external mean-budgeted method would optimize over a parameterized policy class, so the comparison would confound the constrained functional with the policy space. Holding the policy space fixed at the candidate set used here reduces a mean-budget constraint to Equation~\eqref{eq:naive} exactly. The rules therefore share a candidate set, a predictor, and an injected error realization, and differ only in the functional of the cost distribution each one constrains.

\begin{table}[pos=tbp]
\centering
\small
\caption{The five evaluation environments, ordered by increasing realism, with the source of each outcome and cost. Scale and range are in Table~\ref{tab:envscale}.}
\label{tab:environments}
\begin{adjustbox}{max width=\textwidth}
\begin{tabular}{@{}lccccc@{}}
\toprule
 & Sepsis & Tumor & MIMIC-IV & C-MAPSS & Drink Less \\
\midrule
Covariates & synthetic & synthetic & real & real & real \\
Outcome & synthetic & synthetic & synthetic & real degradation & real (observed) \\
Exact oracle & yes & yes & yes & yes & no (off-policy) \\
Counterfactuals & exact & exact & exact & exact (specified) & IPW \\
Cost unit & treatments & doses & doses & crew-hours & prompts \\
Outcome measure & survival & tumor control & survival & no failure & engagement \\
\bottomrule
\end{tabular}
\end{adjustbox}
\begin{flushleft}\footnotesize
Every synthetic component is an established benchmark or a standard modeling approach from the counterfactual-inference and reliability literatures. In the maintenance environment, the source trajectories run to failure and contain no maintenance actions. The overhaul effect is therefore a specified model component, not one identified from the benchmark.
\end{flushleft}
\end{table}

\subsection{Simulation environments (sepsis and tumor)}
\label{sec:exp-env}

The sepsis simulator of \citet{oberst2019gumbel} is a discrete Markov decision process over $720$ physiological states and $8$ actions. It is adopted as a benchmark for counterfactual and off-policy reasoning in sequential clinical decision making. A latent diabetes indicator drawn once per patient and held fixed makes treatment confounded with prognosis in the observational trajectories used later. The simulator is the primary setting because its transition and reward structure is fully specified over a finite state space. We compute the survival probability and the cumulative treatment cost of any strategy exactly by value iteration and rollout, which supplies the oracle decision quality is measured against.

The tumor-growth model of \citet{geng2017tumor} is the de facto benchmark for time-varying treatment-effect work. In it, chemotherapy and radiotherapy reduce tumor volume over a horizon. A candidate strategy is a volume-threshold rule, the cost is the cumulative treatment count, and the outcome is a final tumor volume below a fixed level. All three are computed exactly by simulating the dynamics. Its uncertainty arises from patient heterogeneity in the growth and treatment-response parameters rather than from stochastic transitions. The cost-distribution spread on which the chance constraint operates therefore has a different origin in each of the two. Both are specified in full in~\ref{app:envspec}.

\subsection{Real-covariate semi-synthetic environment (MIMIC-IV)}
\label{sec:exp-semisynth}

The semi-synthetic environment pairs the real covariate distributions of an intensive-care cohort with a synthetic outcome with counterfactuals known by construction. This retains the exact-regret measurement while removing the artificiality of simulator covariates. We draw the covariates from MIMIC-IV \citep{johnson2023mimiciv}, $3{,}000$ patient trajectories of up to twenty-four hourly steps from a sepsis cohort extracted by a standard pipeline. They therefore carry the severity structure of a real intensive-care population. The outcome follows the benchmark model of the Causal Transformer \citep{melnychuk2022ct}, a standard construction for semi-synthetic time-varying-treatment benchmarks on intensive-care data. Being a deterministic function of the covariates and the treatment sequence, it makes the outcome and cost of any strategy exactly computable, with no off-policy estimation. The candidate strategies are an escalation ladder indexed by a single aggressiveness parameter. Because the synthetic outcome is a design choice, we vary the outcome threshold and the physiological group driving the outcome. Section~\ref{sec:res-generalization} reports the decision-rule behavior under each. The cohort composition, the outcome construction and the ladder are given in~\ref{app:envspec}. We average all reported quantities over seeds that vary both the sampled cohort and the random components of the outcome.

\subsection{Cross-domain environment: predictive maintenance (C-MAPSS)}
\label{sec:exp-maintenance}

The maintenance environment carries the objective outside clinical care. We build it from the NASA C-MAPSS turbofan benchmark \citep{saxena2008cmapss}, a standard collection of run-to-failure sensor trajectories for a fleet of engines under a single operating condition. In a deployment, a remaining-useful-life predictor fit to such trajectories \citep{lei2018prognostics, babu2016cnn, zheng2017lstm, li2018rul} would play the role the g-computation model plays in the clinical environments, with the optimization step again agnostic to it. We summarize each engine's condition by a scalar health index rising from zero at installation toward one at failure, and a candidate strategy is an overhaul threshold. The cost is the cumulative number of maintenance crew-hours consumed over a fixed operating horizon, a hard budget that cannot be traded against the benefit. The outcome is availability, completion of the horizon without an unplanned failure. The benchmark records engines run to failure and contains no maintenance actions, so the overhaul effect is specified rather than estimated from it. This environment tests whether the optimization step transfers to an operations objective under a hard resource limit. It does not test whether the identification assumptions of Section~\ref{sec:method-setup} can be met from degradation data. The spread that makes the chance constraint informative arises from heterogeneity in time-to-failure across engines. The index construction, the overhaul reset and the averaging over sampled fleets are given in~\ref{app:envspec}.

\subsection{Real-outcome environment with known propensities (Drink Less)}
\label{sec:exp-realdata}

This environment replaces modeled outcomes with observed ones, at the price of the oracle. We build it from the Drink Less micro-randomized trial \citep{bell2023notifications}. In this design, the intervention is randomized repeatedly within each participant at every decision point \citep{klasnja2015mrt}. A behavior-change app for alcohol reduction either did or did not send a daily engagement notification, with a probability fixed by the trial design and so known exactly. We treat the notification as the action and its presence as the per-step cost. The budgeted cost is the cumulative number of notifications over the horizon. The outcome to be maximized is engagement with the app in the hour after the decision point. Candidate policies are global rates and state-aware rules that notify more often where the trial shows the notification to be most effective. Because the randomization probabilities are known, we estimate the value of a candidate without outcome-model assumptions, by per-decision inverse probability weighting of the proximal outcome \citep{bao2025perdecision}. This estimator avoids the trajectory-length weight degeneracy that afflicts importance sampling for long-horizon outcomes. We obtain the cost distribution directly, by accumulating a policy's notifications over each participant's observed state sequence. The chance constraint then acts on the tail of the realized cost induced by its randomization over the horizon. Both quantities are evaluated along the histories the trial generated, the standard form of micro-randomized-trial analysis \citep{bao2025perdecision}. For the global policies the cost is exact, since their notifications do not depend on the state. Elsewhere the estimates coincide with the counterfactual quantities of Section~\ref{sec:method-quantities} up to the effect of a notification on the subsequent state. With no specified model there is no exact oracle, so Sections~\ref{sec:exp-oracle} and~\ref{sec:exp-noise} do not apply and regret is not reported. We evaluate the decision rules instead on the estimated value and the estimated violation probability of the recommended policy, summarized over bootstrap resamples of the participants. The policy family and the softening are given in~\ref{app:envspec}.

\subsection{Candidate strategies}
\label{sec:exp-strategies}

The rule selects among a set of candidate strategies rather than optimizing over a parameterized policy class. The finite set is a property of the problem the framework is posed for. First, a cumulative-budget decision of this kind is usually made over a small number of standing alternatives. Examples are a treatment escalation ladder, a maintenance schedule at a few inspection thresholds, and a notification policy at a few intensities. In a deployment, those alternatives are given, fixed by clinical guideline, operating procedure, or regulation. The question put to the analyst is which of them to run under the budget, not what the space of all measurable regimes contains. Second, restricting to a finite set keeps the constraint exactly the object the method is about. The cost tail is evaluated on each strategy's own cumulative-cost distribution. Optimizing over a policy class would require a stage-wise decomposition, and the tail does not admit one. Third, the price of the restriction is bounded, since the guarantees of Section~\ref{sec:method-decision} depend on the set only through $\log\lvert\mathcal{G}\rvert$, so enlarging it costs little. Extending to a parameterized class instead would require a complexity measure in place of that term.

The evaluation cannot take the candidate set as given, because measuring regret against a computable optimum requires the counterfactual outcome and cost of every candidate. A set drawn from practice would not come with those. We therefore construct the set in each environment. The construction is an experimental design choice instead of a component of the method. \ref{app:placement} reports what turns on it. Three constructions that place the candidates differently along the cost axis leave the exchange the chance constraint makes intact. In the micro-randomized trial of Section~\ref{sec:exp-realdata}, the candidates are not solved for but written down from the trial's own randomization structure. They are fixed-rate and state-aware notification policies. This is closer to how a deployed set would arise.

In the simulators, we generate this set by solving for a range of treatment intensities; the semi-synthetic environment instead uses the escalation ladder of Section~\ref{sec:exp-semisynth}. For the sepsis simulator we apply value iteration with a per-step penalty $\lambda$ on the quantity of treatment administered. Sweeping $\lambda$ produces a family of deterministic strategies running from aggressive at $\lambda=0$ to conservative at large $\lambda$, each a well-defined dynamic treatment regime. The solver settings and the ways the candidates can be spaced along the cost axis are given in~\ref{app:placement}.

\subsection{Exact counterfactual computation}
\label{sec:exp-oracle}

For each candidate strategy we compute its outcome value and its cumulative-treatment-cost distribution by rolling the strategy out in the environment. The resulting outcome probability and cost distribution are exact up to Monte Carlo error, which we control by using a large number of rollout episodes. The outcome value is the fraction of episodes ending in discharge, and the cost distribution is the empirical distribution of cumulative treatments over episodes. These exact quantities define the oracle the evaluation uses. They give the oracle optimum at each budget, the outcome-maximizing strategy with expected cost within the budget, against which the regret of every rule is measured,
\begin{equation}
g^{\mathrm{orc}}(B) = \arg\max_{g \in \mathcal{G}} \, V(g) \quad \text{subject to} \quad \mathbb{E}\big[C(g)\big] \le B,
\qquad
\mathrm{Reg}(g) = V\big(g^{\mathrm{orc}}(B)\big) - V(g).
\label{eq:oracle}
\end{equation}
The word oracle is used in two ways below, both referring to this computation. The oracle cost distribution of a strategy is the exact counterfactual distribution it returns. The oracle at a budget is the mean-feasible optimum of Equation~\eqref{eq:oracle} computed from those exact quantities. The constraint in Equation~\eqref{eq:oracle} is on the mean cost instead of on its tail. The oracle at a budget is therefore risk-neutral and admits strategies the chance-constraint rule excludes. The regret of that rule against it then carries the intrinsic price of bounding the cost tail, not only the effect of estimation error. \ref{app:perbudget} reports the regret at zero injected error as a separate column for that reason. A rule can also record a negative regret under Equation~\eqref{eq:oracle}. It does so by selecting a higher-outcome strategy that the mean-feasible optimum excludes, at the price of budget violations. The exact quantities also give the true expected cost of a recommended strategy, against which budget violations are checked.

\subsection{Standing in for a predictor}
\label{sec:exp-noise}

In deployment, the outcome and cost of a strategy would be produced by an estimator returning the distribution of the cumulative cost, not only its mean. The decision rules differ only when they act on such estimates instead of on exact quantities. To study this regime without tying the conclusions to a particular learned model, we inject controlled estimation error into the exact quantities. We perturb the oracle outcome value of each strategy by relative Gaussian noise with standard deviation $\varepsilon_s$. We perturb the oracle cost samples by relative Gaussian noise with standard deviation $\varepsilon_c$. Each decision rule is then applied to the perturbed quantities, with regret and violation measured against the unperturbed oracle. This injected error is a relative perturbation of the estimated value and cost, distinct from the intrinsic spread of the cost distribution on which the chance constraint operates. We apply symmetric mean-zero perturbations, a systematically biased perturbation that underestimates the cost, and a rescaling of each strategy's cost samples about their own mean by a common factor. The rescaling holds every expected cost fixed and alters only how far the cumulative cost disperses around it. Varying the magnitude measures how the decision degrades as the predictor becomes less accurate. Perturbing the outcome and cost components separately attributes regret and violation to the error sources (Section~\ref{sec:res-predictor}).

\subsection{Decision rules and metrics}
\label{sec:exp-rules}

We compare the decision rules of Section~\ref{sec:method-decision}: the naive point-estimate rule of Equation~\eqref{eq:naive}, the upper-bound rule of Equation~\eqref{eq:ub} with margin parameter $\kappa$, and the chance-constraint rule of Equation~\eqref{eq:cc} with tolerated violation probability $\varepsilon$. We additionally report the conditional-value-at-risk rule, which admits a strategy when $\mathrm{CVaR}_{1-\varepsilon}(C(g)) \le B$, the safe convex approximation of Equation~\eqref{eq:cc}. We evaluate it on the same cost samples and at the same $\varepsilon$. They then differ only in which functional of the upper tail the budget constrains, the tail mean instead of the tail quantile. The naive rule serves as the plug-in benchmark, since it inserts the point estimates into the budget feasibility check. Constrained-Markov-decision-process and safe reinforcement-learning methods are not run as baselines, since they require online interaction or known transition dynamics and so address a different regime (Section~\ref{sec:related}). We compute the chance-constraint probability from the perturbed cost samples of each strategy and sweep $\varepsilon$ to trace the safety-utility frontier.

The outcome regret is $\mathrm{Reg}$ of Equation~\eqref{eq:oracle}, reported in percentage points (pp), with lower values better. The budget violation rate is the fraction of recommendations for which the true expected cost, evaluated under the oracle, exceeds the budget, also lower-is-better. The realized tail probability of a recommendation is $\Pr(C(g) > B)$ for the strategy it selects, computed on the true cost samples of that strategy. It is the quantity Equation~\eqref{eq:cc} constrains. It separates from the budget violation rate whenever a strategy is feasible in expectation while carrying an upper tail beyond the budget. The oracle match rate is the fraction of budget-and-seed combinations for which the selected strategy is the one the risk-neutral oracle of Section~\ref{sec:exp-oracle} selects at that budget. Constrained ranking and selection measures this quantity as the probability of correct selection \citep{hong2015ccsb}. Those procedures assume the best candidate is separated from the rest by a stated margin and introduce an acceptable set where it is not. We read the rate alongside regret: when several candidates attain nearly the same outcome, the rate can be low while regret stays small. Because the oracle is constrained on the mean cost, the rate measures how closely a rule recovers the risk-neutral choice. We report it for the point-estimate rule alone. The chance-constraint rule refuses strategies the oracle admits whenever their cost tail exceeds the tolerance, so its agreement is not informative. The safety-utility frontier reports the achievable outcome against the tolerated violation as $\varepsilon$ varies, and the budget frontier reports the oracle-optimal outcome as the budget varies. Counterfactual forecasting error is not reported, since what is claimed is a better decision from a given set of counterfactual estimates, not a better estimate.

\section{Results}
\label{sec:results}

The evaluation is outcome regret against the oracle optimum and the budget violation rate, and summary values are unweighted averages over each environment's budget grid. A budget enters an average only where every rule that average reports is defined in all $10$ seeds. A quantity comparing two rules is therefore taken over the budgets both admit at. A quantity reporting one rule alone is taken over the budgets that rule admits at. Per-budget tables are in~\ref{app:perbudget}. Sections~\ref{sec:res-frontier} to \ref{sec:res-feasibility} establish the frontier and the decision rules under controlled error. Section~\ref{sec:res-predictor} replaces controlled error with estimated predictors. Sections~\ref{sec:res-generalization} to \ref{sec:res-realdata} test transfer across environments, operating-point selection, and real outcomes.

\subsection{The exact budget-outcome frontier}
\label{sec:res-frontier}

For each budget $B$ we compute the strategy maximizing the outcome among those for which the expected cumulative cost does not exceed it. The computation uses the exact distributions obtained by rolling each candidate out in the simulator. These numbers therefore involve no estimation and characterize the trade-off itself. Table~\ref{tab:frontier} reports the result.

The frontier is monotone in the budget, as it must be, and its curvature is the informative part. The outcome gained per unit of cost rises to $8.6$ percentage points over the step from $B=6$ to $B=8$. It then falls to $3.2$ by $B=12$, to $1.9$ by $B=20$, and to $0.2$ once the budget is removed. The value of an additional unit of budget is therefore not constant, and a recommender reporting only the best strategy at a given budget does not convey it.

\subsection{Decision rules under estimation uncertainty}
\label{sec:res-rules}

The oracle quantities are perturbed as Section~\ref{sec:exp-noise} sets out, and regret and violation are measured against the oracle optimum at the same budget. Table~\ref{tab:app-sepsis} reports the rules at every budget at an injected error of $15\%$ and $\varepsilon = 0.2$, since what each rule gives up varies sharply with the budget.

The naive rule is never far from the oracle. Its regret stays between $1.4$ and $3.5$ percentage points across the whole grid, well below the $9.3$ to $26.6$ points the chance-constraint rule gives up. From $B=4$ upward, the naive rule selects strategies with true expected cost above the budget, in $18.7\%$ of cases at $B=4$ and in $4.1\%$ at $B=20$. The chance-constraint rule violates the budget at no budget in the grid, with no seed-to-seed spread. The comparison is therefore not which rule attains more outcome. It is what the guarantee costs.

The violation is also unstable where it is largest, averaging $18.7\%$ with a standard deviation of $19.7$ at $B=4$. A few seeds overrun in most cases while others never do, so a decision maker deploying the rule once would not know which draw had been taken.

\begin{figure}[pos=tbp]
\centering
\begin{minipage}[t]{0.49\linewidth}\centering
\includegraphics[width=0.85\linewidth,height=0.32\textheight,keepaspectratio]{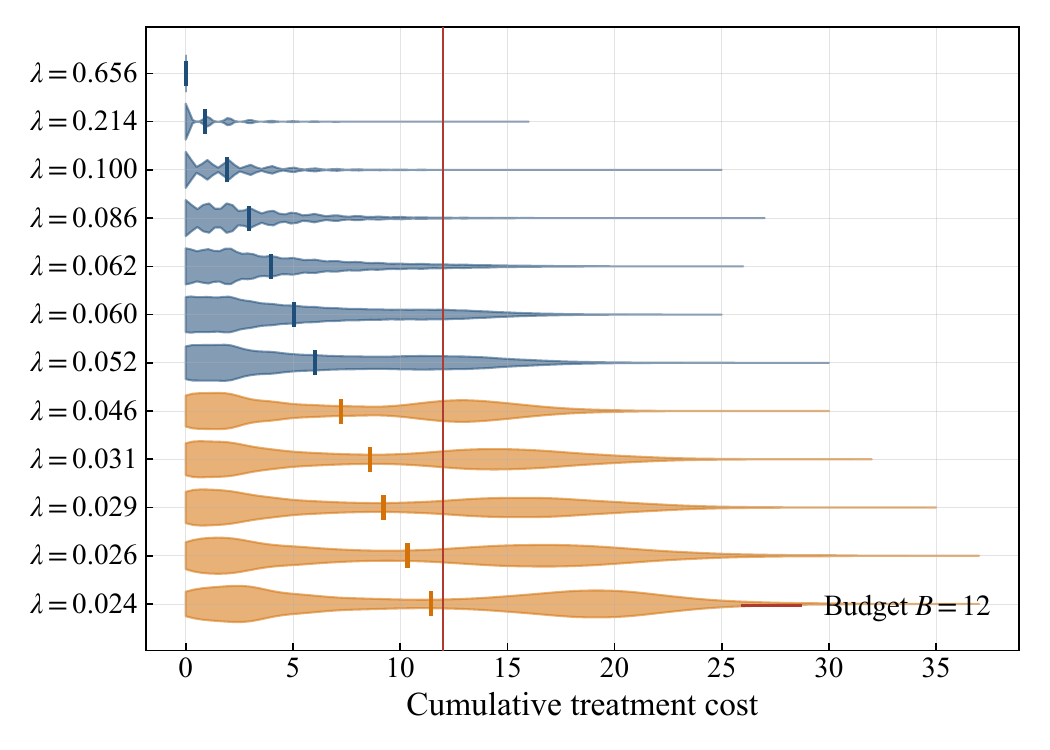}\\[0.4ex]{\small (a)}
\end{minipage}\hfill
\begin{minipage}[t]{0.49\linewidth}\centering
\includegraphics[width=0.85\linewidth,height=0.32\textheight,keepaspectratio]{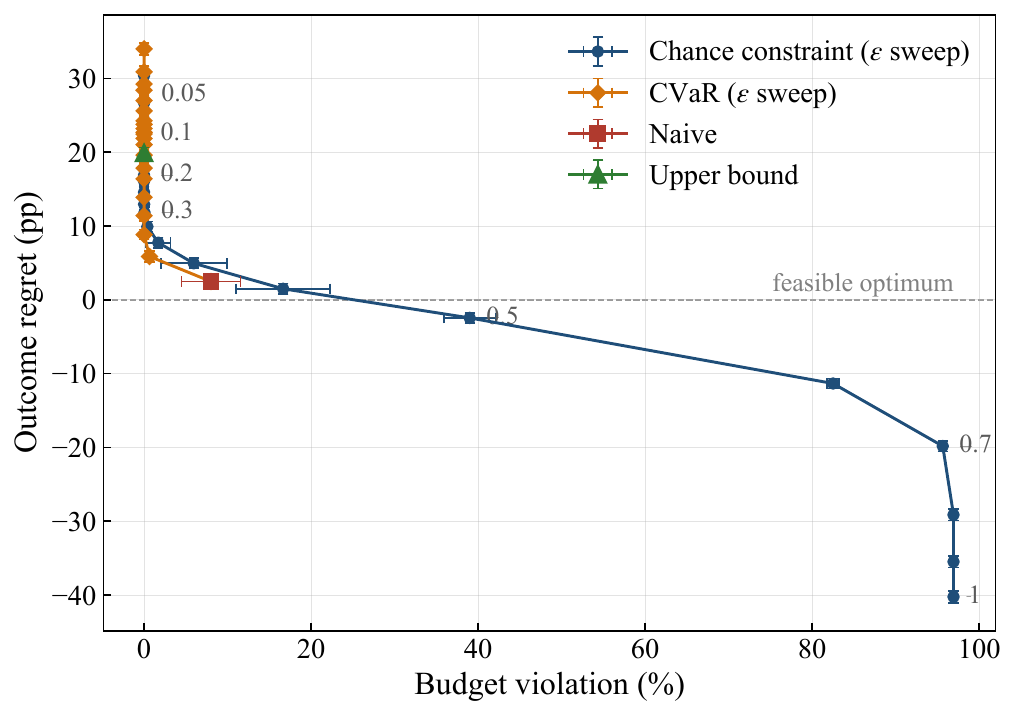}\\[0.4ex]{\small (b)}
\end{minipage}
\caption{Sepsis environment at budget $B=12$. (a) Cumulative-cost distributions of the $12$ of $21$ candidates feasible in expectation, with the five refused at $\varepsilon = 0.2$ in orange; (b) safety-utility frontier over $\varepsilon$ at $15\%$ injected error, $10$ seeds, naive and upper-bound rules as reference points.}
\label{fig:sepsis}
\end{figure}

The realized cost tail locates where the naive rule's outcome advantage comes from. At $B=12$, the strategy it selects places $42.7\%$ of its cumulative-cost mass above the budget, against $12.6\%$ for the strategy the chance-constraint rule selects. That fraction grows with the budget for the naive rule, from $20.5\%$ at $B=1$ to $47.7\%$ at $B=20$. For the chance-constraint rule, it never exceeds the tolerance of $0.2$ at any budget. Figure~\ref{fig:sepsis}(a) shows why the two rules can disagree while both respect the budget in expectation. Twelve of the twenty-one candidates are feasible in expectation at $B=12$. Five of those twelve place more than a fifth of their cost mass above it, and the chance constraint refuses exactly those five.

The two tightest budgets qualify the picture. At $B=1$ and $B=2$, the naive rule violates nothing and gives up less than the chance-constraint rule. The reason is that only two or three strategies are feasible in expectation and the constraint removes options without averting an overrun. Its selection at $B=1$ nonetheless carries a cost tail of $20.5\%$, which the budget violation rate does not register because the true mean stays inside the budget. Section~\ref{sec:res-cvar} evaluates the rules on the realized tail for that reason.

The price of the guarantee is not constant, largest at $B=8$ and $B=12$ at $25.6$ and $26.6$ points and smallest at $B=4$ at $9.3$. A decision maker cannot know in advance which budget will be in force. The same estimation noise that leaves the naive rule mean-feasible at $B=2$ makes it overrun at $B=4$. The chance-constraint rule, by contrast, converts a stated tolerance into a strategy regardless of which budget applies.

Sweeping the tolerance traces that exchange directly. Table~\ref{tab:epssweep} reports the sweep at $B=12$, the budget at which the naive rule's regret is largest. Figure~\ref{fig:sepsis}(b) plots regret against violation with the naive and upper-bound rules as reference points.

\begin{table}[pos=tbp]
\centering
\small
\caption{Chance-constraint frontier in the sepsis environment, swept over $\varepsilon$ at budget $B=12$.}
\label{tab:epssweep}
\begin{adjustbox}{max width=\textwidth}
\begin{tabular}{@{}lrrrrrrrrr@{}}
\toprule
$\varepsilon$ & 0.02 & 0.05 & 0.10 & 0.20 & 0.30 & 0.40 & 0.50 & 0.70 & 1.00 \\
\midrule
Outcome regret (pp) & $\num{eps_budget.sepsis.eps0.02.B12.regret_pp}{46.4} \pm \num{eps_budget.sepsis.eps0.02.B12.regret_pp_std}{0.5}$ & $\num{eps_budget.sepsis.eps0.05.B12.regret_pp}{42.0} \pm \num{eps_budget.sepsis.eps0.05.B12.regret_pp_std}{0.6}$ & $\num{eps_budget.sepsis.eps0.1.B12.regret_pp}{36.4} \pm \num{eps_budget.sepsis.eps0.1.B12.regret_pp_std}{0.7}$ & $\num{eps_budget.sepsis.eps0.2.B12.regret_pp}{26.6} \pm \num{eps_budget.sepsis.eps0.2.B12.regret_pp_std}{0.5}$ & $\num{eps_budget.sepsis.eps0.3.B12.regret_pp}{16.9} \pm \num{eps_budget.sepsis.eps0.3.B12.regret_pp_std}{1.0}$ & $\num{eps_budget.sepsis.eps0.4.B12.regret_pp}{8.9} \pm \num{eps_budget.sepsis.eps0.4.B12.regret_pp_std}{0.6}$ & $\num{eps_budget.sepsis.eps0.5.B12.regret_pp}{2.5} \pm \num{eps_budget.sepsis.eps0.5.B12.regret_pp_std}{0.4}$ & $\num{eps_budget.sepsis.eps0.7.B12.regret_pp}{-12.9} \pm \num{eps_budget.sepsis.eps0.7.B12.regret_pp_std}{0.5}$ & $\num{eps_budget.sepsis.eps1.B12.regret_pp}{-15.1} \pm \num{eps_budget.sepsis.eps1.B12.regret_pp_std}{0.4}$ \\
Budget violation (\%) & $\num{eps_budget.sepsis.eps0.02.B12.violation_pct}{0.0} \pm \num{eps_budget.sepsis.eps0.02.B12.violation_pct_std}{0.0}$ & $\num{eps_budget.sepsis.eps0.05.B12.violation_pct}{0.0} \pm \num{eps_budget.sepsis.eps0.05.B12.violation_pct_std}{0.0}$ & $\num{eps_budget.sepsis.eps0.1.B12.violation_pct}{0.0} \pm \num{eps_budget.sepsis.eps0.1.B12.violation_pct_std}{0.0}$ & $\num{eps_budget.sepsis.eps0.2.B12.violation_pct}{0.0} \pm \num{eps_budget.sepsis.eps0.2.B12.violation_pct_std}{0.0}$ & $\num{eps_budget.sepsis.eps0.3.B12.violation_pct}{0.0} \pm \num{eps_budget.sepsis.eps0.3.B12.violation_pct_std}{0.0}$ & $\num{eps_budget.sepsis.eps0.4.B12.violation_pct}{0.1} \pm \num{eps_budget.sepsis.eps0.4.B12.violation_pct_std}{0.1}$ & $\num{eps_budget.sepsis.eps0.5.B12.violation_pct}{19.4} \pm \num{eps_budget.sepsis.eps0.5.B12.violation_pct_std}{3.4}$ & $\num{eps_budget.sepsis.eps0.7.B12.violation_pct}{97.3} \pm \num{eps_budget.sepsis.eps0.7.B12.violation_pct_std}{0.4}$ & $\num{eps_budget.sepsis.eps1.B12.violation_pct}{99.2} \pm \num{eps_budget.sepsis.eps1.B12.violation_pct_std}{0.2}$ \\
\bottomrule
\end{tabular}
\end{adjustbox}
\begin{flushleft}\footnotesize
Mean $\pm$ s.d. over $10$ seeds at $15\%$ injected error. At $\varepsilon = 1$, the constraint is vacuous.
\end{flushleft}
\end{table}

The exchange is monotone over the $21$ tolerances in the grid, with no reversal at any step. Regret falls from $46.4$ points at $\varepsilon = 0.02$ to $-15.1$ at $\varepsilon = 1$ while violation rises from zero to $99.2\%$. Violation stays at exactly zero up to $\varepsilon = 0.33$, by which point half the outcome available on this frontier has been recovered. It stays under a tenth of a percent through $\varepsilon = 0.40$. Beyond $\varepsilon = 0.5$ the rule gains outcome almost entirely by overrunning.

The separation is not an artifact of how densely the candidate set is populated. \ref{app:placement} repeats the comparison on the practice-sized set of seven a deployment would more plausibly enumerate. The naive rule violates under both constructions, at $4.6\%$ against $8.0\%$ under $21$. The chance-constraint rule holds at zero under both. A denser set supplies more strategies with cost tail near the budget. The opportunity to misjudge feasibility from a point estimate therefore grows with the candidate count rather than shrinking.

\subsection{Comparison with conservative decision rules}
\label{sec:res-cvar}

The chance constraint bounds how often the cumulative cost exceeds the budget. A conditional-value-at-risk constraint bounds the average of the costs that do exceed it. The upper-bound rule bounds neither.

The comparison cannot be read on the budget violation rate. The budget violation rate counts a recommendation as violating when the true expected cost exceeds the budget. Both conservative rules admit only mean-feasible strategies by construction, so their rate is zero at every tolerance short of one. Matching on it would measure one rule against a criterion the other satisfies as an identity. We evaluate each rule on the realized tail probability of Section~\ref{sec:exp-rules}, the quantity the chance constraint is stated over.

Table~\ref{tab:calibration} reports the ratio of the selected strategy's realized tail probability to the tolerance that produced it, over the whole tolerance grid. Under the chance constraint, the fraction rises with the tolerance in every environment and reaches between $0.75$ and $0.91$ at the loose end of the grid. A decision maker who raises the tolerated violation obtains a recommendation with an overrun probability that moves with it. Under the conditional value at risk, it saturates below $0.31$ everywhere and below $0.10$ in the maintenance environment. That tolerance is a conservatism setting whose realized effect is three to ten times smaller than its nominal value and varies threefold across environments. The upper-bound rule cannot be entered in the table at all. Its $\kappa$ is a multiple of a standard deviation, and no tolerance exists against which a realized tail could be read as a fraction. Choosing between $\kappa = 0.25$ and $\kappa = 0.5$ requires measuring decision quality against the true optimum. A deployment has no such optimum.

\begin{figure}[pos=tbp]
\centering
\includegraphics[width=0.85\linewidth,height=0.24\textheight,keepaspectratio]{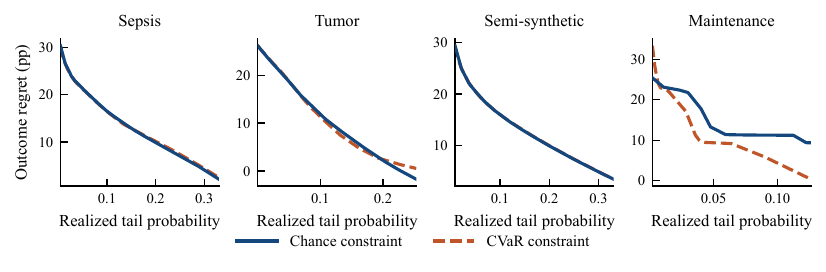}
\caption{Outcome regret against realized tail probability, chance constraint (blue) against conditional value at risk (orange), by environment, $10$ seeds.}
\label{fig:cvar}
\end{figure}

\begin{table}[pos=tbp]
\centering
\small
\caption{Ratio of the selected strategy's realized tail probability to the tolerance that produced it.}
\label{tab:calibration}
\begin{adjustbox}{max width=\textwidth}
\begin{tabular}{@{}lcccccccc@{}}
\toprule
& \multicolumn{2}{c}{Sepsis} & \multicolumn{2}{c}{Tumor} & \multicolumn{2}{c}{Semi-synthetic} & \multicolumn{2}{c}{Maintenance} \\
\cmidrule(lr){2-3} \cmidrule(lr){4-5} \cmidrule(lr){6-7} \cmidrule(lr){8-9}
Tolerance & chance & CVaR & chance & CVaR & chance & CVaR & chance & CVaR \\
\midrule
$\varepsilon = 0.10$ & \num{calibration.sepsis.chance.eps0.1}{0.32} & \num{calibration.sepsis.cvar.eps0.1}{0.08} & \num{calibration.tumor.chance.eps0.1}{0.24} & \num{calibration.tumor.cvar.eps0.1}{0.02} & \num{calibration.mimic.chance.eps0.1}{0.48} & \num{calibration.mimic.cvar.eps0.1}{0.16} & \num{calibration.cmapss.chance.eps0.1}{0.30} & \num{calibration.cmapss.cvar.eps0.1}{0.05} \\
$\varepsilon = 0.20$ & \num{calibration.sepsis.chance.eps0.2}{0.48} & \num{calibration.sepsis.cvar.eps0.2}{0.12} & \num{calibration.tumor.chance.eps0.2}{0.53} & \num{calibration.tumor.cvar.eps0.2}{0.07} & \num{calibration.mimic.chance.eps0.2}{0.58} & \num{calibration.mimic.cvar.eps0.2}{0.17} & \num{calibration.cmapss.chance.eps0.2}{0.30} & \num{calibration.cmapss.cvar.eps0.2}{0.06} \\
$\varepsilon = 0.50$ & \num{calibration.sepsis.chance.eps0.5}{0.79} & \num{calibration.sepsis.cvar.eps0.5}{0.20} & \num{calibration.tumor.chance.eps0.5}{0.87} & \num{calibration.tumor.cvar.eps0.5}{0.20} & \num{calibration.mimic.chance.eps0.5}{0.74} & \num{calibration.mimic.cvar.eps0.5}{0.23} & \num{calibration.cmapss.chance.eps0.5}{0.64} & \num{calibration.cmapss.cvar.eps0.5}{0.08} \\
maximum & \num{calibration.sepsis.chance.max}{0.89} & \num{calibration.sepsis.cvar.max}{0.31} & \num{calibration.tumor.chance.max}{0.91} & \num{calibration.tumor.cvar.max}{0.25} & \num{calibration.mimic.chance.max}{0.75} & \num{calibration.mimic.cvar.max}{0.31} & \num{calibration.cmapss.chance.max}{0.90} & \num{calibration.cmapss.cvar.max}{0.10} \\
\bottomrule
\end{tabular}
\end{adjustbox}
\begin{flushleft}\footnotesize
Ten seeds. A value of one means the realized tail equals the tolerance exactly; values below one are conservative, the direction Proposition~3 permits. The maximum is over the tolerances below one, since at $\varepsilon = 1$ the constraint is vacuous. The upper-bound rule has no tolerance to form the ratio against.
\end{flushleft}
\end{table}

The same ordering appears in how often each rule declines to recommend. The conditional-value-at-risk rule declines more often at every tolerance in the environments where either rule declines;~\ref{app:maintenance} gives the rates. Declining is informative rather than a failure, since it reports that the budget is infeasible at the stated risk appetite, which a mean-constrained rule cannot express. A declined recommendation is excluded from that rule's averages rather than counted as zero. No rule is therefore charged for the outcome it forgoes, and these rates are read alongside the regret.

Proposition~3 bounds the realized tail probability at each budget separately, so the check is made at that level. Across the four environments, $21$ tolerances and every budget, the mean realized tail probability of the admitted strategy stays at or below the tolerance. This holds in all $520$ combinations at which the rule admits a strategy. At the level of the individual recommendation, the realized tail exceeds the nominal tolerance in up to $47.9\%$ of cases. It exceeds the certified ceiling $\varepsilon + \delta$ of Equation~\eqref{eq:l0feas} in at most $0.40\%$, against the $5\%$ the Dvoretzky-Kiefer-Wolfowitz bound allows. Those exceedances sit entirely at tolerances of $0.22$ and above. The rule therefore approaches the ceiling only where the constraint is loose enough that it operates at the boundary.

\begin{table}[pos=tbp]
\centering
\small
\caption{Mean absolute outcome margin of each conservative rule against the chance-constraint rule at matched realized tail probability, in percentage points.}
\label{tab:matchedtail}
\begin{tabular}{@{}lrr@{}}
\toprule
 & CVaR & Upper bound \\
\midrule
Sepsis & $0.21$ & $0.24$ \\
Tumor & $0.58$ & $0.76$ \\
Semi-synthetic & $0.04$ & $0.12$ \\
Maintenance & $4.9$ & $5.3$ \\
\bottomrule
\end{tabular}
\begin{flushleft}\footnotesize
At a matched realized tail probability of $0.15$ the chance-constraint and conditional-value-at-risk rules both reach $12.9$ points of regret in sepsis and $12.8$ in the semi-synthetic environment.
\end{flushleft}
\end{table}

Table~\ref{tab:matchedtail} reports the outcome margin of each conservative rule against the chance constraint, read at a common realized tail probability (Figure~\ref{fig:cvar}). The rules are built on different functionals, one on the tail mean and one on a multiple of the standard deviation. They nonetheless trace close to the same efficient set as the chance constraint in these environments.

\paragraph{The maintenance environment} At a matched realized tail probability, both conservative rules reach less regret than the chance-constraint rule there. The margin is $4.9$ percentage points for the conditional value at risk and $5.3$ for the upper-bound rule. One cause accounts for both margins. The outcome takes three levels across the $25$ candidates, and $23$ tie at the top. At budgets of four and above, the regret is therefore near zero whichever strategy is selected. At $B=2$, the chance constraint admits no strategy until the tolerance reaches $0.13$, and the conditional value at risk none until $0.30$. The comparison is a single binary choice at $B=3$, where choosing wrongly costs $56.24$ points. The averaged regret is that one error spread over the budgets the average retains. Regret counts only the outcome. A rule that declines an overrunning strategy is therefore charged for what it gives up and credited with nothing for the overruns it avoids. At that budget, the chance constraint refuses the overrunning strategy and accepts a realized tail of $0.006$. The conditional value at risk at a vacuous tolerance accepts $0.238$. The tumor environment shares the structure at a smaller scale, with $20$ of $23$ candidates tied. \ref{app:maintenance} works through the arithmetic and the per-budget breakdown.

The two constraints answer different questions and the framework accommodates either. Where an overrun is an administrative event with cost largely independent of its size, the frequency is the quantity to bound. A crew-hour budget exceeded by one shift or by three is an overrun of this kind. The chance constraint states the frequency in units a decision maker can set. Where the harm grows with the size of the overrun, as for a cumulative-exposure limit, the average severity is the quantity to bound. The conditional value at risk is then the appropriate choice, at the cost of a tolerance that must be calibrated environment by environment.

\subsection{Finite-sample feasibility across sample sizes}
\label{sec:res-feasibility}

Proposition~3 bounds the realized tail probability of an admitted strategy by $\varepsilon + \delta(n)$ once the cost distribution is estimated from $n$ independent samples. The bound is distribution-free, so how much of it the data uses is an open question at deployment sample sizes. We apply the empirical rule to resamples of every candidate's oracle cost distribution; Table~\ref{tab:e2} reports the protocol and the result.

The observed deviation is a roughly constant fraction of $\delta(n)$ across a fortyfold range of sample size, running from $0.65$ to $0.72$ depending on the environment. The bound is therefore correct in its dependence on $n$ and loose by about a third in its constant. Premise and conclusion both hold inside their nominal levels.

These deviations are the sampling term alone: the cost distributions are the exact ones, and $\hat{F}_g$ differs from $F_g$ only because it is built from finitely many draws. Section~\ref{sec:res-predictor} measures the same quantity when the cost distribution is itself estimated from observational trajectories, which is the model term. A deployment carries both, and Proposition~3 then applies with $\delta$ replaced by their sum.

\subsection{Estimation error and an estimated predictor}
\label{sec:res-predictor}

\begin{figure}[pos=tbp]
\centering
\includegraphics[width=\linewidth,height=0.40\textheight,keepaspectratio]{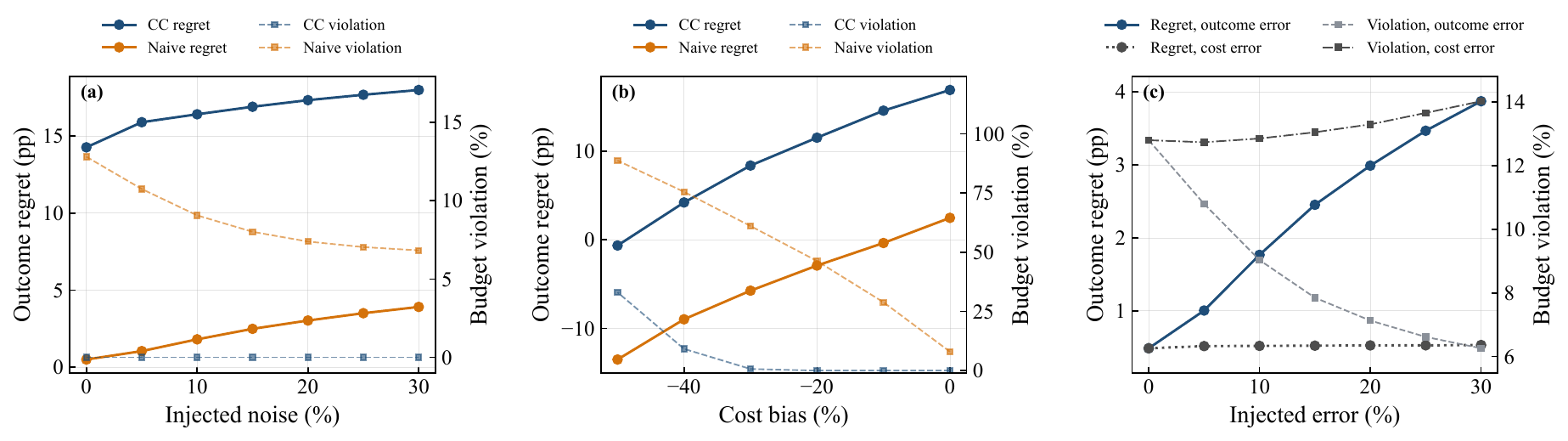}
\caption{Sensitivity of the decision to estimation error, sepsis environment, averaged over the budget grid. (a) Mean-zero relative noise; (b) one-sided cost underestimate; (c) outcome-only against cost-only error, naive rule. Means over $10$ seeds; we report seed-to-seed spread per budget in~\ref{app:perbudget}.}
\label{fig:predictor}
\end{figure}

The decision rests on estimates of outcome and cost, so its behavior as those degrade determines what a deployment can rely on. We inject controlled error into the exact quantities and average the response over the budget grid. Figure~\ref{fig:predictor} reports mean-zero noise, a one-sided cost underestimate, and the decomposition of the injected error into its outcome and cost components.

Under mean-zero noise, the chance-constraint violation is exactly zero at every injected-error level from zero to $30\%$, on every seed and every budget. Its regret rises from $14.3$ to $18.0$ points. Degraded estimates cost outcome without ever breaching the budget. The naive rule moves the other way on both axes, its violation falling from $12.8\%$ to $6.8\%$ as its regret rises from $0.5$ to $3.9$ points. Outcome noise redirects the selection away from the boundary strategy its own sampling error would otherwise admit. At zero injected error, its violation is that sampling effect alone. The chance-constraint rule's exposure does not depend on the noise level at all.

A trained predictor does not err symmetrically. Because units in worse states receive more intervention in observational data, a predictor fit to such data can \emph{underestimate} cost. Underestimation is the error mode most consequential for a tail constraint. We inject a one-sided cost bias on top of the mean-zero noise and sweep it past what the estimated predictors below exhibit. The sweep therefore locates the breaking point rather than an operating point.

Figure~\ref{fig:predictor}(b) reports it. The chance-constraint violation stays at zero through a $10\%$ underestimate, reaches $0.70\%$ at $30\%$, then $9.2\%$ at $40\%$ and $33.1\%$ at $50\%$. The naive rule runs at $28.9\%$ to $88.7\%$ under the same bias, a factor of $87$ apart at a $30\%$ underestimate. A modest bias moves the chance-constraint rule's selection to a different strategy that is still truly feasible. The violation rises only once the bias pulls a truly infeasible strategy into the set the biased estimate judges feasible. The upper-bound rule absorbs more bias, holding zero through $30\%$, with higher regret at every level. The level at which the guarantee fails is a requirement on deployment rather than a property of the rule, here between a $30\%$ and a $40\%$ underestimate. We read that band on the aggregate bias, and the estimated predictors below include one that satisfies it and breaks the rule anyway.

The error sources act on different objectives and in opposite directions, as Figure~\ref{fig:predictor}(c) shows. Outcome error raises the regret from $0.5$ to $3.9$ points and lowers the violation from $12.8\%$ to $6.3\%$. Cost error does the reverse, leaving the regret at $0.5$ points and raising the violation to $14.0\%$. Accuracy on the cost distribution and on the outcome are therefore not interchangeable, and no single forecast-accuracy figure determines decision quality.

A predictor can place the expected cost correctly and still understate how widely the cost disperses around it, which no perturbation above reaches. Rescaling each strategy's cost samples about their own mean by a common factor $s$ holds every expected cost fixed. The point-estimate rule's admitted set is therefore identical across the sweep, and everything the chance-constraint rule does across $s$ is attributable to the spread alone. Understating it raises the outcome and the realized tail together. As $s$ falls from one to one half, the chance-constraint regret falls by $5.7$, $5.5$ and $5.3$ outcome percentage points in sepsis, tumor and the semi-synthetic environment. This is because an understated tail admits strategies the correct one refuses. The realized tail probability of the selected strategy meanwhile rises to $0.184$, $0.154$ and $0.206$ against a nominal tolerance of $0.2$. The largest of these exceeds the tolerance, because samples with an understated spread no longer satisfy the premise of Proposition~3. The budget violation moves far less, staying at zero in two environments and rising only from $0.02\%$ to $0.08\%$ in the tumor environment. The effect of an understated spread is therefore recorded mostly on the realized tail probability, not on the budget violation rate. \ref{app:spread} reports the sweep, the maintenance environment, and the region of $s$ it excludes.

\paragraph{An estimated predictor end to end}
The injected error above is a controlled substitute for a predictor. We replace it with an estimated predictor. The sepsis state space is finite. The transition kernel can therefore be estimated by counting transitions rather than by function approximation, without any tuning choice that would confound the reading. We generate observational trajectories under a behavior policy that conditions on severity, so treatment is confounded with prognosis. We fit the kernel by categorical maximum likelihood and roll each of the $21$ candidate strategies forward under it. State-action pairs the behavior policy never visits receive a uniform transition over states. The estimated outcome value and cost distribution go to both rules unchanged, and the oracle enters only to score the result. \ref{app:backbone} records the validation of the rollout against the simulator's own implementation. Estimation and decision quality move together between $1{,}000$ and $5{,}000$ trajectories. The uniform deviation falls from $0.294$ to $0.097$, the point-estimate rule's violation from $18.6\%$ to $2.9\%$ and its regret from $4.85$ to $2.89$ points. The fraction of evaluation episodes traversing a state-action pair the behavior data never visited falls from $26.5\%$ to $7.7\%$. The last is the mechanism: where the observational data does not cover the region a strategy traverses, the uniform fallback supplies the transition. The estimated cost distribution is then wrong in the tail.

At $1{,}000$ trajectories the point-estimate rule overruns the budget in $18.6\%$ of recommendations. The chance-constraint rule overruns in none, at that sample size and at every larger one. Its regret is the price, $16.37$ points against $4.85$. The regrets then behave differently as the estimate improves. The point-estimate rule's regret falls from $4.85$ to $0.57$ points, which is estimation error dissolving. The chance-constraint rule's stays between $15.36$ and $16.37$, because it is not estimation error but the outcome conceded by refusing strategies the mean-feasible oracle admits. A deployment cannot tell from a single run which sample-size regime it is in. The exposure of the point-estimate rule depends on that regime and the exposure of the chance-constraint rule does not.

The estimator errs in the level of cost, not in its order;~\ref{app:pathb-diag} reports the rank correlations and the bias. A rule selecting on cost ranking alone would be unaffected. The chance constraint, however, compares an estimated cost distribution against an absolute budget, so the level governs it. Its realized cost tail stays between $0.579$ and $0.697$ of the nominal tolerance across the four sample sizes. The rule is therefore conservative against its own stated level under model error as well as under sampling error.

\paragraph{Function approximation} Counting is viable only because the sepsis state space is finite. \begin{table}[pos=tbp]
\centering
\small
\caption{Function approximation at $20{,}000$ trajectories, chance-constraint rule at $\varepsilon = 0.2$.}
\label{tab:fa}
\begin{tabular}{@{}llrrrrr@{}}
\toprule
& & & \multicolumn{2}{c}{Deviation} & \multicolumn{2}{c}{Violation (\%)} \\
\cmidrule(lr){4-5}\cmidrule(lr){6-7}
Environment & Estimator & Cost bias (\%) & Largest & $> \delta$ & Point estimate & Chance \\
\midrule
Tumor          & ensemble & $-14.30$ & $0.252$ & $6/23$  & $10.0$ & $0.0$ \\
               & ridge    & $+8.86$  & $0.103$ & $0/23$  & $0.0$  & $0.0$ \\
Semi-synthetic & ensemble & $+2.51$  & $0.147$ & $4/25$  & $6.7$  & $0.0$ \\
               & ridge    & $-5.30$  & $0.261$ & $16/25$ & $96.7$ & $11.7$ \\
\bottomrule
\end{tabular}
\begin{flushleft}\footnotesize
The certified slack is $\delta = 0.131$ in both environments. The deviation column is the largest per-strategy deviation over seed means, as in Tables~\ref{tab:tumor-fa} and~\ref{tab:mimic-fa}, which report all sample sizes.
\end{flushleft}
\end{table}

On the tumor environment, a gradient-boosted ensemble and a ridge regression of deliberately different flexibility err in opposite directions (Table~\ref{tab:fa};~\ref{app:tumor-fa}). Neither bias contracts as the sample grows, which places it in the model rather than in the sampling. The bias is residual confounding the fitted model does not remove.

That changes what the point-estimate rule is exposed to. In the sepsis environment, its violation fell to zero as the estimate improved, so its exposure was a small-sample condition. Here it does not fall at any of four sample sizes spanning two orders of magnitude (Table~\ref{tab:fa}). Under the ridge, it is zero throughout, not because the rule is safe but because that estimator happens to err upward. The safety of a rule reading only the mean is a property of which way its predictor errs.

The chance-constraint rule holds the violation at zero under both estimators at every sample size. It does so even where the premise of Proposition~3 fails. The guarantee is a sufficient condition on the deviation, not a necessary one, and the rule is observed to hold on the far side of it. The regret is $8.1$ to $10.3$ points under the ensemble and $13.9$ to $14.4$ under the ridge. Regret is small for either rule in absolute terms because $20$ of the $23$ candidates tie at an outcome of $99.9\%$.

The semi-synthetic environment has the tightest cost-underestimate band of the four, at $6.6\%$ violation for a $30\%$ underestimate against $0.70\%$ in sepsis. We fit the same estimators there, in~\ref{app:mimic-fa}. Both aggregate biases are small, the ensemble overestimating the cost by $2.5\%$ and the ridge underestimating it by $5.3\%$. The band therefore admits either with a wide margin. The aggregate bias conceals the ridge's per-strategy errors. These run from $-28\%$ on four adjacent strategies to $+79\%$ on one and cancel in the mean. A band read on the aggregate does not detect this, and the deviation does. The ridge exceeds the certified slack $\delta = 0.131$ at $16$ of the $25$ strategies, at a largest per-strategy value of $0.261$.

The chance-constraint rule does not hold the budget here. This is the only configuration of Table~\ref{tab:fa} in which it fails. It admits a violating strategy in $11.7$ to $15.0\%$ of recommendations, and the realized tail of what it selects reaches $0.324$ against a nominal tolerance of $0.2$. The certified ceiling still holds. Proposition~3 bounds the realized tail probability at $\varepsilon + \delta = 0.331$ under its premise, and the realized tail stays under that at every sample size, at $0.323$ to $0.329$. The point-estimate rule on the same estimator overruns in $93$ to $97\%$ of recommendations, at a negative regret of about $-5.7$ points. The rule gains outcome the mean-feasible optimum excludes, with the corresponding overruns.

The cause is the estimator, not the covariates. \ref{app:mimic-fa} reports the coverage correlations and the overlap between the estimators' failure sets. The contrast with the spread sweep above locates the mechanism. There the mis-statement is common to every strategy, and almost everything the widened admitted set takes in is still feasible in expectation. The violation therefore stays below one percent. Here it differs across strategies by more than a hundred percentage points, and the rule admits infeasible strategies in more than one recommendation in ten. Uniform mis-statement of the spread is the benign form and dispersion of it across strategies is the damaging one. The requirement therefore transfers, and the number does not. A predictor supplying the cost distribution must keep its deviation inside the certified slack. The deviation column of Table~\ref{tab:fa} checks this, and a bias band read on the aggregate is not a substitute.

\paragraph{The share of the regret that is estimation error} The exact counterfactual distribution of each strategy bounds what any predictor could deliver to the rule. Running the same rule on that distribution at $20{,}000$ trajectories and comparing within seed separates the price of bounding the tail from the price of estimating it. \ref{app:ceiling} reports the three arms, the paired intervals and the scale factors. In sepsis, the estimated predictor is ahead of the exact-distribution arm at every tolerance up to $0.3$. The sign follows from the benchmark. The oracle at a budget is mean-feasible, so a predictor reading the tails exactly refuses strategies the oracle admits, while one understating a tail admits them. Under function approximation, both estimators leave a gap. The gap reaches $9.95$ outcome percentage points for the ridge at a tolerance of $0.2$, with every paired interval excluding zero. Correcting the level does not recover that part. A single pooled rescaling to the correct mean leaves one estimator's selection unchanged and moves the other the wrong way. The estimators misstate the shape of the upper tail rather than its position, and a scalar applied after the fact does not reach it.

\subsection{Generalization across environments}
\label{sec:res-generalization}

The remaining environments differ in mechanism, in the source of their uncertainty, and in one case in domain. In the tumor and maintenance environments, the budget stops binding early and the frontier is flat thereafter; in sepsis and the semi-synthetic environment it binds throughout the grid. Where the frontier saturates, every budget above the saturation point admits a strategy at the top outcome level and the decision is trivial there. That makes the regret comparisons in those two environments turn on a single step. Full per-budget frontiers are in~\ref{app:perbudget}.

Panel (a) of Table~\ref{tab:three-env} reports the naive and chance-constraint rules averaged over each environment's budget grid, the two the violation axis separates. Section~\ref{sec:res-cvar} places the conservative rules on the axis that separates them. The naive rule overruns in every environment, between $5.3\%$ and $15.2\%$ of the time. The chance-constraint rule holds the violation at or below $0.21\%$ at a regret between $9.5$ and $16.9$ points. Sweeping the tolerance reproduces the exchange of outcome for violation in each of them;~\ref{app:frontiers} plots the frontiers.

\begin{table}[pos=tbp]
\centering
\small
\caption{Decision rules at $\varepsilon=0.2$, $10$ seeds. (a) Exact-counterfactual environments; (b) real-outcome environment.}
\label{tab:three-env}
\begin{minipage}[c]{0.63\linewidth}\centering
\textit{(a) Exact counterfactuals}\\[0.4ex]
\begin{adjustbox}{max width=\linewidth}
\begin{tabular}{@{}lcccc@{}}
\toprule
& \multicolumn{2}{c}{Naive} & \multicolumn{2}{c}{Chance constraint} \\
\cmidrule(lr){2-3} \cmidrule(lr){4-5}
Environment & Regret & Viol. & Regret & Viol. \\
\midrule
Sepsis                    & $\num{three_rule.sepsis.naive.regret_pp.mean}{2.5}{\pm}\num{three_rule.sepsis.naive.regret_pp.std}{0.7}$ & $\num{three_rule.sepsis.naive.violation_pct.mean}{8.0}{\pm}\num{three_rule.sepsis.naive.violation_pct.std}{3.5}$ & $\num{three_rule.sepsis.cc.regret_pp.mean}{16.9}{\pm}\num{three_rule.sepsis.cc.regret_pp.std}{0.7}$ & $\num{three_rule.sepsis.cc.violation_pct.mean}{0.00}{\pm}\num{three_rule.sepsis.cc.violation_pct.std}{0.00}$ \\
Tumor                     & $\num{three_rule.tumor.naive.regret_pp.mean}{0.5}{\pm}\num{three_rule.tumor.naive.regret_pp.std}{0.3}$ & $\num{three_rule.tumor.naive.violation_pct.mean}{15.2}{\pm}\num{three_rule.tumor.naive.violation_pct.std}{2.5}$ & $\num{three_rule.tumor.cc.regret_pp.mean}{11.1}{\pm}\num{three_rule.tumor.cc.regret_pp.std}{0.8}$ & $\num{three_rule.tumor.cc.violation_pct.mean}{0.02}{\pm}\num{three_rule.tumor.cc.violation_pct.std}{0.01}$ \\
Semi-synthetic & $\num{three_rule.mimic.naive.regret_pp.mean}{3.5}{\pm}\num{three_rule.mimic.naive.regret_pp.std}{0.4}$ & $\num{three_rule.mimic.naive.violation_pct.mean}{7.3}{\pm}\num{three_rule.mimic.naive.violation_pct.std}{3.2}$ & $\num{three_rule.mimic.cc.regret_pp.mean}{14.9}{\pm}\num{three_rule.mimic.cc.regret_pp.std}{0.7}$ & $\num{three_rule.mimic.cc.violation_pct.mean}{0.00}{\pm}\num{three_rule.mimic.cc.violation_pct.std}{0.00}$ \\
Maintenance    & $\num{kappa_v3.cmapss.kappa_sweep.0.regret_pp_mean}{0.4}{\pm}\num{kappa_v3.cmapss.kappa_sweep.0.regret_pp_std}{0.3}$ & $\num{kappa_v3.cmapss.kappa_sweep.0.violation_pct_mean}{5.3}{\pm}\num{kappa_v3.cmapss.kappa_sweep.0.violation_pct_std}{1.2}$ & $\num{kappa_v3.cmapss.eps_sweep_chance_only.0.2.regret_pp_mean}{9.5}{\pm}\num{kappa_v3.cmapss.eps_sweep_chance_only.0.2.regret_pp_std}{0.2}$ & $\num{kappa_v3.cmapss.eps_sweep_chance_only.0.2.violation_pct_mean}{0.21}{\pm}\num{kappa_v3.cmapss.eps_sweep_chance_only.0.2.violation_pct_std}{0.13}$ \\
\bottomrule
\end{tabular}
\end{adjustbox}
\end{minipage}\hfill
\begin{minipage}[c]{0.35\linewidth}\centering
\textit{(b) Real outcomes, Drink Less}\\[0.4ex]
\begin{adjustbox}{max width=\linewidth}
\begin{tabular}{@{}lcc@{}}
\toprule
Rule & Value & Est.\ tail \\
\midrule
Naive             & $\num{drinkless.three_rule.naive.value.mean}{2.66}{\pm}\num{drinkless.three_rule.naive.value.sd}{0.16}$ & $\num{drinkless.three_rule.naive.violation.mean}{0.149}{\pm}\num{drinkless.three_rule.naive.violation.sd}{0.081}$ \\
Upper-bound       & $\num{drinkless.three_rule.ub.value.mean}{2.28}{\pm}\num{drinkless.three_rule.ub.value.sd}{0.16}$ & $\num{drinkless.three_rule.ub.violation.mean}{0.099}{\pm}\num{drinkless.three_rule.ub.violation.sd}{0.005}$ \\
Chance-constraint & $\num{drinkless.three_rule.cc.value.mean}{2.63}{\pm}\num{drinkless.three_rule.cc.value.sd}{0.18}$ & $\num{drinkless.three_rule.cc.violation.mean}{0.115}{\pm}\num{drinkless.three_rule.cc.violation.sd}{0.010}$ \\
\bottomrule
\end{tabular}
\end{adjustbox}
\end{minipage}
\begin{flushleft}\footnotesize
Panel (a) averages over each environment's budget grid at $15\%$ injected error, regret in outcome percentage points and violation in percent. The semi-synthetic environment is built on MIMIC-IV and the maintenance one on C-MAPSS. Panel (b) is at $B=15$ and reports value, since no oracle exists there.
\end{flushleft}
\end{table}

The chance-constraint regret rises and falls with the steepness of the oracle frontier. It peaks at $24.6$ points at the single step where the tumor outcome rises and at $56.3$ at the corresponding step in the maintenance environment. The semi-synthetic environment has no frontier step and stays within a $12.6$ to $17.8$ point band. Where a strategy with a materially better outcome first becomes affordable, the constraint must refuse its cost tail, and the guarantee is expensive there. The naive rule meanwhile overruns at nearly every budget rather than at isolated ones. It reaches $18.7\%$ in sepsis, $22.8\%$ in the tumor environment, $9.9\%$ in the semi-synthetic environment and $17.0\%$ in the maintenance environment. The sepsis simulator's uncertainty is in its transitions, and in the others it is in the units. The behavior is therefore not tied to a particular mechanism, to synthetic covariates, or to the setting in which exact evaluation is most readily available.

The semi-synthetic outcome is a design choice, so we vary the benchmark construction through the outcome-threshold quantile and the real covariates across disjoint physiological groups. The chance-constraint rule holds the violation at zero in every configuration and the naive rule overruns in between $8.4\%$ and $13.2\%$ of cases (Table~\ref{tab:mimic-robust}). The success criterion in the tumor environment and the outcome threshold in the semi-synthetic environment qualify this. They are modeling choices, not intrinsic terminal events as the outcome is in the sepsis simulator. Still, a volume target is clinically meaningful, and we report the robustness to the threshold above. The value of the chance constraint is also not a simple function of how widely the candidate cost distributions are dispersed within an environment. That dispersion and the naive-versus-chance gap were not monotonically related. The claim is therefore restricted to the consistency of the behavior across environments rather than to a single quantitative driver.

\subsection{Structure of the frontier and operating-point selection}
\label{sec:res-frontier-char}

The frontier is monotone in the tolerance, as the operating-point selection of Section~\ref{sec:method-decision} requires. Sweeping $\varepsilon$ over the $21$ tolerances of the grid, the outcome regret is non-increasing and the budget violation is non-decreasing in every one of the four environments. There is no reversal at any step. Relaxing the tolerance can therefore only gain outcome and can only raise exposure. The curvature of the frontier carries no comparable regularity: the second difference of regret changes sign $8$ to $12$ times depending on the environment. The frontier is not convex in any of them. That is expected rather than incidental. The constrained functional is a quantile, and a quantile constraint does not deliver a convex efficient set even where the feasible set at a fixed tolerance is convex. An operating point is therefore selected by reading the frontier, not by locating a knee.

The density of the frontier follows the spread of the candidate outcomes. Where candidates realize outcome values across the whole range, as in sepsis and the semi-synthetic environment, regret falls at every step of the tolerance grid. Where most are tied at one value, as in the tumor and maintenance environments, the frontier is sparse. A tightened target may then land on a conservative operating point that a slightly looser one would improve on.

Table~\ref{tab:operating-points} reports the operating point the target-based selection returns at two values of the target $\rho$. At a target of $10\%$, it recommends a strategy with regret between $0.6$ and $4.9$ outcome percentage points in three environments. The realized tail probability is below the target in each. Tightening the target to five percent costs between $1.7$ and $2.8$ further points. The decision maker states a rate and receives a strategy, without choosing $\varepsilon$ directly. The maintenance environment is the exception. Relaxing the target there gains $0.02$ points: all of its regret comes from the single binary choice at $B=3$. That choice does not turn over until the tolerance passes $0.43$. Where a frontier is that sparse, a target expressed as a violation rate has little to select over.

The certified slack grows only in $\log\lvert\mathcal{G}\rvert$, so densifying the candidate set is cheap in the guarantee. A sweep over set size on the tumor environment measures what it costs in computation. From $50$ to $10{,}000$ candidates, the slack rises by a factor of $1.30$ at $\eta = 0.05$ and the budget violation stays at zero at every size. Generating a candidate's cost distribution is a rollout and takes $0.18$ seconds however many there are. The optimization step takes $0.12$ seconds plus $0.22$ milliseconds per candidate. At $10{,}000$ candidates, a rollout pass therefore runs for half an hour and the selection for two seconds. The optimization step accounts for $0.13$ percent of the runtime, at unchanged peak memory. The outcome the chance constraint attains rises from $0.861$ to $0.900$ as the grid fills in. Its margin over the point-estimate rule narrows from $7.9$ to $4.8$ percentage points, most of that by $1{,}000$ candidates. We measure regret against the mean-feasible optimum of the set being swept, so it moves with that set and is not read across sizes.

\subsection{Real outcomes under known propensities}
\label{sec:res-realdata}

\begin{figure}[pos=tbp]
\centering
\includegraphics[width=0.70\linewidth,height=0.32\textheight,keepaspectratio]{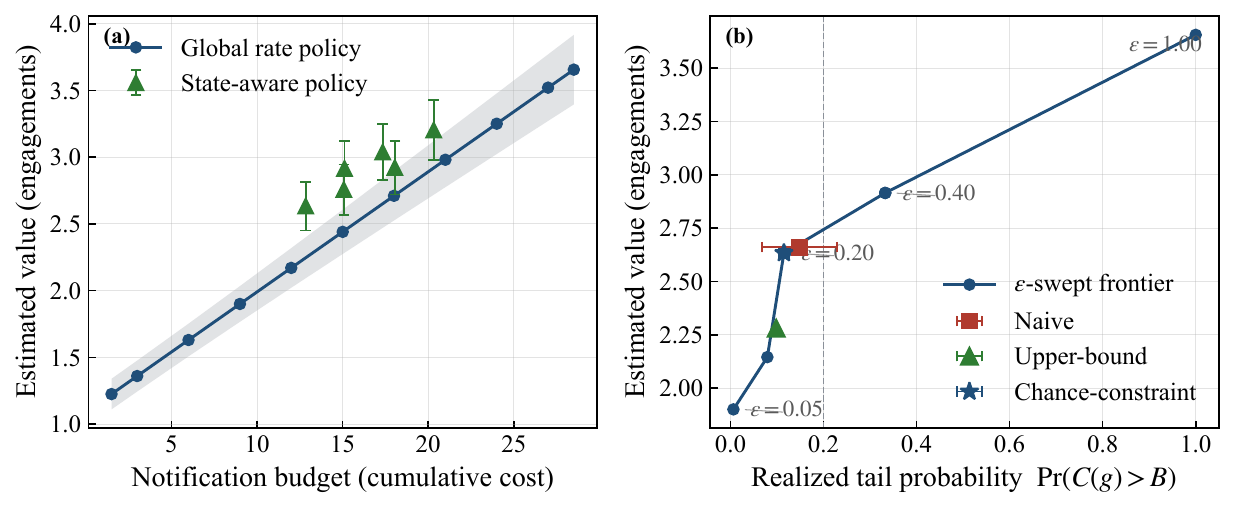}
\caption{Drink Less real-outcome environment, bootstrap mean${\pm}$s.d. (a) Estimated value against notification budget, global-rate policies as a banded line and state-aware policies as markers; (b) safety-utility frontier over $\varepsilon$ at $B=15$.}
\label{fig:drinkless}
\end{figure}

On the Drink Less environment, the framework operates on real engagement outcomes. The per-decision weighting stays well conditioned: the effective sample size ranges from about $4{,}650$ to the full $10{,}470$ decision points. No importance weight exceeds $2.375$, the largest ratio the softened candidate policies can form against the trial's randomization probabilities. The value comes directly from the trial data without an oracle, and per-policy diagnostics are in~\ref{app:ess}. The estimated value increases monotonically with the notification budget, from about $1.2$ engagements over the horizon to about $3.7$. This traces a real-outcome budget frontier with the shape seen on the model-based environments. The best state-aware policy attains $0.22{\pm}0.04$ more engagements than the global policy of the same cost, as Figure~\ref{fig:drinkless}(a) shows.

The three decision rules separate on real outcomes as the controlled study predicts, in panel (b) of Table~\ref{tab:three-env}. At a budget of $15$ notifications, the naive rule attains the highest value, but the estimated tail probability of its selection is large and highly variable across resamples. The chance-constraint rule attains nearly the same value while holding the estimated tail probability to a tightly controlled level; the upper-bound rule is the most conservative. Sweeping the tolerance traces the frontier directly, as Figure~\ref{fig:drinkless}(b) shows. The estimated value rises from $1.90{\pm}0.13$ to $3.66{\pm}0.26$ as $\varepsilon$ relaxes from $0.05$ to $1$, while the realized tail probability rises from $0.007{\pm}0.001$ to one. The pattern is unchanged when the outcome is engagement within the following day rather than the following hour.

Because this environment has no oracle, the quantities are estimated value and estimated violation rather than regret against a known optimum. Its evidence is therefore of a different kind from the preceding sections and complements them. Selecting a policy by its estimated value and reporting that estimate would inflate the value if the selection were chasing sampling noise. Over $50$ random splits of the participants, we choose the recommended policy on one half and re-estimate it on the other. The split estimates sit at or above the single-sample values, and the chance-constraint rule selects the same policy under every split. The half-sample standard deviations are $0.08$ to $0.12$ on a value scale spanning $1.2$ to $3.7$. They bound any optimism at roughly a tenth of an engagement rather than excluding it.

\section{Conclusion, Limitations, and Further Research}
\label{sec:conclusion}
\label{sec:discussion}

We presented a predict-then-optimize framework for sequential intervention selection under a cumulative budget. In the prediction step, any estimator returning an outcome value and a cumulative-cost distribution for each candidate strategy supplies what the decision rule consumes. Both are counterfactual quantities identified from observational trajectories. In the optimization step, the rule returns the outcome-maximizing strategy subject to a chance constraint on the probability that the cost exceeds the budget. That tail does not decompose across stages. The constraint is therefore evaluated on the full-horizon distribution of each candidate, and the decision is a selection over a finite set. Distribution-free finite-sample bounds accompany the rule on both sides. They limit how far an admitted strategy can exceed the tolerated violation and how far its outcome can fall short of what is achievable at a tightened tolerance.

The method traces the budget-outcome frontier and locates where an added unit of budget is worth most. It eliminates the budget violations the naive rule incurs, at a quantified outcome cost that varies with the budget. The decision is robust to estimation error in the predictor, including moderate systematic cost bias. The chance constraint does not improve on the naive rule at every budget, and we report the exchange directly. It costs outcome where a high-outcome strategy carries a heavy cost tail and controls violation where the naive rule overruns.

The method returns the safety-utility frontier, and a stated operational target selects a single recommendation on it. The frontier shows the outcome achievable at each level of tolerated budget violation. Its single control, the tolerated violation probability, has an operational rather than statistical meaning. It sets, at the level of policy, not of the individual case, how often a cumulative-toxicity limit or a maintenance-hour budget may be overrun across a population. The per-case recommendation is computed within that envelope. The level can therefore be set against the operational cost of an overrun in the domain at hand. It is tightened where an overrun is expensive and relaxed where outcome is at a premium.

Where to invest in a better model follows from the error decomposition of Section~\ref{sec:res-predictor}. Outcome error and cost error act on different objectives, regret and violation respectively. Budget overruns observed in deployment reflect the cost distribution the predictor implies. In a g-computation model, that distribution is a functional of the same simulated trajectory that produces the outcome. The check that matters is calibration of that distribution against the absolute budget, not accuracy on the outcome alone.

Several limitations bound the scope of the claims. First, the decision rules and the frontier are established against exact or known outcomes. The four exact-counterfactual environments supply a ground-truth optimum, and the micro-randomized trial supplies real outcomes under known randomization probabilities. Those four span real covariates and real degradation data, so the consistency across them is not an artifact of one setting. None supplies a clinical real-outcome deployment in which the outcome is a long-horizon terminal event and the propensities must be estimated. That case is harder for a tail-constrained method specifically. Importance-sampling estimators recover the upper tail of a cumulative cost far less reliably than its mean once a strategy departs from the behavior policy. Such a study is therefore better suited to a qualitative check that the conservative rules reduce overruns than to a precise reconstruction of the frontier. It is the natural next step.

Second, the frontier and the decision rules are established under exact counterfactuals, and the effect of estimation error is studied by injecting controlled noise. This isolates the optimization step from the accuracy of any one predictor. Where the state space is finite the substitution can be checked directly, and Section~\ref{sec:res-predictor} does so on the sepsis environment. A continuous-state environment requires function approximation. The same section repeats the check with estimators of opposite cost bias. The bias is architecture-dependent, and the violation the chance constraint admits is not. A deployment would still inherit the accuracy of whatever predictor it fits. Where the cost distribution is fit by function approximation rather than counted, a measurable part of the regret is estimation error rather than the price of the constraint. Recalibrating the estimated distribution to the correct mean recovers none of it. Closing that requires a predictor that gets the upper tail right.

Third, like any counterfactual method, the validity of the recommendations rests on the identification assumptions, in particular sequential ignorability. Sequential ignorability is not testable from observational data alone and would have to be argued from domain knowledge about what drives treatment decisions. Fourth, the method selects a strategy once at the start of the horizon and does not re-select as the remaining budget is consumed. Fifth, the candidate sets used here are constructed rather than drawn from practice, since measuring regret against a computable optimum requires counterfactuals for every candidate. \ref{app:placement} shows the reported exchange survives both a practice-sized and a dense construction, but a deployment would inherit whatever set its guidelines supply. How the method behaves on a set chosen for reasons other than cost spacing is not something these experiments can settle.

These limitations indicate the directions the work should take. Integrating an end-to-end predictor is the first, and the cost estimate on which the violation guarantee depends is the component to validate. The deviation against the certified slack, not an aggregate bias band, is the check a deployment should run. A rolling-horizon rule that updates the recommendation on the realized cost so far would address the single-selection limitation. Settings with several simultaneous resource or toxicity limits would call for a vector of budgets and a joint chance constraint. Where the dynamics are estimated rather than known, the strategy set could be derived from operating guidelines or learned directly, in place of the sweep across treatment intensities used here. This would change how the frontier is populated without changing the decision component. Finally, which features of the history drive a switch between strategies is a natural complement to the decision-quality evaluation and would address the interpretability expectations of high-stakes deployment.

\section*{CRediT authorship contribution statement}
\textbf{Minkyoung Kim:} Conceptualization, Methodology, Software, Formal analysis, Investigation, Data curation, Visualization, Writing -- original draft. \textbf{Beakcheol Jang:} Conceptualization, Supervision, Funding acquisition, Writing -- review and editing.

\section*{Declaration of competing interest}
The authors declare that they have no known competing financial interests or personal relationships that could have appeared to influence the work reported in this paper.

\section*{Funding}
This study was supported by the National Research Foundation of Korea Fund (grant number RS-2023-00273751).

\section*{Data availability}
This study uses four publicly available sources and one credentialed-access database. The environments are built on the Gumbel-max sepsis simulator \citep{oberst2019gumbel}, the tumor-growth model of \citet{geng2017tumor}, the NASA C-MAPSS turbofan degradation dataset \citep{saxena2008cmapss}, the Drink Less micro-randomized trial data \citep{bell2023notifications}, available at \url{https://osf.io/w3szp/}, and MIMIC-IV \citep{johnson2023mimiciv}, accessed through PhysioNet under credentialed access. \ref{app:envspec} gives the construction of each environment. The code implementing the evaluation environments, the decision rule, and all reported analyses is available at \url{https://github.com/mfriendly/counterfactual-chance-selection}.

\section*{Declaration of generative AI and AI-assisted technologies in the writing process}
During the preparation of this work the authors used Claude in order to refine the phrasing of the manuscript. After using this tool, the authors reviewed and edited the content as needed and take full responsibility for the content of the published article.

\bibliographystyle{cas-model2-names}
\bibliography{references}

\begin{thebibliography}{62}
\expandafter\ifx\csname natexlab\endcsname\relax\def\natexlab#1{#1}\fi
\providecommand{\url}[1]{\texttt{#1}}
\providecommand{\href}[2]{#2}
\providecommand{\path}[1]{#1}
\providecommand{\DOIprefix}{doi:}
\providecommand{\ArXivprefix}{arXiv:}
\providecommand{\URLprefix}{URL: }
\providecommand{\Pubmedprefix}{pmid:}
\providecommand{\doi}[1]{\href{http://dx.doi.org/#1}{\path{#1}}}
\providecommand{\Pubmed}[1]{\href{pmid:#1}{\path{#1}}}
\providecommand{\bibinfo}[2]{#2}
\ifx\xfnm\relax \def\xfnm[#1]{\unskip,\space#1}\fi
\bibitem[{Altman(1999)}]{altman1999constrained}
\bibinfo{author}{Altman, E.}, \bibinfo{year}{1999}.
\newblock \bibinfo{title}{Constrained Markov Decision Processes}.
\newblock \bibinfo{publisher}{Chapman and Hall/CRC}.
\bibitem[{Angelotti et~al.(2026)Angelotti, Drougard and
  Chanel}]{angelotti2026offline}
\bibinfo{author}{Angelotti, G.}, \bibinfo{author}{Drougard, N.},
  \bibinfo{author}{Chanel, C.P.C.}, \bibinfo{year}{2026}.
\newblock \bibinfo{title}{An offline risk-aware policy selection method for
  {Bayesian} {Markov} decision processes}.
\newblock \bibinfo{journal}{Artificial Intelligence} \bibinfo{volume}{354},
  \bibinfo{pages}{104519}.
\newblock \DOIprefix\doi{10.1016/j.artint.2026.104519}.
\bibitem[{Athey and Wager(2021)}]{atheywager2021policy}
\bibinfo{author}{Athey, S.}, \bibinfo{author}{Wager, S.}, \bibinfo{year}{2021}.
\newblock \bibinfo{title}{Policy learning with observational data}.
\newblock \bibinfo{journal}{Econometrica} \bibinfo{volume}{89},
  \bibinfo{pages}{133--161}.
\newblock \DOIprefix\doi{10.3982/ECTA15732}.
\bibitem[{Bao et~al.(2025)Bao, Bell, Williamson, Garnett and
  Qian}]{bao2025perdecision}
\bibinfo{author}{Bao, Y.}, \bibinfo{author}{Bell, L.},
  \bibinfo{author}{Williamson, E.}, \bibinfo{author}{Garnett, C.},
  \bibinfo{author}{Qian, T.}, \bibinfo{year}{2025}.
\newblock \bibinfo{title}{Estimating causal effects for binary outcomes using
  per-decision inverse probability weighting}.
\newblock \bibinfo{journal}{Biostatistics} \bibinfo{volume}{26},
  \bibinfo{pages}{kxae025}.
\newblock \DOIprefix\doi{10.1093/biostatistics/kxae025}.
\bibitem[{B\"{a}uerle and Ott(2011)}]{bauerleott2011avar}
\bibinfo{author}{B\"{a}uerle, N.}, \bibinfo{author}{Ott, J.},
  \bibinfo{year}{2011}.
\newblock \bibinfo{title}{Markov decision processes with average-value-at-risk
  criteria}.
\newblock \bibinfo{journal}{Mathematical Methods of Operations Research}
  \bibinfo{volume}{74}, \bibinfo{pages}{361--379}.
\newblock \DOIprefix\doi{10.1007/s00186-011-0367-0}.
\bibitem[{Bell et~al.(2023)Bell, Garnett, Bao, Cheng, Qian, Perski, Potts and
  Williamson}]{bell2023notifications}
\bibinfo{author}{Bell, L.}, \bibinfo{author}{Garnett, C.},
  \bibinfo{author}{Bao, Y.}, \bibinfo{author}{Cheng, Z.},
  \bibinfo{author}{Qian, T.}, \bibinfo{author}{Perski, O.},
  \bibinfo{author}{Potts, H.W.W.}, \bibinfo{author}{Williamson, E.},
  \bibinfo{year}{2023}.
\newblock \bibinfo{title}{How notifications affect engagement with a behavior
  change app: Results from a micro-randomized trial}.
\newblock \bibinfo{journal}{JMIR mHealth and uHealth} \bibinfo{volume}{11},
  \bibinfo{pages}{e38342}.
\newblock \DOIprefix\doi{10.2196/38342}.
\bibitem[{Bertsimas and Kallus(2020)}]{bertsimas2020prescriptive}
\bibinfo{author}{Bertsimas, D.}, \bibinfo{author}{Kallus, N.},
  \bibinfo{year}{2020}.
\newblock \bibinfo{title}{From predictive to prescriptive analytics}.
\newblock \bibinfo{journal}{Management Science} \bibinfo{volume}{66},
  \bibinfo{pages}{1025--1044}.
\newblock \DOIprefix\doi{10.1287/mnsc.2018.3253}.
\bibitem[{Bertsimas and Van~Parys(2022)}]{bertsimas2022bootstrap}
\bibinfo{author}{Bertsimas, D.}, \bibinfo{author}{Van~Parys, B.},
  \bibinfo{year}{2022}.
\newblock \bibinfo{title}{Bootstrap robust prescriptive analytics}.
\newblock \bibinfo{journal}{Mathematical Programming} \bibinfo{volume}{195},
  \bibinfo{pages}{39--78}.
\newblock \DOIprefix\doi{10.1007/s10107-021-01679-2}.
\bibitem[{Bica et~al.(2020)Bica, Alaa, Jordon and van~der Schaar}]{bica2020crn}
\bibinfo{author}{Bica, I.}, \bibinfo{author}{Alaa, A.M.},
  \bibinfo{author}{Jordon, J.}, \bibinfo{author}{van~der Schaar, M.},
  \bibinfo{year}{2020}.
\newblock \bibinfo{title}{Estimating counterfactual treatment outcomes over
  time through adversarially balanced representations}, in:
  \bibinfo{booktitle}{International Conference on Learning Representations
  (ICLR)}.
\bibitem[{Caljon et~al.(2026)Caljon, Van~Belle, Berrevoets and
  Verbeke}]{caljon2026interference}
\bibinfo{author}{Caljon, D.}, \bibinfo{author}{Van~Belle, J.},
  \bibinfo{author}{Berrevoets, J.}, \bibinfo{author}{Verbeke, W.},
  \bibinfo{year}{2026}.
\newblock \bibinfo{title}{Optimizing treatment allocation in the presence of
  interference}.
\newblock \bibinfo{journal}{European Journal of Operational Research}
  \bibinfo{volume}{328}, \bibinfo{pages}{620--632}.
\newblock \DOIprefix\doi{10.1016/j.ejor.2025.09.015}.
\bibitem[{Chakraborty and Moodie(2013)}]{chakraborty2013dtr}
\bibinfo{author}{Chakraborty, B.}, \bibinfo{author}{Moodie, E.E.M.},
  \bibinfo{year}{2013}.
\newblock \bibinfo{title}{Statistical Methods for Dynamic Treatment Regimes:
  Reinforcement Learning, Causal Inference, and Personalized Medicine}.
\newblock Statistics for Biology and Health, \bibinfo{publisher}{Springer},
  \bibinfo{address}{New York, NY}.
\newblock \DOIprefix\doi{10.1007/978-1-4614-7428-9}.
\bibitem[{Charnes and Cooper(1959)}]{charnes1959ccp}
\bibinfo{author}{Charnes, A.}, \bibinfo{author}{Cooper, W.W.},
  \bibinfo{year}{1959}.
\newblock \bibinfo{title}{Chance-constrained programming}.
\newblock \bibinfo{journal}{Management Science} \bibinfo{volume}{6},
  \bibinfo{pages}{73--79}.
\newblock \DOIprefix\doi{10.1287/mnsc.6.1.73}.
\bibitem[{Chow et~al.(2018)Chow, Ghavamzadeh, Janson and Pavone}]{chow2017risk}
\bibinfo{author}{Chow, Y.}, \bibinfo{author}{Ghavamzadeh, M.},
  \bibinfo{author}{Janson, L.}, \bibinfo{author}{Pavone, M.},
  \bibinfo{year}{2018}.
\newblock \bibinfo{title}{Risk-constrained reinforcement learning with
  percentile risk criteria}.
\newblock \bibinfo{journal}{Journal of Machine Learning Research}
  \bibinfo{volume}{18}, \bibinfo{pages}{1--51}.
\newblock \URLprefix \url{http://jmlr.org/papers/v18/15-636.html}.
\bibitem[{De~Vos et~al.(2026)De~Vos, Bockel-Rickermann, Lessmann and
  Verbeke}]{devos2026uplift}
\bibinfo{author}{De~Vos, S.}, \bibinfo{author}{Bockel-Rickermann, C.},
  \bibinfo{author}{Lessmann, S.}, \bibinfo{author}{Verbeke, W.},
  \bibinfo{year}{2026}.
\newblock \bibinfo{title}{Uplift modeling with continuous treatments: A
  predict-then-optimize approach}.
\newblock \bibinfo{journal}{European Journal of Operational Research}
  \bibinfo{volume}{330}, \bibinfo{pages}{230--244}.
\newblock \DOIprefix\doi{10.1016/j.ejor.2025.10.025}.
\bibitem[{Dud{\'\i}k et~al.(2011)Dud{\'\i}k, Langford and Li}]{dudik2011dr}
\bibinfo{author}{Dud{\'\i}k, M.}, \bibinfo{author}{Langford, J.},
  \bibinfo{author}{Li, L.}, \bibinfo{year}{2011}.
\newblock \bibinfo{title}{Doubly robust policy evaluation and learning}, in:
  \bibinfo{booktitle}{Proceedings of the 28th International Conference on
  Machine Learning (ICML)}, pp. \bibinfo{pages}{1097--1104}.
\bibitem[{Elmachtoub and Grigas(2022)}]{elmachtoub2022spo}
\bibinfo{author}{Elmachtoub, A.N.}, \bibinfo{author}{Grigas, P.},
  \bibinfo{year}{2022}.
\newblock \bibinfo{title}{Smart ``predict, then optimize''}.
\newblock \bibinfo{journal}{Management Science} \bibinfo{volume}{68},
  \bibinfo{pages}{9--26}.
\newblock \DOIprefix\doi{10.1287/mnsc.2020.3922}.
\bibitem[{Fern\'andez-Lor\'ia and Provost(2022)}]{fernandezloria2022causal}
\bibinfo{author}{Fern\'andez-Lor\'ia, C.}, \bibinfo{author}{Provost, F.},
  \bibinfo{year}{2022}.
\newblock \bibinfo{title}{Causal decision making and causal effect estimation
  are not the same and why it matters}.
\newblock \bibinfo{journal}{INFORMS Journal on Data Science}
  \bibinfo{volume}{1}, \bibinfo{pages}{4--16}.
\newblock \DOIprefix\doi{10.1287/ijds.2021.0006}.
\bibitem[{Geng et~al.(2017)Geng, Paganetti and Grassberger}]{geng2017tumor}
\bibinfo{author}{Geng, C.}, \bibinfo{author}{Paganetti, H.},
  \bibinfo{author}{Grassberger, C.}, \bibinfo{year}{2017}.
\newblock \bibinfo{title}{Prediction of treatment response for combined chemo-
  and radiation therapy for non-small cell lung cancer patients using a
  bio-mathematical model}.
\newblock \bibinfo{journal}{Scientific Reports} \bibinfo{volume}{7},
  \bibinfo{pages}{13542}.
\newblock \DOIprefix\doi{10.1038/s41598-017-13646-z}.
\bibitem[{Haskell and Jain(2015)}]{haskelljain2015convex}
\bibinfo{author}{Haskell, W.B.}, \bibinfo{author}{Jain, R.},
  \bibinfo{year}{2015}.
\newblock \bibinfo{title}{A convex analytic approach to risk-aware {Markov}
  decision processes}.
\newblock \bibinfo{journal}{SIAM Journal on Control and Optimization}
  \bibinfo{volume}{53}, \bibinfo{pages}{1569--1598}.
\newblock \DOIprefix\doi{10.1137/140969221}.
\bibitem[{Hoeffding(1963)}]{hoeffding1963}
\bibinfo{author}{Hoeffding, W.}, \bibinfo{year}{1963}.
\newblock \bibinfo{title}{Probability inequalities for sums of bounded random
  variables}.
\newblock \bibinfo{journal}{Journal of the American Statistical Association}
  \bibinfo{volume}{58}, \bibinfo{pages}{13--30}.
\newblock \DOIprefix\doi{10.1080/01621459.1963.10500830}.
\bibitem[{Hong et~al.(2015)Hong, Luo and Nelson}]{hong2015ccsb}
\bibinfo{author}{Hong, L.J.}, \bibinfo{author}{Luo, J.},
  \bibinfo{author}{Nelson, B.L.}, \bibinfo{year}{2015}.
\newblock \bibinfo{title}{Chance constrained selection of the best}.
\newblock \bibinfo{journal}{INFORMS Journal on Computing} \bibinfo{volume}{27},
  \bibinfo{pages}{317--334}.
\newblock \DOIprefix\doi{10.1287/ijoc.2014.0628}.
\bibitem[{Johnson et~al.(2023)Johnson, Bulgarelli, Shen, Gayles, Shammout,
  Horng, Pollard, Hao, Moody, Gow, Lehman, Celi and Mark}]{johnson2023mimiciv}
\bibinfo{author}{Johnson, A.E.W.}, \bibinfo{author}{Bulgarelli, L.},
  \bibinfo{author}{Shen, L.}, \bibinfo{author}{Gayles, A.},
  \bibinfo{author}{Shammout, A.}, \bibinfo{author}{Horng, S.},
  \bibinfo{author}{Pollard, T.J.}, \bibinfo{author}{Hao, S.},
  \bibinfo{author}{Moody, B.}, \bibinfo{author}{Gow, B.},
  \bibinfo{author}{Lehman, L.w.H.}, \bibinfo{author}{Celi, L.A.},
  \bibinfo{author}{Mark, R.G.}, \bibinfo{year}{2023}.
\newblock \bibinfo{title}{{MIMIC-IV}, a freely accessible electronic health
  record dataset}.
\newblock \bibinfo{journal}{Scientific Data} \bibinfo{volume}{10},
  \bibinfo{pages}{1}.
\newblock \DOIprefix\doi{10.1038/s41597-022-01899-x}.
\bibitem[{Kallus and Zhou(2022)}]{kallus2022stateful}
\bibinfo{author}{Kallus, N.}, \bibinfo{author}{Zhou, A.}, \bibinfo{year}{2022}.
\newblock \bibinfo{title}{Stateful offline contextual policy evaluation and
  learning}, in: \bibinfo{booktitle}{Proceedings of the 25th International
  Conference on Artificial Intelligence and Statistics (AISTATS), PMLR}, pp.
  \bibinfo{pages}{11169--11194}.
\bibitem[{Kannan et~al.(2025)Kannan, Bayraksan and Luedtke}]{kannan2025saa}
\bibinfo{author}{Kannan, R.}, \bibinfo{author}{Bayraksan, G.},
  \bibinfo{author}{Luedtke, J.R.}, \bibinfo{year}{2025}.
\newblock \bibinfo{title}{Data-driven sample average approximation with
  covariate information}.
\newblock \bibinfo{journal}{Operations Research} \bibinfo{volume}{73},
  \bibinfo{pages}{3245--3259}.
\newblock \DOIprefix\doi{10.1287/opre.2020.0533}.
\bibitem[{Killian et~al.(2023)Killian, Parbhoo and
  Ghassemi}]{killian2023distded}
\bibinfo{author}{Killian, T.W.}, \bibinfo{author}{Parbhoo, S.},
  \bibinfo{author}{Ghassemi, M.}, \bibinfo{year}{2023}.
\newblock \bibinfo{title}{Risk sensitive dead-end identification in
  safety-critical offline reinforcement learning}.
\newblock \bibinfo{journal}{Transactions on Machine Learning Research (TMLR)} .
\bibitem[{Klasnja et~al.(2015)Klasnja, Hekler, Shiffman, Boruvka, Almirall,
  Tewari and Murphy}]{klasnja2015mrt}
\bibinfo{author}{Klasnja, P.}, \bibinfo{author}{Hekler, E.B.},
  \bibinfo{author}{Shiffman, S.}, \bibinfo{author}{Boruvka, A.},
  \bibinfo{author}{Almirall, D.}, \bibinfo{author}{Tewari, A.},
  \bibinfo{author}{Murphy, S.A.}, \bibinfo{year}{2015}.
\newblock \bibinfo{title}{Microrandomized trials: An experimental design for
  developing just-in-time adaptive interventions}.
\newblock \bibinfo{journal}{Health Psychology} \bibinfo{volume}{34S},
  \bibinfo{pages}{1220--1228}.
\newblock \DOIprefix\doi{10.1037/hea0000305}.
\bibitem[{Komorowski et~al.(2018)Komorowski, Celi, Badawi, Gordon and
  Faisal}]{komorowski2018aiclinician}
\bibinfo{author}{Komorowski, M.}, \bibinfo{author}{Celi, L.A.},
  \bibinfo{author}{Badawi, O.}, \bibinfo{author}{Gordon, A.C.},
  \bibinfo{author}{Faisal, A.A.}, \bibinfo{year}{2018}.
\newblock \bibinfo{title}{The artificial intelligence clinician learns optimal
  treatment strategies for sepsis in intensive care}.
\newblock \bibinfo{journal}{Nature Medicine} \bibinfo{volume}{24},
  \bibinfo{pages}{1716--1720}.
\newblock \DOIprefix\doi{10.1038/s41591-018-0213-5}.
\bibitem[{Kumar et~al.(2023)Kumar, Bhat, Kavitha and
  Hemachandra}]{kumar2023crsmdp}
\bibinfo{author}{Kumar, U.M.}, \bibinfo{author}{Bhat, S.P.},
  \bibinfo{author}{Kavitha, V.}, \bibinfo{author}{Hemachandra, N.},
  \bibinfo{year}{2023}.
\newblock \bibinfo{title}{Approximate solutions to constrained risk-sensitive
  {Markov} decision processes}.
\newblock \bibinfo{journal}{European Journal of Operational Research}
  \bibinfo{volume}{310}, \bibinfo{pages}{249--267}.
\newblock \DOIprefix\doi{10.1016/j.ejor.2023.02.039}.
\bibitem[{Laber et~al.(2018)Laber, Wu, Munera, Lipkovich, Colucci and
  Ripa}]{laber2018safety}
\bibinfo{author}{Laber, E.B.}, \bibinfo{author}{Wu, F.},
  \bibinfo{author}{Munera, C.}, \bibinfo{author}{Lipkovich, I.},
  \bibinfo{author}{Colucci, S.}, \bibinfo{author}{Ripa, S.},
  \bibinfo{year}{2018}.
\newblock \bibinfo{title}{Identifying optimal dosage regimes under safety
  constraints: An application to long term opioid treatment of chronic pain}.
\newblock \bibinfo{journal}{Statistics in Medicine} \bibinfo{volume}{37},
  \bibinfo{pages}{1407--1418}.
\newblock \DOIprefix\doi{10.1002/sim.7566}.
\bibitem[{Lei et~al.(2018)Lei, Li, Guo, Li, Yan and Lin}]{lei2018prognostics}
\bibinfo{author}{Lei, Y.}, \bibinfo{author}{Li, N.}, \bibinfo{author}{Guo, L.},
  \bibinfo{author}{Li, N.}, \bibinfo{author}{Yan, T.}, \bibinfo{author}{Lin,
  J.}, \bibinfo{year}{2018}.
\newblock \bibinfo{title}{Machinery health prognostics: A systematic review
  from data acquisition to {RUL} prediction}.
\newblock \bibinfo{journal}{Mechanical Systems and Signal Processing}
  \bibinfo{volume}{104}, \bibinfo{pages}{799--834}.
\newblock \DOIprefix\doi{10.1016/j.ymssp.2017.11.016}.
\bibitem[{Li et~al.(2021)Li, Hu, Lu, Utsumi, Chakraborty, Sow, Madan, Li,
  Ghalwash, Shahn and Lehman}]{li2021gnet}
\bibinfo{author}{Li, R.}, \bibinfo{author}{Hu, S.}, \bibinfo{author}{Lu, M.},
  \bibinfo{author}{Utsumi, Y.}, \bibinfo{author}{Chakraborty, P.},
  \bibinfo{author}{Sow, D.M.}, \bibinfo{author}{Madan, P.},
  \bibinfo{author}{Li, J.}, \bibinfo{author}{Ghalwash, M.},
  \bibinfo{author}{Shahn, Z.}, \bibinfo{author}{Lehman, L.w.H.},
  \bibinfo{year}{2021}.
\newblock \bibinfo{title}{{G-Net}: a recurrent network approach to
  {g}-computation for counterfactual prediction under a dynamic treatment
  regime}, in: \bibinfo{booktitle}{Proceedings of Machine Learning for Health
  (ML4H), PMLR}, pp. \bibinfo{pages}{282--299}.
\bibitem[{Li et~al.(2018)Li, Ding and Sun}]{li2018rul}
\bibinfo{author}{Li, X.}, \bibinfo{author}{Ding, Q.}, \bibinfo{author}{Sun,
  J.Q.}, \bibinfo{year}{2018}.
\newblock \bibinfo{title}{Remaining useful life estimation in prognostics using
  deep convolution neural networks}.
\newblock \bibinfo{journal}{Reliability Engineering \& System Safety}
  \bibinfo{volume}{172}, \bibinfo{pages}{1--11}.
\newblock \DOIprefix\doi{10.1016/j.ress.2017.11.021}.
\bibitem[{Lim et~al.(2018)Lim, Alaa and van~der Schaar}]{lim2018rmsn}
\bibinfo{author}{Lim, B.}, \bibinfo{author}{Alaa, A.M.},
  \bibinfo{author}{van~der Schaar, M.}, \bibinfo{year}{2018}.
\newblock \bibinfo{title}{Forecasting treatment responses over time using
  recurrent marginal structural networks}, in: \bibinfo{booktitle}{Advances in
  Neural Information Processing Systems (NeurIPS)}, pp.
  \bibinfo{pages}{7493--7503}.
\bibitem[{Liu and Zhu(2024)}]{liu2024newsvendor}
\bibinfo{author}{Liu, C.}, \bibinfo{author}{Zhu, W.}, \bibinfo{year}{2024}.
\newblock \bibinfo{title}{Newsvendor conditional value-at-risk minimisation: A
  feature-based approach under adaptive data selection}.
\newblock \bibinfo{journal}{European Journal of Operational Research}
  \bibinfo{volume}{313}, \bibinfo{pages}{548--564}.
\newblock \DOIprefix\doi{10.1016/j.ejor.2023.08.043}.
\bibitem[{Liu et~al.(2024a)Liu, Wang, Fu and Zeng}]{liu2024cbr}
\bibinfo{author}{Liu, M.}, \bibinfo{author}{Wang, Y.}, \bibinfo{author}{Fu,
  H.}, \bibinfo{author}{Zeng, D.}, \bibinfo{year}{2024}a.
\newblock \bibinfo{title}{Controlling cumulative adverse risk in learning
  optimal dynamic treatment regimens}.
\newblock \bibinfo{journal}{Journal of the American Statistical Association}
  \bibinfo{volume}{119}, \bibinfo{pages}{2622--2633}.
\newblock \DOIprefix\doi{10.1080/01621459.2023.2270637}.
\bibitem[{Liu et~al.(2024b)Liu, Wang, Fu and Zeng}]{liu2024brdtr}
\bibinfo{author}{Liu, M.}, \bibinfo{author}{Wang, Y.}, \bibinfo{author}{Fu,
  H.}, \bibinfo{author}{Zeng, D.}, \bibinfo{year}{2024}b.
\newblock \bibinfo{title}{Learning optimal dynamic treatment regimens subject
  to stagewise risk controls}.
\newblock \bibinfo{journal}{Journal of Machine Learning Research}
  \bibinfo{volume}{25}, \bibinfo{pages}{1--64}.
\bibitem[{Luedtke and Ahmed(2008)}]{luedtke2008saa}
\bibinfo{author}{Luedtke, J.}, \bibinfo{author}{Ahmed, S.},
  \bibinfo{year}{2008}.
\newblock \bibinfo{title}{A sample approximation approach for optimization with
  probabilistic constraints}.
\newblock \bibinfo{journal}{SIAM Journal on Optimization} \bibinfo{volume}{19},
  \bibinfo{pages}{674--699}.
\newblock \DOIprefix\doi{10.1137/070702928}.
\bibitem[{Massart(1990)}]{massart1990dkw}
\bibinfo{author}{Massart, P.}, \bibinfo{year}{1990}.
\newblock \bibinfo{title}{The tight constant in the
  {Dvoretzky-Kiefer-Wolfowitz} inequality}.
\newblock \bibinfo{journal}{The Annals of Probability} \bibinfo{volume}{18},
  \bibinfo{pages}{1269--1283}.
\newblock \DOIprefix\doi{10.1214/aop/1176990746}.
\bibitem[{McFowland~III et~al.(2021)McFowland~III, Gangarapu, Bapna and
  Sun}]{mcfowland2021prescriptive}
\bibinfo{author}{McFowland~III, E.}, \bibinfo{author}{Gangarapu, S.},
  \bibinfo{author}{Bapna, R.}, \bibinfo{author}{Sun, T.}, \bibinfo{year}{2021}.
\newblock \bibinfo{title}{A prescriptive analytics framework for optimal policy
  deployment using heterogeneous treatment effects}.
\newblock \bibinfo{journal}{MIS Quarterly} \bibinfo{volume}{45},
  \bibinfo{pages}{1807--1832}.
\newblock \DOIprefix\doi{10.25300/MISQ/2021/15684}.
\bibitem[{Melnychuk et~al.(2022)Melnychuk, Frauen and
  Feuerriegel}]{melnychuk2022ct}
\bibinfo{author}{Melnychuk, V.}, \bibinfo{author}{Frauen, D.},
  \bibinfo{author}{Feuerriegel, S.}, \bibinfo{year}{2022}.
\newblock \bibinfo{title}{Causal transformer for estimating counterfactual
  outcomes}, in: \bibinfo{booktitle}{International Conference on Machine
  Learning (ICML), PMLR}, pp. \bibinfo{pages}{15293--15329}.
\bibitem[{Murphy(2003)}]{murphy2003optimal}
\bibinfo{author}{Murphy, S.A.}, \bibinfo{year}{2003}.
\newblock \bibinfo{title}{Optimal dynamic treatment regimes}.
\newblock \bibinfo{journal}{Journal of the Royal Statistical Society: Series B
  (Statistical Methodology)} \bibinfo{volume}{65}, \bibinfo{pages}{331--355}.
\newblock \DOIprefix\doi{10.1111/1467-9868.00389}.
\bibitem[{Nemirovski and Shapiro(2006)}]{nemirovski2006convex}
\bibinfo{author}{Nemirovski, A.}, \bibinfo{author}{Shapiro, A.},
  \bibinfo{year}{2006}.
\newblock \bibinfo{title}{Convex approximations of chance constrained
  programs}.
\newblock \bibinfo{journal}{SIAM Journal on Optimization} \bibinfo{volume}{17},
  \bibinfo{pages}{969--996}.
\newblock \DOIprefix\doi{10.1137/050622328}.
\bibitem[{Oberst and Sontag(2019)}]{oberst2019gumbel}
\bibinfo{author}{Oberst, M.}, \bibinfo{author}{Sontag, D.},
  \bibinfo{year}{2019}.
\newblock \bibinfo{title}{Counterfactual off-policy evaluation with
  {Gumbel}-max structural causal models}, in: \bibinfo{booktitle}{International
  Conference on Machine Learning (ICML), PMLR}, pp.
  \bibinfo{pages}{4881--4890}.
\bibitem[{Rahimian and Pagnoncelli(2023)}]{rahimian2023contextual}
\bibinfo{author}{Rahimian, H.}, \bibinfo{author}{Pagnoncelli, B.},
  \bibinfo{year}{2023}.
\newblock \bibinfo{title}{Data-driven approximation of contextual
  chance-constrained stochastic programs}.
\newblock \bibinfo{journal}{SIAM Journal on Optimization} \bibinfo{volume}{33},
  \bibinfo{pages}{2248--2274}.
\newblock \DOIprefix\doi{10.1137/22M1528045}.
\bibitem[{Robins(1986)}]{robins1986}
\bibinfo{author}{Robins, J.M.}, \bibinfo{year}{1986}.
\newblock \bibinfo{title}{A new approach to causal inference in mortality
  studies with a sustained exposure period---application to control of the
  healthy worker survivor effect}.
\newblock \bibinfo{journal}{Mathematical Modelling} \bibinfo{volume}{7},
  \bibinfo{pages}{1393--1512}.
\newblock \DOIprefix\doi{10.1016/0270-0255(86)90088-6}.
\bibitem[{Robins et~al.(2000)Robins, Hern\'an and Brumback}]{robins2000msm}
\bibinfo{author}{Robins, J.M.}, \bibinfo{author}{Hern\'an, M.A.},
  \bibinfo{author}{Brumback, B.}, \bibinfo{year}{2000}.
\newblock \bibinfo{title}{Marginal structural models and causal inference in
  epidemiology}.
\newblock \bibinfo{journal}{Epidemiology} \bibinfo{volume}{11},
  \bibinfo{pages}{550--560}.
\newblock \DOIprefix\doi{10.1097/00001648-200009000-00011}.
\bibitem[{Rockafellar and Uryasev(2000)}]{rockafellar2000cvar}
\bibinfo{author}{Rockafellar, R.T.}, \bibinfo{author}{Uryasev, S.},
  \bibinfo{year}{2000}.
\newblock \bibinfo{title}{Optimization of conditional value-at-risk}.
\newblock \bibinfo{journal}{Journal of Risk} \bibinfo{volume}{2},
  \bibinfo{pages}{21--41}.
\newblock \DOIprefix\doi{10.21314/JOR.2000.038}.
\bibitem[{Sadana et~al.(2025)Sadana, Chenreddy, Delage, Forel, Frejinger and
  Vidal}]{sadana2025contextual}
\bibinfo{author}{Sadana, U.}, \bibinfo{author}{Chenreddy, A.},
  \bibinfo{author}{Delage, E.}, \bibinfo{author}{Forel, A.},
  \bibinfo{author}{Frejinger, E.}, \bibinfo{author}{Vidal, T.},
  \bibinfo{year}{2025}.
\newblock \bibinfo{title}{A survey of contextual optimization methods for
  decision-making under uncertainty}.
\newblock \bibinfo{journal}{European Journal of Operational Research}
  \bibinfo{volume}{320}, \bibinfo{pages}{271--289}.
\newblock \DOIprefix\doi{10.1016/j.ejor.2024.03.020}.
\bibitem[{Sakaguchi(2025)}]{sakaguchi2025dewm}
\bibinfo{author}{Sakaguchi, S.}, \bibinfo{year}{2025}.
\newblock \bibinfo{title}{Estimation of optimal dynamic treatment assignment
  rules under policy constraints}.
\newblock \bibinfo{journal}{Quantitative Economics} \bibinfo{volume}{16},
  \bibinfo{pages}{981--1022}.
\newblock \DOIprefix\doi{10.3982/QE2288}.
\bibitem[{Sateesh~Babu et~al.(2016)Sateesh~Babu, Zhao and Li}]{babu2016cnn}
\bibinfo{author}{Sateesh~Babu, G.}, \bibinfo{author}{Zhao, P.},
  \bibinfo{author}{Li, X.L.}, \bibinfo{year}{2016}.
\newblock \bibinfo{title}{Deep convolutional neural network based regression
  approach for estimation of remaining useful life}, in:
  \bibinfo{booktitle}{Database Systems for Advanced Applications (DASFAA),
  LNCS}, pp. \bibinfo{pages}{214--228}.
\newblock \DOIprefix\doi{10.1007/978-3-319-32025-0_14}.
\bibitem[{Saxena et~al.(2008)Saxena, Goebel, Simon and
  Eklund}]{saxena2008cmapss}
\bibinfo{author}{Saxena, A.}, \bibinfo{author}{Goebel, K.},
  \bibinfo{author}{Simon, D.}, \bibinfo{author}{Eklund, N.},
  \bibinfo{year}{2008}.
\newblock \bibinfo{title}{Damage propagation modeling for aircraft engine
  run-to-failure simulation}, in: \bibinfo{booktitle}{International Conference
  on Prognostics and Health Management (PHM)}, pp. \bibinfo{pages}{1--9}.
\newblock \DOIprefix\doi{10.1109/PHM.2008.4711414}.
\bibitem[{Smith and Winkler(2006)}]{smithwinkler2006}
\bibinfo{author}{Smith, J.E.}, \bibinfo{author}{Winkler, R.L.},
  \bibinfo{year}{2006}.
\newblock \bibinfo{title}{The optimizer's curse: Skepticism and postdecision
  surprise in decision analysis}.
\newblock \bibinfo{journal}{Management Science} \bibinfo{volume}{52},
  \bibinfo{pages}{311--322}.
\newblock \DOIprefix\doi{10.1287/mnsc.1050.0451}.
\bibitem[{Sun(2026)}]{sun2026ewm}
\bibinfo{author}{Sun, L.}, \bibinfo{year}{2026}.
\newblock \bibinfo{title}{Empirical welfare maximization with constraints}.
\newblock \bibinfo{journal}{Journal of Econometrics} \bibinfo{volume}{253},
  \bibinfo{pages}{106169}.
\newblock \DOIprefix\doi{10.1016/j.jeconom.2025.106169}.
\bibitem[{Thomas and Brunskill(2016)}]{thomas2016ope}
\bibinfo{author}{Thomas, P.S.}, \bibinfo{author}{Brunskill, E.},
  \bibinfo{year}{2016}.
\newblock \bibinfo{title}{Data-efficient off-policy policy evaluation for
  reinforcement learning}, in: \bibinfo{booktitle}{Proceedings of the 33rd
  International Conference on Machine Learning (ICML), PMLR}, pp.
  \bibinfo{pages}{2139--2148}.
\bibitem[{Thomas et~al.(2015)Thomas, Theocharous and
  Ghavamzadeh}]{thomas2015hcope}
\bibinfo{author}{Thomas, P.S.}, \bibinfo{author}{Theocharous, G.},
  \bibinfo{author}{Ghavamzadeh, M.}, \bibinfo{year}{2015}.
\newblock \bibinfo{title}{High-confidence off-policy evaluation}, in:
  \bibinfo{booktitle}{Proceedings of the Twenty-Ninth AAAI Conference on
  Artificial Intelligence (AAAI)}, pp. \bibinfo{pages}{3000--3006}.
\newblock \DOIprefix\doi{10.1609/aaai.v29i1.9541}.
\bibitem[{Vanderschueren et~al.(2024)Vanderschueren, Baesens, Verdonck and
  Verbeke}]{vanderschueren2024allocating}
\bibinfo{author}{Vanderschueren, T.}, \bibinfo{author}{Baesens, B.},
  \bibinfo{author}{Verdonck, T.}, \bibinfo{author}{Verbeke, W.},
  \bibinfo{year}{2024}.
\newblock \bibinfo{title}{A new perspective on classification: Optimally
  allocating limited resources to uncertain tasks}.
\newblock \bibinfo{journal}{Decision Support Systems} \bibinfo{volume}{179},
  \bibinfo{pages}{114151}.
\newblock \DOIprefix\doi{10.1016/j.dss.2024.114151}.
\bibitem[{Wang et~al.(2025)Wang, Mehrotra and Peng}]{wang2025robust}
\bibinfo{author}{Wang, S.}, \bibinfo{author}{Mehrotra, S.},
  \bibinfo{author}{Peng, C.}, \bibinfo{year}{2025}.
\newblock \bibinfo{title}{Robust concave utility maximization over chance
  constraints}.
\newblock \bibinfo{journal}{European Journal of Operational Research}
  \bibinfo{volume}{321}, \bibinfo{pages}{800--813}.
\newblock \DOIprefix\doi{10.1016/j.ejor.2024.10.007}.
\bibitem[{Wu et~al.(2024)Wu, Zhou, Chen and Zhu}]{wu2024counterfactual}
\bibinfo{author}{Wu, S.}, \bibinfo{author}{Zhou, W.}, \bibinfo{author}{Chen,
  M.}, \bibinfo{author}{Zhu, S.}, \bibinfo{year}{2024}.
\newblock \bibinfo{title}{Counterfactual generative models for time-varying
  treatments}, in: \bibinfo{booktitle}{Proceedings of the 30th ACM SIGKDD
  Conference on Knowledge Discovery and Data Mining (KDD)}, pp.
  \bibinfo{pages}{3402--3413}.
\newblock \DOIprefix\doi{10.1145/3637528.3671950}.
\bibitem[{Xiong et~al.(2024)Xiong, Wu, Deng, Su and
  Lehman}]{xiong2024gtransformer}
\bibinfo{author}{Xiong, H.}, \bibinfo{author}{Wu, F.}, \bibinfo{author}{Deng,
  L.}, \bibinfo{author}{Su, M.}, \bibinfo{author}{Lehman, L.w.H.},
  \bibinfo{year}{2024}.
\newblock \bibinfo{title}{{G-Transformer}: counterfactual outcome prediction
  under dynamic and time-varying treatment regimes}, in:
  \bibinfo{booktitle}{Proceedings of the 9th Machine Learning for Healthcare
  Conference (MLHC), PMLR}.
\bibitem[{Zhao et~al.(2012)Zhao, Zeng, Rush and Kosorok}]{zhao2012owl}
\bibinfo{author}{Zhao, Y.}, \bibinfo{author}{Zeng, D.}, \bibinfo{author}{Rush,
  A.J.}, \bibinfo{author}{Kosorok, M.R.}, \bibinfo{year}{2012}.
\newblock \bibinfo{title}{Estimating individualized treatment rules using
  outcome weighted learning}.
\newblock \bibinfo{journal}{Journal of the American Statistical Association}
  \bibinfo{volume}{107}, \bibinfo{pages}{1106--1118}.
\newblock \DOIprefix\doi{10.1080/01621459.2012.695674}.
\bibitem[{Zheng et~al.(2017)Zheng, Ristovski, Farahat and
  Gupta}]{zheng2017lstm}
\bibinfo{author}{Zheng, S.}, \bibinfo{author}{Ristovski, K.},
  \bibinfo{author}{Farahat, A.}, \bibinfo{author}{Gupta, C.},
  \bibinfo{year}{2017}.
\newblock \bibinfo{title}{Long short-term memory network for remaining useful
  life estimation}, in: \bibinfo{booktitle}{2017 IEEE International Conference
  on Prognostics and Health Management (ICPHM)}, pp. \bibinfo{pages}{88--95}.
\newblock \DOIprefix\doi{10.1109/ICPHM.2017.7998311}.
\bibitem[{Zhong et~al.(2026)Zhong, Zhao, Zhang and Jiang}]{zhong2026ccsb}
\bibinfo{author}{Zhong, Y.}, \bibinfo{author}{Zhao, B.},
  \bibinfo{author}{Zhang, K.}, \bibinfo{author}{Jiang, G.},
  \bibinfo{year}{2026}.
\newblock \bibinfo{title}{Chance-constrained selection of the best: An
  indifference-zone-free procedure}.
\newblock \bibinfo{journal}{European Journal of Operational Research}
  \bibinfo{volume}{335}, \bibinfo{pages}{479--489}.
\newblock \DOIprefix\doi{10.1016/j.ejor.2026.05.042}.

\end{thebibliography}

\clearpage
\appendix
\renewcommand{\thesection}{S\arabic{section}}
\setcounter{section}{0}
\renewcommand{\thetable}{S\arabic{table}}
\setcounter{table}{0}
\renewcommand{\thefigure}{S\arabic{figure}}
\setcounter{figure}{0}
\renewcommand{\thealgorithm}{S\arabic{algorithm}}
\setcounter{algorithm}{0}
\begin{center}{\LARGE\bfseries Supplementary material}\end{center}
\vspace{1em}
\section{Proof of Proposition 4}
\label{app:prop4}

Proposition~4 of Section~\ref{sec:method-guarantees} states the following. With $\delta$ and $\epsilon_V$ set as in Equation~\eqref{eq:l1const}, the strategy $\hat{g}$ returned by the empirical chance-constraint rule at tolerance $\varepsilon$ satisfies Equation~\eqref{eq:l1regret} with probability at least $1-\eta$. The
argument intersects two uniform deviation bounds, each taken at half the confidence budget.

\paragraph{Uniform bounds}
Applying the Dvoretzky-Kiefer-Wolfowitz inequality with the tight constant of \citet{massart1990dkw} to each
strategy separately and taking a union bound over $\mathcal{G}$ at level $\eta/2$ gives
\begin{equation}
\sup_b \big\lvert \hat{F}_g(b) - F_g(b) \big\rvert \le \delta
\qquad \text{for every } g \in \mathcal{G},
\label{eq:p4dkw}
\end{equation}
with $\delta = \sqrt{\log(4\lvert\mathcal{G}\rvert/\eta)/(2n)}$. The per-unit contributions to $\hat{V}(g)$ lie
in an interval of length $R$ by assumption, so Hoeffding's inequality \citep{hoeffding1963} applied to
$\hat{V}(g)$ and a union bound at the same level give
\begin{equation}
\big\lvert \hat{V}(g) - V(g) \big\rvert \le \epsilon_V
\qquad \text{for every } g \in \mathcal{G},
\label{eq:p4hoeff}
\end{equation}
with $\epsilon_V = R\sqrt{\log(4\lvert\mathcal{G}\rvert/\eta)/(2n)}$. Each event fails with probability at most
$\eta/2$, so both hold simultaneously with probability at least $1-\eta$. The remainder of the argument works on
that intersection, which makes the conclusions of Equation~\eqref{eq:l1regret} simultaneous rather than separate.

\paragraph{Feasibility}
The empirical rule admits $\hat{g}$ only if $1 - \hat{F}_{\hat{g}}(B) \le \varepsilon$. Evaluating
Equation~\eqref{eq:p4dkw} at $b = B$ gives $F_{\hat{g}}(B) \ge \hat{F}_{\hat{g}}(B) - \delta \ge (1-\varepsilon) -
\delta$, and rearranging gives $\Pr(C(\hat{g}) > B) \le \varepsilon + \delta$. This is the argument of
Proposition~3 at the constant of Equation~\eqref{eq:l1const}.

\paragraph{Regret}
Write $\tilde{g} = g^{\ast}(\varepsilon-\delta)$ for the best strategy truly feasible at the tightened tolerance.
Since $\Pr(C(\tilde{g}) > B) \le \varepsilon - \delta$, Equation~\eqref{eq:p4dkw} evaluated at $b = B$ gives
$1 - \hat{F}_{\tilde{g}}(B) \le \varepsilon$, so $\tilde{g}$ is admitted by the empirical rule. The rule returns
the admitted strategy of greatest estimated value, and it maximizes over a set containing $\tilde{g}$, so
$\hat{V}(\hat{g}) \ge \hat{V}(\tilde{g})$. Applying Equation~\eqref{eq:p4hoeff} to $\tilde{g}$ and $\hat{g}$,
\begin{equation}
V(\tilde{g}) - V(\hat{g})
\;\le\; \big(\hat{V}(\tilde{g}) + \epsilon_V\big) - \big(\hat{V}(\hat{g}) - \epsilon_V\big)
\;\le\; 2\epsilon_V ,
\label{eq:p4regret}
\end{equation}
which is the second half of Equation~\eqref{eq:l1regret}. The tightened tolerance enters only through the admission of $\tilde{g}$: at the nominal tolerance $\varepsilon$ a strategy with a true tail exactly at $\varepsilon$ need not be admitted. The argument would then have no benchmark inside the empirical feasible set.

\section{Notation}
\label{app:notation}
Table~\ref{tab:notation} collects the notation and Table~\ref{tab:glossary} the terms used throughout.

\begin{table}[pos=h!]
\centering
\caption{Notation used throughout.}
\label{tab:notation}
\small
\begin{adjustbox}{max width=\textwidth}
\begin{tabular}{@{}ll@{}}
\toprule
Symbol & Meaning \\
\midrule
$t \in \{0, \ldots, K\}$ & Discrete decision steps over the horizon \\
$L_t$, $A_t$ & Time-varying covariates and treatment decision at step $t$ \\
$\bar{L}_m$, $\bar{A}_{m-1}$ & Covariate and treatment histories up to step $m$ \\
$H_m = (\bar{L}_m, \bar{A}_{m-1})$ & Observed history at step $m$ \\
$L_t(g)$ & Potential covariate vector at step $t$ under strategy $g$ \\
$g = (g_m, \ldots, g_K)$ & Dynamic treatment strategy, with $g_t : H_t \mapsto A_t$ \\
$\mathcal{G}$ & Finite candidate set of strategies \\
\midrule
$V(g)$, $\hat{V}(g)$ & Outcome value of strategy $g$, and its estimate \\
$c(\cdot)$ & Cost of the treatment administered at a step \\
$C(g) = \sum_t c(A_t)$ & Cumulative cost of strategy $g$ over the horizon \\
$S_t = \sum_{s=m}^{t} c(A_s)$ & Partial cost sum through step $t$ \\
$\hat{C}(g)$, $\hat{\sigma}_C(g)$ & Estimated mean and standard deviation of the cost \\
$F_g$, $\hat{F}_g$ & True and empirical distribution function of $C(g)$ \\
\midrule
$B$ & Cumulative budget \\
$\varepsilon$ & Tolerated probability that the cost exceeds the budget \\
$\kappa$ & Margin parameter of the upper-bound rule \\
$g^{\ast}_{\mathrm{naive}}$, $g^{\ast}_{\mathrm{ub}}$, $g^{\ast}_{\mathrm{cc}}$ & Strategies selected by the three decision rules \\
$\mathcal{A}(\varepsilon)$ & Admissible set at tolerance $\varepsilon$ \\
$V^{\ast}(\varepsilon)$ & Achievable outcome at tolerance $\varepsilon$ \\
$\rho$ & Target ceiling on the realized tail probability \\
$\mathcal{P}_r(\hat{F}_g)$, $r$ & Kolmogorov ambiguity set and its radius \\
\midrule
$n$ & Number of independent samples per strategy \\
$\delta$ & Uniform deviation of $\hat{F}_g$ from $F_g$ \\
$\epsilon_V$, $R$ & Value deviation, and the range of per-unit value contributions \\
$\eta$ & Failure probability of the finite-sample guarantees \\
\midrule
$g^{\mathrm{orc}}(B)$, $\mathrm{Reg}(g)$ & Oracle optimum at budget $B$, and regret against it \\
$\varepsilon_s$, $\varepsilon_c$ & Standard deviation of the injected outcome and cost error \\
$b_c$ & Relative systematic bias in the estimated cost \\
$\lambda$ & Per-step treatment penalty used to generate candidates \\
\bottomrule
\end{tabular}
\end{adjustbox}
\end{table}

\begin{table}[pos=h!]
\centering
\caption{Terms used throughout, with the section that defines each.}
\label{tab:glossary}
\small
\begin{adjustbox}{max width=\textwidth}
\begin{tabular}{@{}p{0.24\textwidth}p{0.58\textwidth}l@{}}
\toprule
Term & Definition & Defined in \\
\midrule
Outcome regret & Outcome of the oracle at a budget minus the outcome of the selected strategy, in percentage points & \ref{sec:exp-oracle} \\
Budget violation rate & Fraction of budget-and-seed combinations in which the selected strategy's true expected cost exceeds the budget & \ref{sec:exp-rules} \\
Realized tail probability & $\Pr(C(g) > B)$ for the selected strategy, on its true cost samples & \ref{sec:exp-rules} \\
Oracle match rate & Fraction of budget-and-seed combinations in which the selected strategy is the one the oracle selects at that budget & \ref{sec:exp-rules} \\
Oracle cost distribution & Exact counterfactual distribution of the cumulative cost of a strategy & \ref{sec:exp-oracle} \\
Oracle at a budget & Mean-feasible optimum, the outcome-maximizing strategy with true expected cost within the budget & \ref{sec:exp-oracle} \\
Tolerance $\varepsilon$ & Tolerated probability that the cumulative cost exceeds the budget & \ref{sec:method-rules} \\
Certified slack $\delta$ & Uniform deviation of the empirical cost distribution from the true one, at confidence $1-\eta$ & \ref{sec:method-guarantees} \\
Certified ceiling & $\varepsilon + \delta$, the bound Proposition~3 places on the realized tail probability & \ref{sec:method-guarantees} \\
Admissible set $\mathcal{A}(\varepsilon)$ & Candidate strategies satisfying the chance constraint at a given tolerance and budget & \ref{sec:method-operating} \\
Operating point & The tolerance the decision maker adopts, and the recommendation it produces & \ref{sec:method-operating} \\
Safety-utility frontier & Achievable outcome against the tolerated violation as $\varepsilon$ varies & \ref{sec:method-operating} \\
Budget frontier & Oracle-optimal outcome as the budget varies & \ref{sec:exp-rules} \\
Coverage gap & Fraction of the discretized state-action cells a strategy traverses in which the behavior policy generated no samples & \ref{app:funcapprox} \\
\bottomrule
\end{tabular}
\end{adjustbox}
\end{table}

\section{Identification of the counterfactual cost distribution}
\label{app:identification}

Section~\ref{sec:method-setup} states Equation~\eqref{eq:gcomp} and attributes it to \citet{robins1986}. The budget constraint is imposed on the full law of the cumulative cost rather than on a summary of it. We set out the argument that carries Assumptions 1 to 3 to that law.

Write $\bar{L}_{m:K}(g)$ for the covariate path generated under strategy $g$. Fix a path $\bar{l}_{m:K}$ and let $\bar{a}_{t-1}$ denote the treatment history the strategy assigns along it, $a_s = g_s(h_s)$ for $s < t$. By the chain rule, the counterfactual path law factorizes into its step-wise conditionals,
\begin{equation}
\Pr\big(\bar{L}_{m:K}(g) = \bar{l}_{m:K}\big) = \prod_{t=m}^{K} \Pr\big(L_t(g) = l_t \mid \bar{L}_{t-1}(g) = \bar{l}_{t-1}\big).
\label{eq:idchain}
\end{equation}
Under Assumption 1, the treatment received at each step is independent of the future potential covariate path given the observed history. Conditioning further on the event that the observed treatment history agrees with what the strategy assigns therefore leaves each factor unchanged,
\begin{equation}
\Pr\big(L_t(g) = l_t \mid \bar{L}_{t-1}(g) = \bar{l}_{t-1}\big) = \Pr\big(L_t(g) = l_t \mid \bar{L}_{t-1} = \bar{l}_{t-1}, \bar{A}_{t-1} = \bar{a}_{t-1}\big).
\label{eq:idignor}
\end{equation}
On that event, Assumption 3 equates the potential covariate with the observed one. This replaces the counterfactual on the right of Equation~\eqref{eq:idignor} by a conditional law of the observed data,
\begin{equation}
\Pr\big(L_t(g) = l_t \mid \bar{L}_{t-1} = \bar{l}_{t-1}, \bar{A}_{t-1} = \bar{a}_{t-1}\big) = p\big(l_t \mid \bar{l}_{t-1}, \bar{a}_{t-1}\big).
\label{eq:idconsist}
\end{equation}
Assumption 2 makes each conditioning event in Equations~\eqref{eq:idignor} and~\eqref{eq:idconsist} occur with positive probability, so every factor is well defined. Substituting Equations~\eqref{eq:idignor} and~\eqref{eq:idconsist} into Equation~\eqref{eq:idchain} gives Equation~\eqref{eq:gcomp}.

The cumulative cost is a deterministic function of the path. Along $\bar{l}_{m:K}$ the strategy assigns $a_t = g_t(h_t)$ at every step, so $C(g) = \sum_{t=m}^{K} c\big(g_t(h_t)\big)$ carries no randomness beyond that of the path itself. Its law is the image of the path law under that map,
\begin{equation}
F_g(b) = \sum_{\bar{l}_{m:K}} \Pr\big(\bar{L}_{m:K}(g) = \bar{l}_{m:K}\big) \, \mathbb{1}\!\left[\sum_{t=m}^{K} c\big(g_t(h_t)\big) \le b\right],
\label{eq:idpushforward}
\end{equation}
which is Equation~\eqref{eq:costdist} once the path law is replaced by its identified form. The outcome value $V(g)$ of Equation~\eqref{eq:value} is an expectation under the same law and is identified by the same substitution. Both quantities the decision rule consumes therefore rest on Equations~\eqref{eq:idchain} to~\eqref{eq:idconsist} and on nothing further.

The three assumptions deliver the whole of $F_g$ and not one of its moments. A rule that constrains $\mathbb{E}[C(g)]$ uses only the first moment of Equation~\eqref{eq:idpushforward}, and Equation~\eqref{eq:meandecomp} shows that this moment is a sum of per-stage terms. The chance constraint of Equation~\eqref{eq:cc} evaluates $F_g$ at the budget, which Equation~\eqref{eq:taildecomp} shows depends on the dependence between stages and not on the per-stage laws alone. The identification argument is indifferent to which functional is taken, but the estimation problem is not: the tail at a single point is estimated less accurately than a mean at the same sample size. Proposition~3 bounds the consequence of that gap for the recommendation.

\section{Selection procedure}
\label{app:algorithm}

Section~\ref{sec:method-decision} states the chance-constrained rule. Algorithm~\ref{alg:selection} writes it out. Selection touches each candidate once, so the cost is $O(\lvert\mathcal{G}\rvert n)$ in the size of the candidate set and the number of cost samples. A sweep over $\varepsilon$ reuses the same per-strategy tail counts.

\begin{algorithm}[h!]
\caption{Chance-constrained selection at a tolerated violation $\varepsilon$.}
\label{alg:selection}
\begin{algorithmic}[1]
\Require candidate strategies $\mathcal{G}$, budget $B$, tolerated violation $\varepsilon$
\Ensure recommended strategy $g^{\ast}_{\mathrm{cc}}$
\For{each $g \in \mathcal{G}$}
  \State obtain $\hat{V}(g)$ and cost samples $C^{(1)}(g), \ldots, C^{(n)}(g)$ from the predictor
  \State $\widehat{v}(g) \gets \frac{1}{n}\sum_{j=1}^{n} \mathbf{1}[\,C^{(j)}(g) > B\,]$ \Comment{empirical tail probability $1 - \hat{F}_g(B)$}
\EndFor
\State $\mathcal{A} \gets \{\, g \in \mathcal{G} : \widehat{v}(g) \le \varepsilon \,\}$ \Comment{admissible set}
\State \Return $g^{\ast}_{\mathrm{cc}} \gets \arg\max_{g \in \mathcal{A}} \hat{V}(g)$, or infeasible if $\mathcal{A} = \emptyset$
\end{algorithmic}
\end{algorithm}

\section{Environment specifications}
\label{app:envspec}

Section~\ref{sec:experiments} summarizes each environment. This appendix records the mechanism in
full, so that every quantity the decision rule consumes can be traced to its generative source.
Table~\ref{tab:envscale} gives the scale of each environment and the range each quantity covers,
completing Table~\ref{tab:environments}.

\begin{table}[pos=h!]
\centering
\small
\caption{Scale and range of the five evaluation environments.}
\label{tab:envscale}
\begin{adjustbox}{max width=\textwidth}
\begin{tabular}{@{}lccccc@{}}
\toprule
 & Sepsis & Tumor & MIMIC-IV & C-MAPSS & Drink Less \\
\midrule
Units & 8000 & 2000 & 3000 & 2000 & 349 \\
Horizon & 20 & 30 & 24 & 500 & 30 \\
Decision points & 20 & 30 & 16.0$^{\ddagger}$ & 25 & 30 \\
State & 720 discrete & continuous & continuous & continuous & continuous \\
Actions & 8 & 4 & 2 & 2 & 2 \\
Candidate strategies & 21 & 23 & 25 & 25 & 17 \\
\midrule
Mean cost range & 0.00--34.03 & 0.00--14.19 & 1.08--15.98 & 0.85--25.00 & 1.51--28.52 \\
Outcome range & 9.9--96.5\% & 65.5--99.8\% & 35.1--90.2\% & 43.6--100.0\% & 1.23--3.64$^{\dagger}$ \\
\bottomrule
\end{tabular}
\end{adjustbox}
\begin{flushleft}\footnotesize
The cost unit and the outcome measure of each column are given in Table~\ref{tab:environments}. $^{\dagger}$ Drink Less values are per-participant IPW estimates of expected cumulative engagement, not an outcome probability. $^{\ddagger}$ Mean over stays; episodes follow the real stay lengths.
\end{flushleft}
\end{table}

\subsection{Sepsis simulator}

A patient state is described by four physiological variables, heart rate, blood pressure, oxygen saturation and glucose. Three binary indicators record whether each of antibiotics, vasopressors and mechanical ventilation is currently active. The simulator carries a
latent binary diabetes indicator, drawn once per patient and held fixed over the episode.
This indicator makes treatment confounded with prognosis in the observational trajectories of
Section~\ref{sec:res-predictor}. At each step, the clinician chooses one of eight actions, the on
or off setting of the three treatments.

Policies are solved on the mixture of the two diabetes-specific transition matrices over the $720$ observed states, while rollouts use the diabetes-specific dynamics. The oracle therefore follows the simulator's generative process rather than the mixture used for planning. Following the simulator's definition, a state is absorbing and fatal when at least three physiological variables are abnormal. It is absorbing and discharging when no variable is abnormal and no treatment is active. All other states are transient. An episode runs for at most twenty steps, with a reward of
$+1$ for discharge and $-1$ for death.

\subsection{Tumor-growth simulator}

The environment implements the pharmacokinetic-pharmacodynamic model of \citet{geng2017tumor}, in which tumor volume evolves under chemotherapy and radiotherapy over a fixed horizon. Growth and treatment-response parameters are drawn per patient, so two patients under the same strategy follow different volume paths and consume different amounts of treatment. That per-patient heterogeneity, rather than stochastic transitions, is the source of the cost spread the chance constraint operates on. This makes the environment complementary to the sepsis simulator.

A candidate strategy is a volume-threshold rule: treatment is administered at a step whenever the tumor volume exceeds the strategy's threshold. A lower threshold is therefore a more aggressive strategy. The cost is the cumulative count of treatments administered along the realized path. The outcome is binary success, taken as a final tumor volume below a fixed level. Because the dynamics are specified in full, the success probability and the cost distribution of every candidate are obtained by simulating the model directly. The oracle and the decision-quality metrics are therefore defined exactly as in the sepsis environment.

\subsection{Maintenance environment}

Each engine's condition is summarized by a scalar health index obtained from a closed-form regression of its informative sensor channels onto the fraction of life elapsed, clipped to the unit interval and made monotone. The index therefore rises from zero at installation toward one at failure. Inspections are spaced a fixed number of cycles apart, and a candidate strategy performs
an overhaul whenever the index exceeds its threshold.

Baseline degradation is taken directly from the real run-to-failure trajectories. The effect of an overhaul is modeled as a virtual-age reset as is standard in reliability analysis. The reset restores the engine to a younger effective age instead of to new. The spread in the cost distribution is induced by a per-engine multiplier on the degradation rate. The time-to-failure heterogeneity that makes the chance constraint informative therefore comes from the fleet rather than from the maintenance policy.

\subsection{Semi-synthetic cohort}

The sampled cohort comprises $47{,}986$ hourly records drawn from an extraction of $27{,}935$ stays. Table~\ref{tab:cohort} reports its composition.

\begin{table}[pos=h!]
\centering
\small
\caption{Semi-synthetic cohort, mean (s.d.) over the sampled trajectories.}
\label{tab:cohort}
\begin{tabular}{@{}lr@{}}
\toprule
Quantity & Value \\
\midrule
Trajectories sampled & $3{,}000$ \\
Sequence length (steps) & $16.0$ $(5.7)$ \\
Age (years) & $65.6$ $(16.2)$ \\
Charlson comorbidity index & $5.2$ $(3.0)$ \\
SOFA score & $5.5$ $(2.7)$ \\
\bottomrule
\end{tabular}
\end{table}

The covariate trajectories therefore carry the severity structure of a real intensive-care
population. The cohort is large enough that the empirical cost distribution on which the chance constraint operates is stable across resamples. The benchmark remains tractable to regenerate under each random seed.

The outcome follows the benchmark model of the Causal Transformer \citep{melnychuk2022ct}. An untreated trajectory is the sum of an exogenous component and an endogenous component. The exogenous component is a random nonlinear function of selected real covariates, and the endogenous component is drawn from a Gaussian process with a spline trend. A
designed treatment effect that decays over an effect window is applied for each of the three
synthetic treatments. The terminal outcome is taken from the trajectory by a soft threshold, so that a more favorable trajectory has a higher probability of the favorable outcome. The threshold is placed at a quantile of the untreated outcome distribution that fixes the baseline difficulty of the benchmark.

\subsection{Semi-synthetic candidate ladder and benchmark variation}

Each candidate strategy applies the three synthetic treatments according to thresholds on the running outcome. A single aggressiveness parameter shifts all three thresholds together, from a setting that treats only the most severe trajectories to one that treats readily. Sweeping it produces a family of state-dependent regimes analogous to the penalty sweep in the simulators. The exact oracle, the injected estimation error and the decision rules of the main text apply to it without change.

Because the synthetic outcome is a design choice, the benchmark is varied. The difficulty is varied by placing the outcome threshold at different quantiles of the untreated outcome distribution. The real covariates that drive the exogenous component are exchanged across disjoint physiological groups. Section~\ref{sec:res-generalization} reports the decision-rule behavior under each.

\subsection{Drink Less policy family}

The candidate policies are written down from the trial's own randomization structure rather than solved for. Global policies notify at a fixed rate over the horizon. State-aware policies notify more often where the trial shows the notification to be most effective. They concentrate notifications early in enrollment and on days when the participant has already engaged. Every policy is softened away from the deterministic boundary, so that no decision point receives a candidate probability of zero or one and the per-decision importance weights remain bounded. \ref{app:ess} reports the resulting effective sample sizes and the largest weight each policy forms against the trial's randomization probabilities.

\section{Per-budget decision-rule tables}
\label{sup:perbudget}

\label{app:perbudget}

Tables~\ref{tab:app-sepsis} to~\ref{tab:app-cmapss} report the three decision rules at every budget in each environment, the per-budget breakdown underlying the summaries of Section~\ref{sec:res-rules} and Table~\ref{tab:three-env}. Regret is in outcome percentage points and violation in percent, each averaged over $10$ seeds. The summary value for a rule is the unweighted mean of its column here, so every summary value quoted in Section~\ref{sec:results} is reproducible from these tables. A budget is included only where every seed admits a strategy under every rule; the excluded cells are shown as a dash.

In the violation columns, the upper-bound rule admits a strategy only when the estimated mean plus a positive margin stays within the budget. The zeros in its violation column therefore record the admission criterion rather than a measured outcome.

\begin{table}[pos=h!]
\centering
\small
\caption{Sepsis: decision rules at each budget ($15\%$ injected error, $\varepsilon=0.2$; mean $\pm$ s.d., 10 seeds). The last column is the chance rule at $0\%$ injected error.}
\label{tab:app-sepsis}
\begin{tabular}{@{}r rr rr rrr@{}}
\toprule
 & \multicolumn{2}{c}{Naive} & \multicolumn{2}{c}{Upper bound} & \multicolumn{3}{c}{Chance ($\varepsilon=0.2$)} \\
\cmidrule(lr){2-3}\cmidrule(lr){4-5}\cmidrule(lr){6-8}
$B$ & Regret & Viol. & Regret & Viol. & Regret & Viol. & Regret ($0\%$) \\
\midrule
$1$ & $1.6{\pm}0.4$ & $0.0{\pm}0.0$ & $12.0{\pm}0.6$ & $0.0{\pm}0.0$ & $12.0{\pm}0.6$ & $0.0{\pm}0.0$ & $10.5$ \\
$2$ & $3.5{\pm}0.7$ & $0.0{\pm}0.0$ & $19.5{\pm}0.6$ & $0.0{\pm}0.0$ & $10.5{\pm}0.6$ & $0.0{\pm}0.0$ & $8.6$ \\
$4$ & $1.4{\pm}3.3$ & $18.7{\pm}19.7$ & $15.2{\pm}3.8$ & $0.0{\pm}0.0$ & $9.3{\pm}3.6$ & $0.0{\pm}0.0$ & $7.6$ \\
$6$ & $2.0{\pm}3.6$ & $17.1{\pm}17.8$ & $19.6{\pm}3.8$ & $0.0{\pm}0.0$ & $17.1{\pm}3.8$ & $0.0{\pm}0.0$ & $14.3$ \\
$8$ & $2.3{\pm}0.3$ & $10.1{\pm}2.7$ & $28.3{\pm}0.9$ & $0.0{\pm}0.0$ & $25.6{\pm}0.9$ & $0.0{\pm}0.0$ & $23.3$ \\
$12$ & $3.5{\pm}0.4$ & $6.2{\pm}1.7$ & $26.8{\pm}0.4$ & $0.0{\pm}0.0$ & $26.6{\pm}0.5$ & $0.0{\pm}0.0$ & $23.5$ \\
$20$ & $3.1{\pm}0.3$ & $4.1{\pm}1.1$ & $18.0{\pm}0.3$ & $0.0{\pm}0.0$ & $17.3{\pm}0.3$ & $0.0{\pm}0.0$ & $12.1$ \\
\bottomrule
\end{tabular}
\end{table}

\begin{table}[pos=h!]
\centering
\small
\caption{Tumor: decision rules at each budget ($15\%$ injected error, $\varepsilon=0.2$; mean $\pm$ s.d., 10 seeds). The last column is the chance rule at $0\%$ injected error.}
\label{tab:app-tumor}
\begin{tabular}{@{}r rr rr rrr@{}}
\toprule
 & \multicolumn{2}{c}{Naive} & \multicolumn{2}{c}{Upper bound} & \multicolumn{3}{c}{Chance ($\varepsilon=0.2$)} \\
\cmidrule(lr){2-3}\cmidrule(lr){4-5}\cmidrule(lr){6-8}
$B$ & Regret & Viol. & Regret & Viol. & Regret & Viol. & Regret ($0\%$) \\
\midrule
$1$ & $2.6{\pm}0.7$ & $2.4{\pm}1.2$ & $5.0{\pm}0.7$ & $0.0{\pm}0.0$ & $4.6{\pm}0.8$ & $0.1{\pm}0.1$ & $4.4$ \\
$2$ & $-0.8{\pm}1.2$ & $19.1{\pm}5.8$ & $14.6{\pm}0.7$ & $0.0{\pm}0.0$ & $11.2{\pm}0.9$ & $0.0{\pm}0.0$ & $1.9$ \\
$3$ & $1.0{\pm}0.2$ & $17.4{\pm}7.8$ & $32.4{\pm}0.9$ & $0.0{\pm}0.0$ & $24.6{\pm}1.1$ & $0.0{\pm}0.0$ & $21.7$ \\
$4$ & $0.2{\pm}0.0$ & $11.0{\pm}9.0$ & $26.5{\pm}1.2$ & $0.0{\pm}0.0$ & $19.4{\pm}1.6$ & $0.0{\pm}0.0$ & $3.7$ \\
$5$ & $0.1{\pm}0.0$ & $18.1{\pm}7.7$ & $22.0{\pm}1.3$ & $0.0{\pm}0.0$ & $5.4{\pm}1.1$ & $0.0{\pm}0.0$ & $3.7$ \\
$6$ & $0.0{\pm}0.0$ & $22.8{\pm}7.7$ & $9.7{\pm}1.9$ & $0.0{\pm}0.0$ & $1.3{\pm}0.2$ & $0.0{\pm}0.0$ & $0.0$ \\
\bottomrule
\end{tabular}
\end{table}

\begin{table}[pos=h!]
\centering
\small
\caption{Semi-synthetic (MIMIC-IV): decision rules at each budget ($15\%$ injected error, $\varepsilon=0.2$; mean $\pm$ s.d., 10 seeds). The last column is the chance rule at $0\%$ injected error.}
\label{tab:app-mimic}
\begin{tabular}{@{}r rr rr rrr@{}}
\toprule
 & \multicolumn{2}{c}{Naive} & \multicolumn{2}{c}{Upper bound} & \multicolumn{3}{c}{Chance ($\varepsilon=0.2$)} \\
\cmidrule(lr){2-3}\cmidrule(lr){4-5}\cmidrule(lr){6-8}
$B$ & Regret & Viol. & Regret & Viol. & Regret & Viol. & Regret ($0\%$) \\
\midrule
$6$ & $3.2{\pm}0.9$ & $8.4{\pm}6.1$ & $17.7{\pm}1.1$ & $0.0{\pm}0.0$ & $15.3{\pm}0.8$ & $0.0{\pm}0.0$ & $11.1$ \\
$8$ & $3.3{\pm}0.6$ & $6.0{\pm}5.7$ & $17.2{\pm}1.8$ & $0.0{\pm}0.0$ & $14.5{\pm}1.0$ & $0.0{\pm}0.0$ & $10.2$ \\
$10$ & $3.1{\pm}0.4$ & $7.2{\pm}4.5$ & $15.1{\pm}2.2$ & $0.0{\pm}0.0$ & $12.6{\pm}1.4$ & $0.0{\pm}0.0$ & $8.1$ \\
$12$ & $3.4{\pm}1.0$ & $9.9{\pm}3.7$ & $14.6{\pm}2.2$ & $0.0{\pm}0.0$ & $12.6{\pm}1.7$ & $0.0{\pm}0.0$ & $8.4$ \\
$14$ & $4.0{\pm}0.7$ & $8.0{\pm}4.7$ & $18.6{\pm}2.7$ & $0.0{\pm}0.0$ & $16.7{\pm}3.2$ & $0.0{\pm}0.0$ & $12.9$ \\
$16$ & $4.1{\pm}0.4$ & $4.0{\pm}7.3$ & $19.7{\pm}1.1$ & $0.0{\pm}0.0$ & $17.8{\pm}1.2$ & $0.0{\pm}0.0$ & $13.2$ \\
\bottomrule
\end{tabular}
\end{table}

\begin{table}[pos=h!]
\centering
\small
\caption{Maintenance (C-MAPSS): decision rules at each budget ($15\%$ injected error, $\varepsilon=0.2$; mean $\pm$ s.d., 10 seeds). The last column is the chance rule at $0\%$ injected error. Budgets at which no strategy is admitted are shown as a dash and excluded from the averages.}
\label{tab:app-cmapss}
\begin{tabular}{@{}r rr rr rrr@{}}
\toprule
 & \multicolumn{2}{c}{Naive} & \multicolumn{2}{c}{Upper bound} & \multicolumn{3}{c}{Chance ($\varepsilon=0.2$)} \\
\cmidrule(lr){2-3}\cmidrule(lr){4-5}\cmidrule(lr){6-8}
$B$ & Regret & Viol. & Regret & Viol. & Regret & Viol. & Regret ($0\%$) \\
\midrule
$2$ & $0.0{\pm}0.0$ & $0.0{\pm}0.0$ & $0.0{\pm}0.0$ & $0.0{\pm}0.0$ & $0.0{\pm}0.0$ & $0.0{\pm}0.0$ & $0.0$ \\
$3$ & $2.1{\pm}1.6$ & $0.0{\pm}0.0$ & $56.3{\pm}1.1$ & $0.0{\pm}0.0$ & $56.3{\pm}1.1$ & $0.0{\pm}0.0$ & $51.4$ \\
$4$ & $0.1{\pm}0.0$ & $8.7{\pm}2.6$ & $16.8{\pm}6.9$ & $0.0{\pm}0.0$ & $0.6{\pm}0.3$ & $0.1{\pm}0.1$ & $0.0$ \\
$5$ & $0.0{\pm}0.0$ & $0.8{\pm}0.4$ & $0.2{\pm}0.1$ & $0.0{\pm}0.0$ & $0.0{\pm}0.0$ & $0.1{\pm}0.1$ & $0.0$ \\
$6$ & $0.0{\pm}0.0$ & $5.2{\pm}9.7$ & $0.1{\pm}0.1$ & $0.0{\pm}0.0$ & $0.0{\pm}0.0$ & $0.4{\pm}0.8$ & $0.0$ \\
$7$ & $0.0{\pm}0.0$ & $17.0{\pm}6.6$ & $0.1{\pm}0.1$ & $0.0{\pm}0.0$ & $0.0{\pm}0.0$ & $0.7{\pm}0.4$ & $0.0$ \\
\bottomrule
\end{tabular}
\end{table}

\section{Supporting results from the main text}
\label{sup:main}

\begin{table}[pos=h!]
\centering
\caption{Oracle budget frontier (sepsis, 21 candidate strategies, 10 seeds). The last column is the outcome gained per unit of realized cost over the preceding row.}
\label{tab:frontier}
\begin{tabular}{@{}rrrr@{}}
\toprule
Budget $B$ & Outcome (\%) & Realized cost & $\Delta$Outcome / $\Delta$Cost \\
\midrule
$1$ & $21.9 \pm 0.6$ & $0.88 \pm 0.01$ & \\
$2$ & $30.5 \pm 0.5$ & $1.95 \pm 0.03$ & $8.1$ \\
$4$ & $39.1 \pm 3.7$ & $3.46 \pm 0.55$ & $5.7$ \\
$6$ & $52.2 \pm 3.7$ & $5.53 \pm 0.46$ & $6.3$ \\
$8$ & $66.6 \pm 0.6$ & $7.21 \pm 0.07$ & $8.6$ \\
$12$ & $79.9 \pm 0.4$ & $11.37 \pm 0.09$ & $3.2$ \\
$20$ & $93.2 \pm 0.3$ & $18.49 \pm 0.15$ & $1.9$ \\
$\infty$ & $96.5$ & $32.73$ & $0.2$ \\
\bottomrule
\end{tabular}
\end{table}

\begin{table}[pos=h!]
\centering
\small
\caption{Mean uniform deviation $\sup_b \lvert \hat{F}_g - F_g \rvert$ as a fraction of the distribution-free bound $\delta(n)$, at $\eta = 0.05$. Candidate-set sizes are $21$, $23$, $25$ and $25$, so $\delta(n)$ differs across environments only in the third decimal.}
\label{tab:e2}
\begin{tabular}{@{}rrrrrr@{}}
\toprule
$n$ & $\delta(n)$ & Sepsis & Tumor & Semi-synthetic & Maintenance \\
\midrule
$50$ & $0.259$ & $0.718$ & $0.681$ & $0.717$ & $0.657$ \\
$100$ & $0.183$ & $0.721$ & $0.679$ & $0.715$ & $0.657$ \\
$200$ & $0.130$ & $0.722$ & $0.679$ & $0.710$ & $0.653$ \\
$500$ & $0.082$ & $0.720$ & $0.683$ & $0.716$ & $0.653$ \\
$1{,}000$ & $0.058$ & $0.721$ & $0.683$ & $0.718$ & $0.659$ \\
$2{,}000$ & $0.041$ & $0.717$ & $0.682$ & $0.711$ & $0.661$ \\
\bottomrule
\end{tabular}
\begin{flushleft}\footnotesize
$\delta(n)$ shown for the sepsis candidate set. Two thousand resamples per cell.
\end{flushleft}
\end{table}

We draw $n \in \{50, 100, 200, 500, 1000, 2000\}$ independent samples from the oracle cost distribution of every candidate, form $\hat{F}_g$ and apply the empirical rule, over $200$ resamples at each of $10$ seeds and at $\eta = 0.05$. This gives both the uniform deviation and the realized tail probability of the admitted strategies. Both the premise and the conclusion hold well inside their nominal level. The uniform deviation falls within $\delta(n)$ in $97.8\%$ to $99.5\%$ of resamples against a nominal $95\%$. The realized tail probability of the admitted strategies stays within the certified excess in $99.6\%$ to $100\%$.

The ratio is flat in $n$: within each environment it varies by less than one percentage point over the fortyfold range of sample size. It runs from $0.717$ to $0.722$ in sepsis and from $0.653$ to $0.661$ in the maintenance environment. A constant ratio confirms the rate: the observed deviation shrinks at exactly the $n^{-1/2}$ rate the bound asserts, and only the constant differs between environments.

The largest single deviation exceeds $\delta(n)$ in every environment, by factors between $1.11$ and
$1.39$. A bound holding with probability $1-\eta$ produces exactly this: at two thousand resamples roughly one hundred are permitted to exceed it, and the maximum over those is above it by construction. The guarantee therefore concerns the satisfaction rate, not the maximum. At the four largest sample sizes, the realized tail probability of the admitted strategies stays
within the certified excess in $100\%$ of resamples in three of the four environments.

\begin{table}[pos=h!]
\centering
\small
\caption{An estimated kernel in place of the exact one, sepsis environment (mean $\pm$ s.d., 10 seeds, 7 budgets). Both rules act on the same estimated quantities and are scored against the same mean-feasible oracle; the chance-constraint rule is at $\varepsilon = 0.2$. $\sup_b |\hat{F}_g - F_g|$ is the maximum over the $21$ strategies and $10$ seeds.}
\label{tab:pathb}
\begin{tabular}{@{}rrrrrrrr@{}}
\toprule
& & & \multicolumn{2}{c}{Point-estimate rule} & \multicolumn{3}{c}{Chance-constraint rule} \\
\cmidrule(lr){4-5}\cmidrule(lr){6-8}
Traj. & Cost bias (\%) & $\sup_b |\hat{F}_g - F_g|$ & Regret (pp) & Viol. (\%) & Regret (pp) & Viol. (\%) & Tail$/\varepsilon$ \\
\midrule
$1{,}000$   & $-9.03 \pm 9.71$ & $0.294$ & $4.85 \pm 4.45$ & $18.6$ & $16.37$ & $0.0$ & $0.697$ \\
$5{,}000$   & $+2.89 \pm 3.88$ & $0.097$ & $2.89 \pm 0.96$ & $2.9$  & $15.36$ & $0.0$ & $0.645$ \\
$20{,}000$  & $-0.48 \pm 1.01$ & $0.064$ & $0.72 \pm 1.12$ & $0.0$  & $15.65$ & $0.0$ & $0.628$ \\
$100{,}000$ & $-0.06 \pm 0.63$ & $0.037$ & $0.57 \pm 0.61$ & $0.0$  & $16.33$ & $0.0$ & $0.579$ \\
\bottomrule
\end{tabular}
\end{table}

\begin{table}[pos=h!]
\centering
\small
\caption{Robustness to benchmark construction (MIMIC-IV semi-synthetic at budget $B=12$; $15\%$ injected error; mean $\pm$ s.d., 10 instances). Difficulty set by the outcome-threshold quantile $q$; covariate rows use disjoint physiological groups.}
\label{tab:mimic-robust}
\begin{tabular}{@{}lrrrrr@{}}
\toprule
& \multicolumn{2}{c}{Naive} & \multicolumn{2}{c}{Chance ($\varepsilon=0.2$)} & \\
\cmidrule(lr){2-3}\cmidrule(lr){4-5}
Configuration & Regret & Viol. (\%) & Regret & Viol. (\%) & Zero-viol.\ regret \\
\midrule
$q = 0.25$ (baseline) & $3.4{\pm}1.0$ & $9.9$ & $12.6{\pm}1.7$ & $0.0$ & $8.9$ \\
$q = 0.15$ & $2.9{\pm}1.1$ & $13.2$ & $14.0{\pm}2.2$ & $0.0$ & $9.9$ \\
$q = 0.35$ & $3.3{\pm}0.7$ & $8.4$ & $10.4{\pm}1.3$ & $0.0$ & $7.5$ \\
Covariates: RR, MAP, creatinine & $3.5{\pm}1.1$ & $8.6$ & $12.6{\pm}1.8$ & $0.0$ & $9.0$ \\
Covariates: SpO$_2$, lactate, WBC & $3.7{\pm}0.8$ & $9.3$ & $13.0{\pm}1.4$ & $0.0$ & $9.5$ \\
\bottomrule
\end{tabular}
\begin{flushleft}\footnotesize
Regret in outcome percentage points. The last column is the lowest regret reached along the tolerance frontier while the violation remains at zero.
\end{flushleft}
\end{table}

\begin{table}[pos=h!]
\centering
\small
\caption{Target-based operating point at two violation targets $\rho$, averaged over the budget grid (mean${\pm}$s.d., 10 seeds; regret in outcome percentage points, violation in percent).}
\label{tab:operating-points}
\begin{adjustbox}{max width=\textwidth}
\begin{tabular}{@{}lcccccc@{}}
\toprule
& \multicolumn{3}{c}{$\rho \le 5\%$} & \multicolumn{3}{c}{$\rho \le 10\%$} \\
\cmidrule(lr){2-4} \cmidrule(lr){5-7}
Environment & $\varepsilon$ & Regret & Viol. & $\varepsilon$ & Regret & Viol. \\
\midrule
Sepsis                     & $0.36$ & $7.7{\pm}0.7$ & $1.7{\pm}1.5$ & $0.40$ & $4.9{\pm}0.7$ & $6.0{\pm}4.0$ \\
Tumor                      & $0.28$ & $2.3{\pm}0.6$ & $4.9{\pm}1.3$ & $0.30$ & $0.6{\pm}0.5$ & $9.7{\pm}1.9$ \\
Semi-synthetic (MIMIC-IV)  & $0.40$ & $5.7{\pm}0.6$ & $0.8{\pm}0.8$ & $0.45$ & $3.6{\pm}0.6$ & $8.2{\pm}4.9$ \\
Maintenance (C-MAPSS)      & $0.22$ & $9.4{\pm}0.2$ & $1.0{\pm}0.4$ & $0.25$ & $9.4{\pm}0.2$ & $5.3{\pm}1.3$ \\
\bottomrule
\end{tabular}
\end{adjustbox}
\end{table}

\section{The maintenance environment at a budget of three}
\label{app:maintenance}

Section~\ref{sec:res-cvar} reports that the whole matched-tail gap in the maintenance
environment comes from a single binary choice at a budget of three. The per-budget breakdown
accounts for it.

The averaged regret is the mean of the per-budget regrets of Table~\ref{tab:perbudget}. At a tolerance of $0.30$ the budget of three contributes $56.24$ and the four larger budgets contribute $0.08$ between them. Averaging over the six budgets that admit a strategy therefore gives $9.39$, the observed value at that tolerance and the top of the plateau running from $9.33$ to $9.39$. Below $0.30$ a budget of two admits nothing and five budgets are averaged
instead, raising the figure to $11.27$ against an observed $11.27$ to $11.31$. The plateau
rises across that range because the residual at the larger budgets grows as the tolerance
tightens, reaching $0.70$ at a tolerance of $0.20$. Setting that residual to zero and reading
the average as one error divided by the number of budgets would understate it by the same
amount.

The tolerance at which the rule begins to admit \texttt{t0.65} is displaced from its exact
value by estimation noise. Under the exact cost distribution \texttt{t0.65} exceeds a budget of three with probability $0.248$, so a rule with perfect information would admit it above that tolerance. The observed regret instead collapses between $0.40$ and $0.45$. A fraction $0.371$
of \texttt{t0.65}'s cumulative cost falls at exactly three crew-hours, where the $15\%$
multiplicative perturbation of Section~\ref{sec:exp-noise} pushes it above the budget half
the time. The tail the rule estimates is therefore about $0.248 + 0.371/2 = 0.43$, and the
observed transition brackets that value. Discretely valued cost concentrated at the budget
makes the rule conservative, the direction Proposition~3 permits.

\begin{table}[pos=h!]
\centering
\small
\caption{Per-budget regret in the maintenance environment, chance constraint at $\varepsilon = 0.30$ against the conditional-value-at-risk constraint at $\varepsilon = 1$.}
\label{tab:perbudget}
\begin{tabular}{@{}lcccccc@{}}
\toprule
Budget & 2 & 3 & 4 & 5 & 6 & 7 \\
\midrule
Mean-feasible strategies & 1 & 2 & 3 & 4 & 5 & 5 \\
Chance-constraint regret (pp) & 0.00 & 56.24 & 0.05 & 0.02 & 0.01 & 0.01 \\
CVaR regret (pp) & 0.00 & 2.14 & 0.06 & 0.03 & 0.02 & 0.01 \\
Realized tail, chance & 0.080 & 0.006 & 0.088 & 0.168 & 0.168 & 0.225 \\
Realized tail, CVaR & 0.080 & 0.238 & 0.072 & 0.106 & 0.101 & 0.160 \\
\bottomrule
\end{tabular}
\begin{flushleft}\footnotesize
Ten seeds. The two operating points carry almost the same averaged tail, $0.122$ and $0.126$, and allocate it differently across budgets.
\end{flushleft}
\end{table}

Section~\ref{sec:res-cvar} also reports how often each rule declines to recommend. In sepsis and tumor, both always recommend. In the other two, the conditional-value-at-risk rule declines more often at every tolerance, in $8.6\%$ of cases against $2.9\%$ at $\varepsilon = 0.05$ in the semi-synthetic environment and $32.6\%$ against $20.7\%$ in the maintenance environment.

\section{Mis-specification of the cost spread}
\label{app:spread}

Section~\ref{sec:res-predictor} perturbs the level of the estimated cost. This sweep perturbs its spread. It rescales each strategy's oracle cost samples about their own mean by a common factor $s$, selecting on the rescaled samples and evaluating on the true ones. Ten seeds, $2{,}000$ repetitions per budget, $200$ cost draws per repetition, $\varepsilon = 0.2$, and the same budget grids, candidate sets and injected error as the level sweep.

Table~\ref{tab:spread} reports the sweep for $s \le 1$. Cost is bounded below at zero, so the rescaling is clipped there, and for $s > 1$ the clip raises the realized mean by between $8$ and $76$ percent. The point-estimate rule's violation does fall on that side, but it falls because the mean it reads has been inflated and not because it is insensitive to the spread. Those rows therefore do not measure what the sweep is for. The construction cannot raise the spread while preserving the mean on a non-negative cost. Overstating the spread also moves the chance-constraint rule toward conservatism, which the realized tail on that side confirms, so understatement is the direction that bears on the guarantee.

Levels here are not directly comparable with Table~\ref{tab:three-env}. This sweep pools individual recommendations across the budget grid, so each budget carries weight in proportion to the draws at which the rule admits. The main text instead averages the per-budget means equally. The two weightings agree wherever every budget admits in every draw. That covers all four environments except the maintenance budget of two, where the chance rule admits in about half the draws. The maintenance level here is consequently about one point higher. Comparisons across $s$ within the table are unaffected, since the budget set and the weighting are the same at every $s$.

\begin{table}[pos=h!]
\centering
\small
\caption{Cost spread mis-specification, $s \le 1$ ($\varepsilon = 0.2$, 10 seeds).}
\label{tab:spread}
\begin{tabular}{@{}lrrrrrr@{}}
\toprule
& & \multicolumn{2}{c}{Point estimate} & \multicolumn{3}{c}{Chance constraint} \\
\cmidrule(lr){3-4}\cmidrule(lr){5-7}
Environment & $s$ & Reg. & Viol. & Reg. & Viol. & Realized tail \\
\midrule
Sepsis         & 0.50 & 2.50 &  7.91 & 11.24 & 0.00 & 0.184 \\
               & 0.70 & 2.50 &  7.91 & 13.73 & 0.00 & 0.136 \\
               & 0.85 & 2.50 &  7.91 & 15.56 & 0.00 & 0.112 \\
               & 1.00 & 2.50 &  7.91 & 16.92 & 0.00 & 0.095 \\
\midrule
Tumor          & 0.50 & 0.52 & 15.00 &  5.64 & 0.08 & 0.154 \\
               & 0.70 & 0.52 & 15.00 &  7.55 & 0.03 & 0.134 \\
               & 0.85 & 0.52 & 15.00 &  9.38 & 0.02 & 0.119 \\
               & 1.00 & 0.52 & 15.00 & 11.14 & 0.02 & 0.105 \\
\midrule
Semi-synthetic & 0.50 & 3.50 &  7.22 &  9.68 & 0.00 & 0.206 \\
               & 0.70 & 3.50 &  7.22 & 11.96 & 0.00 & 0.164 \\
               & 0.85 & 3.50 &  7.22 & 13.51 & 0.00 & 0.138 \\
               & 1.00 & 3.50 &  7.22 & 14.94 & 0.00 & 0.115 \\
\midrule
Maintenance    & 0.50 & 0.39 &  5.23 &  9.39 & 0.96 & 0.075 \\
               & 0.70 & 0.39 &  5.23 &  9.41 & 0.48 & 0.070 \\
               & 0.85 & 0.39 &  5.23 &  9.44 & 0.31 & 0.066 \\
               & 1.00 & 0.39 &  5.23 & 10.38 & 0.24 & 0.061 \\
\bottomrule
\end{tabular}
\begin{flushleft}\footnotesize
Regret in outcome percentage points, violation in percent. The realized mean is preserved exactly at every row, so the point-estimate rule's admitted set is identical across $s$ within each environment.
\end{flushleft}
\end{table}

\section{Function approximation on the continuous-state environments}
\label{app:funcapprox}
\label{app:tumor-fa}
\subsection{Tumor}

Two estimators of different flexibility are fit by g-computation on observational tumor
trajectories, and both decision rules act on the resulting estimates. The oracle is mean-feasible
and scores both. Six budgets and $10$ seeds give $60$ observations per cell, reported in Table~\ref{tab:tumor-fa}. Cost bias is the mean over strategies of the relative error in each strategy's estimated mean cost.

\begin{table}[pos=h!]
\centering
\small
\caption{Two estimators under function approximation, tumor environment (mean over 10 seeds; chance-constraint rule at $\varepsilon = 0.2$). The deviation column is the largest, over strategies, of each strategy's mean deviation over seeds. Per-seed maxima are not retained for this environment, so the column understates the supremum. The certified slack is $\delta = 0.131$.}
\label{tab:tumor-fa}
\begin{tabular}{@{}llrrrrrrr@{}}
\toprule
& & & \multicolumn{2}{c}{Deviation} & \multicolumn{2}{c}{Point-estimate rule} & \multicolumn{2}{c}{Chance rule} \\
\cmidrule(lr){4-5}\cmidrule(lr){6-7}\cmidrule(lr){8-9}
Traj. & Estimator & Cost bias (\%) & Largest & $> \delta$ & Regret (pp) & Viol. (\%) & Regret (pp) & Viol. (\%) \\
\midrule
$1{,}000$   & ensemble & $-12.56$ & $0.210$ & $5/23$ & $0.59$ & $15.0$ & $10.34$ & $0.0$ \\
            & ridge    & $+9.25$  & $0.107$ & $0/23$ & $0.73$ & $0.0$  & $14.06$ & $0.0$ \\
$5{,}000$   & ensemble & $-13.24$ & $0.226$ & $5/23$ & $0.73$ & $18.3$ & $9.39$  & $0.0$ \\
            & ridge    & $+8.57$  & $0.103$ & $0/23$ & $0.73$ & $0.0$  & $13.92$ & $0.0$ \\
$20{,}000$  & ensemble & $-14.30$ & $0.252$ & $6/23$ & $0.73$ & $10.0$ & $8.10$  & $0.0$ \\
            & ridge    & $+8.86$  & $0.103$ & $0/23$ & $0.73$ & $0.0$  & $13.92$ & $0.0$ \\
$100{,}000$ & ensemble & $-14.73$ & $0.276$ & $7/23$ & $0.73$ & $16.7$ & $10.04$ & $0.0$ \\
            & ridge    & $+8.97$  & $0.104$ & $0/23$ & $0.73$ & $0.0$  & $14.36$ & $0.0$ \\
\bottomrule
\end{tabular}
\end{table}

The rank correlation between estimated and exact mean cost is $1.000$ for both estimators at every sample size. The estimators therefore disagree on the level of the cost, not on the ordering.
The certified slack at this cost-sample size and candidate-set size is $\delta = 0.131$. The ridge stays below it at every strategy. The ensemble exceeds it at five to seven of the $23$, entirely at the strategies where its cost bias is largest.

\subsection{Semi-synthetic}
\label{app:mimic-fa}

Table~\ref{tab:mimic-fa} reports the same estimators on the semi-synthetic environment, where the cost-underestimate band is the tightest of the four. The certified slack is $\delta = 0.131$ and the certified ceiling
$\varepsilon + \delta = 0.331$. The deviation column is again the largest per-strategy value and
understates the supremum. This grid runs to $20{,}000$ trajectories rather than $100{,}000$.

\begin{table}[pos=h!]
\centering
\small
\caption{Two estimators under function approximation, semi-synthetic environment (mean over 10 seeds; chance-constraint rule at $\varepsilon = 0.2$). Realized tail is that of the strategy the chance rule selects, on oracle cost samples.}
\label{tab:mimic-fa}
\begin{tabular}{@{}llrrrrrrr@{}}
\toprule
& & & \multicolumn{2}{c}{Deviation} & \multicolumn{2}{c}{Point-estimate rule} & \multicolumn{2}{c}{Chance rule} \\
\cmidrule(lr){4-5}\cmidrule(lr){6-7}\cmidrule(lr){8-9}
Traj. & Estimator & Cost bias (\%) & Largest & $> \delta$ & Regret (pp) & Viol. (\%) & Viol. (\%) & Realized tail \\
\midrule
$1{,}000$  & ensemble & $+2.17$ & $0.150$ & $4/25$  & $2.62$  & $6.7$  & $0.0$  & $0.171$ \\
           & ridge    & $-5.30$ & $0.260$ & $16/25$ & $-5.73$ & $95.0$ & $15.0$ & $0.329$ \\
$5{,}000$  & ensemble & $+2.52$ & $0.147$ & $4/25$  & $2.43$  & $6.7$  & $0.0$  & $0.168$ \\
           & ridge    & $-4.99$ & $0.262$ & $15/25$ & $-5.55$ & $93.3$ & $13.3$ & $0.323$ \\
$20{,}000$ & ensemble & $+2.51$ & $0.147$ & $4/25$  & $2.56$  & $6.7$  & $0.0$  & $0.169$ \\
           & ridge    & $-5.30$ & $0.261$ & $16/25$ & $-5.67$ & $96.7$ & $11.7$ & $0.324$ \\
\bottomrule
\end{tabular}
\end{table}

The ridge aggregate bias of $-5.3\%$ is the mean of per-strategy biases running from $-28\%$ to
$+79\%$, which cancel. Its cost error does not track how well the observational data covers the states a strategy visits.
The coverage gap of a strategy is the fraction of the discretized state-action cells it traverses in which the behavior policy generated no samples. The cost error of a strategy is the ratio of its estimated to its true mean cost, minus one, in absolute value. Under $k$-means discretization at $b=5$, the Spearman correlation between the absolute relative cost error and the coverage gap is $0.05$ for the ridge and $0.75$ for the ensemble. Repeating it under uniform and quantile discretization at the same cell count separates the two estimators by sign rather than by magnitude. The ensemble stays positive at $0.81$ and $0.33$, while the ridge moves to $0.26$ and $-0.20$. Coverage predicts where the ensemble errs under every partition and predicts nothing about
the ridge under any. The association is in rank rather than in level, since a few strategies at the sparsely covered ends of the grid account for it.
The two estimators exceed the slack on different strategies, at a Jaccard index of $0.18$. The ridge's worst strategies sit in the best-covered region of the grid.

\section{Safety-utility frontiers in the remaining environments}
\label{app:frontiers}

\begin{figure}[pos=h!]
\centering
\includegraphics[width=\linewidth,height=0.45\textheight,keepaspectratio]{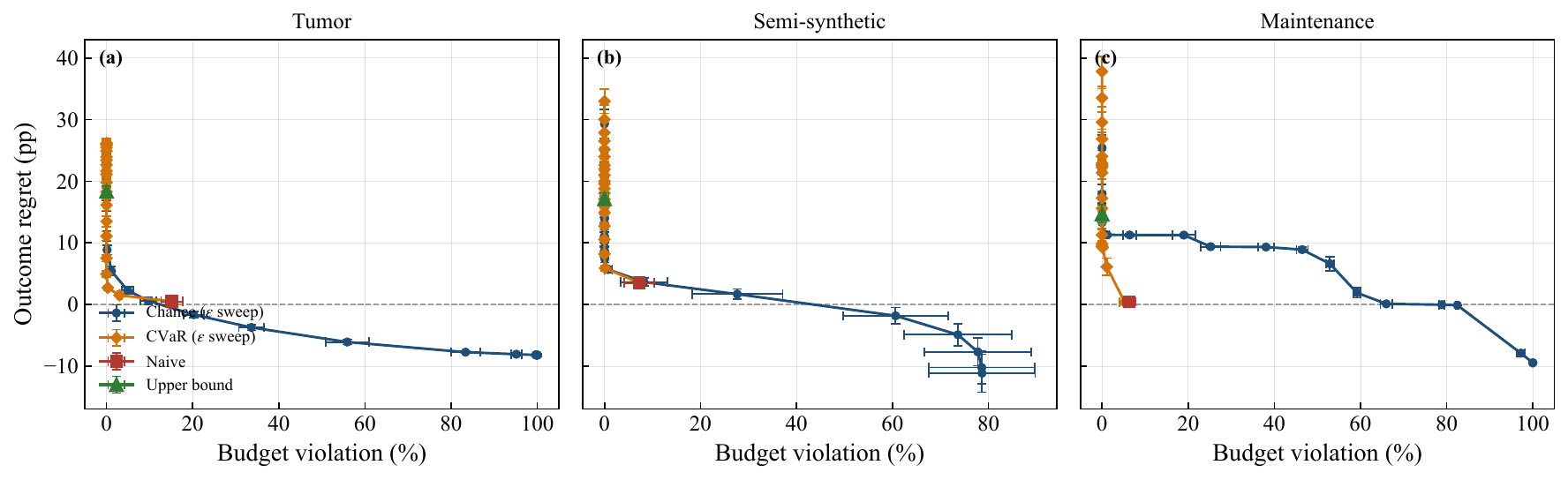}
\caption{Safety-utility frontiers outside the sepsis environment, traced by sweeping the tolerated violation probability $\varepsilon$ under $15\%$ injected estimation error, with the naive and upper-bound rules as single points for reference. The exchange is monotone in all three. The tumor and maintenance frontiers step rather than curve, because most of their candidates are tied at a single outcome value. (a)~Tumor, 10 seeds; (b)~Semi-synthetic, 10 benchmark instances; (c)~Maintenance, 10 instances.}
\label{fig:env-frontiers}
\end{figure}

Figure~\ref{fig:env-frontiers} traces the frontier of Section~\ref{sec:res-generalization} in the three environments other than sepsis. Each panel is obtained by sweeping the tolerated violation probability $\varepsilon$ under $15\%$ injected estimation error, with the naive and upper-bound rules placed as reference points.

\section{Candidate-set placement}
\label{app:placement}

The candidate sets in the simulators are solved rather than enumerated. For the sepsis simulator, value iteration is applied with a per-step penalty $\lambda$ on the quantity of treatment administered. The solved policy therefore maximizes survival net of a treatment cost weighted by $\lambda$.
At $\lambda=0$, the policy maximizes survival without regard to treatment cost, recovering the
aggressive end of the spectrum, while large $\lambda$ yields conservative policies that treat
sparingly. The discount factor is $0.99$, iteration runs to convergence, and absorbing states are
treated as self-absorbing during value iteration. The candidates can then be spaced along the cost axis in three ways, by even spacing in $\lambda$, by even spacing in target cost, or by a denser placement. The rest of this appendix reports what turns on that choice.

Section~\ref{sec:res-rules} reports the decision rules on a candidate set of $21$ value-iteration strategies. This appendix repeats the comparison on the seven-strategy set a deployment would more plausibly enumerate, holding the budget grid, the injected error, the tolerance and the $10$ seeds fixed. The only difference is then how densely the space of strategies is populated. Table~\ref{tab:density} reports the comparison.

\begin{table}[pos=h!]
\centering
\caption{Decision rules under two candidate-set constructions (sepsis, $15\%$ injected error, $\varepsilon=0.2$; mean $\pm$ s.d., 10 seeds). Regret is measured against the oracle optimum of the same construction.}
\label{tab:density}
\begin{tabular}{@{}lrrrrr@{}}
\toprule
Construction & Strategies & Oracle outcome (\%) & Naive regret & Naive viol. (\%) & Chance regret \\
\midrule
Practice-sized & $7$ & $46.2{\pm}0.4$ & $0.4{\pm}0.4$ & $4.6{\pm}2.4$ & $14.3{\pm}0.5$ \\
Dense & $21$ & $54.8{\pm}0.7$ & $2.5{\pm}0.7$ & $8.0{\pm}3.5$ & $16.9{\pm}0.7$ \\
\bottomrule
\end{tabular}
\begin{flushleft}\footnotesize
The chance-constraint rule violates the budget in $0.00\%$ of cases under both constructions. The dense construction yields $21$ distinct strategies from $25$ penalty values.
\end{flushleft}
\end{table}

Three readings follow. The first is that the separation between the rules survives both constructions. The naive rule overruns the budget under each, and the chance-constraint rule holds at exactly zero under each. The result of Section~\ref{sec:res-rules} is therefore not an artifact of how many strategies are enumerated.

The second is that density raises the point-estimate rule's exposure rather than lowering it. Its violation rises from $4.6\%$ to $8.0\%$ as the set grows. A denser set contains more strategies with a cost distribution that places its mean inside the budget and a substantial part of its mass outside. Each of those is an opportunity for a point estimate to admit a strategy that a tail constraint would refuse.

The third concerns how the regret figures should be read. The chance-constraint rule's regret rises from $14.3$ to $16.9$ points across the two constructions, which taken alone would suggest the rule performs worse on the denser set. It does not. Regret is measured against the oracle optimum of the same construction, and that optimum rises from $46.2\%$ to $54.8\%$. In absolute terms, the strategy the chance-constraint rule selects improves from $31.8\%$ to $37.9\%$. The denser set does not refine the same operating points; it contains genuinely better ones, and the rule finds them. The benchmark improves faster than the recommendation does.

\section{Estimated-kernel diagnostics, sepsis}
\label{app:pathb-diag}

This appendix records supporting detail for the estimated-kernel block of Section~\ref{sec:res-predictor}.

The falling unvisited fraction of Section~\ref{sec:res-predictor} shows Assumption 2 failing at small samples and holding at larger ones. The assumption asks for positive probability on the treatments a strategy may assign along histories that can occur, not for coverage of the state space. At a hundred thousand trajectories $58.8\%$ of state-action pairs remain unvisited, and only $0.04\%$ of transitions in the evaluation rollouts are drawn from them. The unvisited region lies outside the candidate strategies' support, and the assumption requires that and no more.

The deviation column of Table~\ref{tab:pathb} is the $\delta$ of Proposition~3, measured under model error rather than assumed. At the cost-sample size used throughout Section~\ref{sec:results}, the certified slack is $\delta = 0.1297$. From $5{,}000$ trajectories the measured deviation stays below it at every strategy and every seed, so model error does not dominate the sampling slack the guarantee already carries. At $1{,}000$ trajectories it exceeds the slack at $17$ of the $21$ strategies, and the $\varepsilon + \delta$ ceiling does not cover that regime. The two terms combine rather than substitute. A deployment that estimates the cost distribution from a fitted model and then evaluates it on $n$ draws carries both. Proposition~3 then applies with $\delta$ replaced by their sum, the sampling slack of $0.1297$ at $200$ draws plus the measured model deviation of $0.0638$ at $20{,}000$ trajectories. Section~\ref{sec:res-feasibility} measures the sampling term on its own, at a mean of $0.72\,\delta$ in this environment.

\begin{figure}[pos=h!]
\centering
\includegraphics[width=0.78\linewidth,height=0.38\textheight,keepaspectratio]{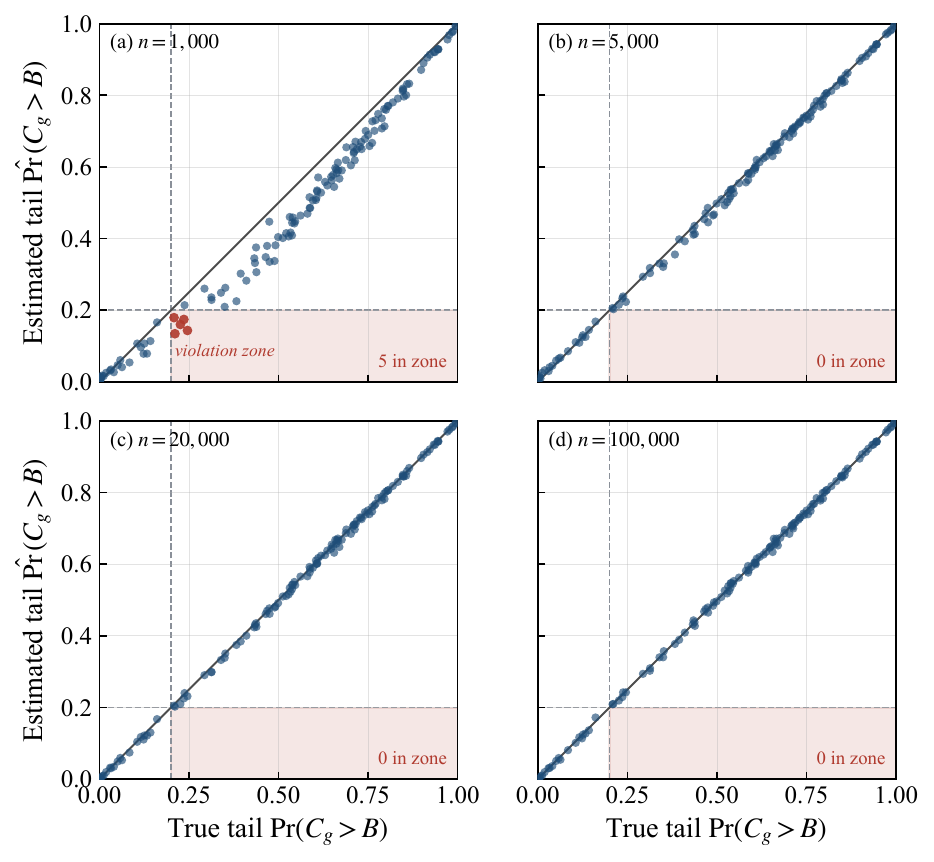}
\caption{Estimated against exact tail probability of the cumulative cost, for each of the $21$ candidate strategies at each budget, under a kernel estimated from observational trajectories. Points in the shaded quadrant (red) have an estimated tail inside the tolerance $\varepsilon = 0.2$ and a true tail outside it. That quadrant is the only route by which the chance constraint admits an infeasible strategy. The quadrant empties between $1{,}000$ and $5{,}000$ trajectories. (a)~$n=1{,}000$; (b)~$n=5{,}000$; (c)~$n=20{,}000$; (d)~$n=100{,}000$.}
\label{fig:pathb-tail}
\end{figure}

Figure~\ref{fig:pathb-tail} shows where the error matters. Points below the identity line are conservative: the rule judges a strategy riskier than it is and refuses it, which costs outcome and breaches nothing. Only the shaded quadrant produces a violation. It contains five of the $147$ strategy-budget pairs at $1{,}000$ trajectories and none at any larger size. The five pairs sit at exact tail probabilities between $0.21$ and $0.25$ against estimated values between $0.13$ and $0.18$. The failure at small samples is therefore a shallow underestimate near the tolerance rather than a gross misordering.

One comparison qualifies the injected-noise design. At $1{,}000$ trajectories the estimated kernel biases cost by $-9.03\%$ on average, and a uniform injected underestimate of $10\%$ leaves the point-estimate rule's violation at zero. The estimator produces $18.6\%$. A fitted model errs unevenly across strategies, with a standard deviation of $9.71$ points around that mean. It errs on the outcome as well, and it distorts the shape of the cost distribution rather than rescaling it. A uniform multiplicative bias of the same average magnitude is therefore an optimistic stand-in. The sweep in Figure~\ref{fig:predictor}(b) locates the breaking point of the rule rather than of a deployment.

The estimator errs in the level of cost and not in its order: the rank correlation between estimated and exact mean cost is $0.999$ at $1{,}000$ trajectories and $1.000$ at every larger size. The level is biased by $-9.03\%$.

\section{Validation of the tabular kernel rollout}
\label{app:backbone}

The estimated-predictor experiment of Section~\ref{sec:res-predictor} rolls candidate strategies forward under a transition kernel estimated by counting. Before any estimated number was produced, the vectorized kernel rollout was checked against the original object-based rollout of the simulator. Two defects were found and corrected in the course of that check, and both would have gone unnoticed in aggregate summaries.

The first is the hidden mixture component. The marginal kernel formed by weighting the two component transition matrices by their mixture probabilities is not the generative process. The latent binary attribute is drawn once per episode and then held fixed. The two components differ by as much as $0.51$ in transition probability. Rolling out under the marginal kernel understated the outcome value by up to $64$ percentage points.

The second is terminal scoring. The simulator terminates an episode on a nonzero reward and treats episodes that reach the horizon without entering an absorbing state as favorable. Scoring the outcome only at absorbing states misclassified every horizon-reaching episode as unfavorable.

After both corrections, Table~\ref{tab:backbone} compares the two implementations at three to four thousand rollouts per strategy.

\begin{table}[pos=h!]
\centering
\caption{Vectorized kernel rollout against the original object-based rollout.}
\label{tab:backbone}
\begin{tabular}{@{}lrrrr@{}}
\toprule
Strategy & Outcome, original & Outcome, vectorized & Cost, original & Cost, vectorized \\
\midrule
\texttt{vi\_l0.000} & $0.9637$ & $0.9595$ & $33.587$ & $34.566$ \\
\texttt{vi\_l0.050} & $0.6428$ & $0.6295$ & $6.898$ & $6.817$ \\
\texttt{vi\_l0.200} & $0.2142$ & $0.2110$ & $0.899$ & $0.838$ \\
\bottomrule
\end{tabular}
\end{table}

The outcome values agree within two standard errors for every strategy. A six-seed repeat at \texttt{vi\_l0.000} gives a cost difference of $+0.156$ against a two-standard-error width of $0.379$, so the cost agreement is within noise. A residual gap of roughly half a percentage point in the outcome persists because the tabular kernel marginalizes the simulator's factored dynamics. That is a property of the tabular representation and not of the estimator.

The observational trajectories used to fit the kernel are generated under a behavior policy with treatment intensity increasing with the number of abnormal physiological variables. This reproduces the confounding pattern that motivates the identification assumptions of Section~\ref{sec:method-setup}. Its mean treatment intensity is $1.601$ interventions per episode.

\section{Importance-weight diagnostics on the real-outcome environment}
\label{app:ess}

Table~\ref{tab:ess} reports, for each candidate policy in the Drink Less environment, the effective sample size and the largest importance weight of the value estimate. The randomization probabilities are fixed by the trial design and the candidate policies are softened away from the deterministic boundary. The weights are therefore bounded above by the ratio of the largest attainable candidate probability to the smallest trial probability. The observed maximum of $2.375$ attains that bound, and no policy approaches degeneracy. The cost distribution is not estimated by importance weighting: it is accumulated directly from each participant's observed state sequence under the candidate policy. The chance constraint therefore does not inherit the weight variance that the value estimate carries.

The state-aware policies sit further from the trial's randomization probabilities than the global ones at comparable cost. This lowers their effective sample size and raises their value at the same time. The policy the chance-constraint rule selects at $B=15$ retains an effective sample size of $58.9\%$ of the decision points. The selection is stable: across the $50$ selection-estimation splits of Section~\ref{sec:res-realdata} it is chosen every time, with a cross-fitted value of $2.627$, matching the full-sample value.

\begin{table}[pos=h!]
\centering
\small
\caption{Per-policy importance-weight diagnostics, Drink Less environment ($10{,}470$ decision points, $349$ participants). $^{\ast}$ marks the policy selected by the chance-constraint rule at $B=15$, $\varepsilon=0.2$.}
\label{tab:ess}
\begin{tabular}{@{}lrrrrr@{}}
\toprule
Policy & Mean cost & Value & ESS & ESS (\%) & Max weight \\
\midrule
global\_r0.0 & 1.5 & 1.23 & 4654 & 44.5 & 2.375 \\
global\_r0.1 & 3.0 & 1.36 & 5148 & 49.2 & 2.250 \\
global\_r0.2 & 6.0 & 1.63 & 6296 & 60.1 & 2.000 \\
global\_r0.3 & 9.1 & 1.90 & 7623 & 72.8 & 1.750 \\
global\_r0.4 & 12.1 & 2.17 & 8977 & 85.7 & 1.500 \\
global\_r0.5 & 15.1 & 2.44 & 10051 & 96.0 & 1.250 \\
global\_r0.6 & 18.0 & 2.70 & 10470 & 100.0 & 1.000 \\
global\_r0.7 & 21.0 & 2.97 & 10050 & 96.0 & 1.167 \\
global\_r0.8 & 24.0 & 3.24 & 8967 & 85.6 & 1.333 \\
global\_r0.9 & 27.0 & 3.51 & 7599 & 72.6 & 1.500 \\
global\_r1.0 & 28.5 & 3.64 & 6912 & 66.0 & 1.583 \\
\midrule
aware\_h0.8\_l0.0$^{\ast}$ & 12.8 & 2.63 & 6168 & 58.9 & 2.375 \\
aware\_h0.8\_l0.2 & 15.1 & 2.75 & 7427 & 70.9 & 2.000 \\
aware\_h1.0\_l0.0 & 15.1 & 2.91 & 5617 & 53.6 & 2.375 \\
aware\_h1.0\_l0.2 & 17.4 & 3.03 & 6636 & 63.4 & 2.000 \\
aware\_h0.8\_l0.4 & 18.1 & 2.91 & 8985 & 85.8 & 1.500 \\
aware\_h1.0\_l0.4 & 20.3 & 3.19 & 7841 & 74.9 & 1.583 \\
\bottomrule
\end{tabular}
\end{table}

\section{Alternative formulations and the distributionally robust form}
\label{app:dro}

A worst-case constraint would require every realized cost to stay within budget. An overrun, however, is a resource event rather than a catastrophe, and the worst realized cost is an unstable target in finite samples. A multi-objective formulation would return a Pareto set, where sweeping $\varepsilon$ traces the same family while keeping the budget as a fixed ceiling.

The distributionally robust form is closer. Suppose the constraint is imposed on the worst case over the ambiguity set $\mathcal{P}_r(\hat{F}_g) = \{F : \lVert F - \hat{F}_g \rVert_\infty \le r\}$ of distributions within Kolmogorov distance $r$ of $\hat{F}_g$, rather than on $\hat{F}_g$ itself. Every such $F$ satisfies $F(B) \ge \hat{F}_g(B) - r$. The bound is attained by the distribution that moves mass $r$ from just below the budget to just above it, so the robust constraint at tolerance $\varepsilon$ reduces to
\begin{equation}
\sup_{F \in \mathcal{P}_r(\hat{F}_g)} \big(1 - F(B)\big) = 1 - \hat{F}_g(B) + r \le \varepsilon
\;\Longleftrightarrow\;
1 - \hat{F}_g(B) \le \varepsilon - r,
\label{eq:droequiv}
\end{equation}
the empirical rule of Equation~\eqref{eq:cc} at the tightened tolerance $\varepsilon - r$. The robust rule is therefore the same rule at a different operating point. Sweeping $\varepsilon$ traces the robust frontier along with the nominal one, and the choice of radius becomes a choice of position on a curve the method already reports. Taking $r = \delta$ recovers Proposition~3 of the main text and its certified ceiling of $\varepsilon + \delta$. The equivalence is specific to an ambiguity set defined in the supremum norm. Balls defined in a transport distance do not reduce to a shift of the tolerance, because they can move mass across the budget by an unbounded amount.

\section{Predictor ceiling: the estimation share of the regret}
\label{app:ceiling}

Section~\ref{sec:res-predictor} reports how much of the chance-constraint rule's regret is attributable to the estimated predictor rather than to the constraint. The comparison runs the same rule on three cost distributions per strategy and differences the results within seed. The estimated arm is the predictor's own distribution and reproduces the reported numbers. The oracle arm is the exact counterfactual distribution, which bounds what any predictor could deliver to the rule. Nothing trained for the decision can improve on the true cost law. The calibrated arm rescales the estimated distribution by a single pooled factor per environment and seed so that its mean matches the truth. It stands in for what a decision-aware level correction could learn. The factor is pooled over candidate strategies and never fitted per strategy. A per-strategy factor would use oracle information, so the arm would no longer stand in for a learnable correction. Feasibility is evaluated against the true cost samples in every arm. A budget-by-tolerance cell is dropped from all three arms whenever any one of them admits nothing, so the arms are always compared on the same cells.

In sepsis, the estimated arm is ahead of the oracle arm at every tolerance up to $0.3$. Its regret is below the exact distribution's by $0.77$ points at a tolerance of $0.2$, with a paired bootstrap interval of $[-1.06, -0.51]$ over $10$ seeds. At the tighter tolerances the gap is $0.29$ to $1.01$ points, with budget violation at zero throughout. A predictor with no error would make the rule more conservative here rather than less. Estimation therefore does not separate the rule from the benchmark in this environment, and integrated training would have nothing to recover.

Under function approximation, the picture reverses. The exact distribution lowers the regret by $2.14$, $4.29$ and $4.76$ points against the ensemble at tolerances of $0.05$, $0.10$ and $0.20$. Against the ridge it lowers the regret by $2.40$, $5.12$ and $9.95$, every interval excluding zero. For the ensemble, five of the thirty budget-by-tolerance cells are undefined in $9$ of the $10$ seeds because one of the arms admits nothing there. The ensemble comparison therefore runs over the remaining $255$ cell-seed pairs.

Rescaling each estimated distribution by a single pooled factor so that its mean matches the truth leaves the ridge selecting the same strategy in every cell. The recovered share is zero. It moves the ensemble the wrong way, raising its regret by $11.51$ points at a tolerance of $0.2$. There its bias is an underestimate of $14$ percent, and the corrected distributions lead the rule to select cheaper strategies with lower outcome. The factors are $0.93$ to $0.95$ and $1.18$ to $1.36$, the sizes the measured biases imply.

\end{document}